\documentclass[11pt,epsf,a4paper]{article}
\DeclareMathAlphabet{\scr}{U}{rsfs}{m}{n}
\usepackage{latexsym,color}
\usepackage{epsfig}
\usepackage[mathscr]{eucal}
\usepackage{amsfonts}
\usepackage{amscd}
\usepackage{cite}
\usepackage{array}
\usepackage{amssymb}
\usepackage{colordvi}
\usepackage[centertags]{amsmath}
\usepackage{enumerate}
\usepackage{graphicx}
\usepackage{booktabs}
\usepackage{theorem}
\usepackage[footnotesize]{caption}
\usepackage{soul}
\usepackage{ulem}

\newcommand{\lsim}{\raisebox{-0.13cm}{~\shortstack{$<$ \\[-0.07cm] $\sim$}}~}

\newcommand{\newc}{\newcommand}
\newc{\be}{\begin{equation}}
\newc{\ee}{\end{equation}}
\newc{\bea}{\begin{eqnarray}}
\newc{\eea}{\end{eqnarray}}
\newc{\ol}{\overline}
\newc{\wt}{\widetilde}
\newc{\bs}{\boldsymbol}
\newc{\m}{\mathcal}
\newc{\la}{\langle}
\newc{\ra}{\rangle}

\newcommand{\non}{\nonumber}
\newcommand{\beq}{\begin{eqnarray}}
\newcommand{\eeq}{\end{eqnarray}}
\newcommand{\s}{\smallskip}

\newcommand{\bpmatrix}{\begin{pmatrix}}
\newcommand{\epmatrix}{\end{pmatrix}}
\newcommand{\diag}{\text{diag}}

\newcommand{\bc}{\begin{center}}
\newcommand{\ec}{\end{center}}

\newcommand{\ba}{\begin{array}}
\newcommand{\ea}{\end{array}}

\begin{document}

\title{
\vspace*{-3cm}
\phantom{h} \hfill\mbox{\small ADP-15-26-T928}\\[-1.1cm]
\phantom{h} \hfill\mbox{\small KA-TP-17-2015}
\\[1cm]
\textbf{Exploring the CP-violating NMSSM:\\
EDM Constraints and Phenomenology}}

\date{}
\author{
S.~F.~King$^{1\,}$\footnote{E-mail: \texttt{king@soton.ac.uk}},
M. M\"{u}hlleitner$^{2\,}$\footnote{E-mail: \texttt{margarete.muehlleitner@kit.edu}},
R.~Nevzorov$^{3\,}$\footnote{E-mail: \texttt{roman.nevzorov@adelaide.edu.au}},
K. Walz$^{2\,}$\footnote{E-mail: \texttt{kathrin.walz@kit.edu}}
\\[9mm]
{\small\it
$^1$Physics and Astronomy,
University of Southampton,}\\
{\small\it Southampton, SO17 1BJ, U.K.}\\[3mm]
{\small\it
$^2$Institute for Theoretical Physics, Karlsruhe Institute of Technology,} \\
{\small\it 76128 Karlsruhe, Germany.}\\[3mm]
{\small\it
$^3$ARC Centre of Excellence for Particle Physics at the Tera-scale,
} \\   
{\small \it School of Chemistry and Physics, University of Adelaide,
Adelaide, SA 5005, Australia.} \\
}

\maketitle

\begin{abstract}
The Next-to-Minimal Supersymmetric extension of the Standard Model
(NMSSM) features extra new sources for CP violation. In
contrast to the MSSM CP violation can already occur at tree level in
the Higgs sector. We investigate the range of possible allowed
CP-violating phases by taking into account the constraints arising
from the measurements of the Electric Dipole Moments (EDMs) and the
latest LHC Higgs data. Our analysis shows that large CP-violating
phases, that are NMSSM-specific, are not in conflict with the
EDMs. They are dominantly constrained by the Higgs data in this
case. We use our results to investigate the prospects of measuring CP
violation through the combined measurement of Higgs rates, on the one
hand, and in observables based on CP-violating Higgs couplings to tau
leptons on the other hand.
\end{abstract}
\thispagestyle{empty}
\vfill
\newpage
\setcounter{page}{1}

\section{Introduction}
The discovery of a Standard Model (SM)-like Higgs boson with mass
around 125~GeV by the Large Hadron Collider experiments ATLAS and
CMS~\cite{:2012gk,:2012gu} represents a milestone for particle physics. 
This discovery supports the Higgs mechanism which allows for the
generation of particle masses without violating the underlying gauge
symmetries of the SM. It has long been realized that the maximum
allowed symmetry compatible with space-time symmetry is supersymmetry
(SUSY) \cite{susy}, which relates bosonic and fermionic degrees of freedom. 
Due to this and many other virtues as well, SUSY has become one of the
most popular and most intensely studied  symmetries beyond the SM
(BSM). While so far no direct sign of new physics has been found, it
remains possible that the SM-like Higgs boson discovered at the LHC is
in fact a SUSY Higgs. \s

The Minimal Supersymmetric Standard Model (MSSM) \cite{mssm}, requires
at least two complex Higgs doublets. In the Next-to-Minimal SUSY model
(NMSSM) \cite{nmssm} another complex singlet superfield is added to
the Higgs sector. The coupling of the singlet field to the MSSM Higgs
doublets allows for a dynamical solution of the $\mu$ problem
\cite{muproblem} when the neutral component of the singlet field
acquires its vacuum expectation value (VEV). The totality of ten
degrees of freedom of the Higgs doublet and singlet fields leads to
seven physical Higgs bosons after electroweak symmetry breaking. 
In the CP-conserving NMSSM these are three CP-even and two
CP-odd neutral Higgs bosons plus two charged ones, whereas in the
CP-violating case all Higgs bosons mix and do not carry a definite CP
quantum number any more. Besides the direct detection of more Higgs
states, an extended Higgs sector can manifest itself in modified
Higgs couplings of the SM-like Higgs boson. These arise from the
mixture with other Higgs states or from new physics effects induced 
through radiative corrections to the Higgs couplings and/or in the
loop mediated Higgs interactions with the photons and
gluons. Furthermore, Higgs decays into other lighter non-SM particles
can be realized leading to modified branching ratios, including also
the possibility of invisible decays. \s

CP violation is one of the three Sakharov conditions
\cite{Sakharov:1967dj} for baryogenesis, leading to matter-antimatter
asymmetry in the universe. In the SM the only source of CP violation
is given by the Cabibbo-Kobayashi-Maskawa (CKM) matrix
\cite{Cabibbo,Kobayashi:1973fv}. While the SM provides the
necessary ingredients for all Sakharov conditions, CP violation based
on the CKM matrix is too small to explain quantitatively the
observed asymmetry.\footnote{Besides the fact, that the 125~GeV Higgs
  boson is too heavy to allow for the strong first order phase transition
in the early universe required for thermal non-equilibrium.}
This motivates studying BSM theories which include additional sources
of CP violation.  SUSY contains many new sources of CP
violation. In the MSSM CP violation in the Higgs sector itself 
cannot occur at tree level and is radiatively induced. In the NMSSM Higgs
sector, however, CP violation can already show up at tree level through
NMSSM specific complex couplings which induce CP-violating
doublet-singlet mixing. \s

\vspace*{-0.4cm}
In this paper we shall be concerned with the CP-violating NMSSM,
including CP-violating effects in the phenomenology of the 125 GeV
Higgs boson as well as other effects in the Higgs sector as follows. 
For example, in the CP-violating NMSSM, a non-zero CP-odd admixture in the 
couplings of the dominantly CP-even non-SM-like light Higgs boson
may weaken its couplings to the weak gauge bosons to such an extent
that it escapes the LEP limits \cite{lep}. Additionally, CP violation in the Higgs
couplings can allow for decays 
of Higgs bosons into Higgs and gauge bosons or a pair of lighter
Higgs bosons in combinations that otherwise would be forbidden. These
could then constitute additional discovery channels for some of the Higgs
states. In general CP mixing alters the Higgs couplings and hence the
production and decay rates with effects on the discovery prospects of
the additional NMSSM Higgs bosons.\footnote{For a recent
  discussion on two Higgs bosons near 125~GeV in the complex NMSSM, see
\cite{Moretti:2015bua}.} Moreover, CP-violating phases influence
the Higgs mass spectrum already at tree level in case of Higgs sector
CP violation and at loop level through radiatively induced CP
violation \cite{Graf:2012hh,Muhlleitner:2014vsa}. 
Some of these effects, however, can also be explained in the CP-conserving
 NMSSM by choosing the parameter 
combinations accordingly. Additionally, without further information on
the CP nature of the Higgs bosons from other observables, genuine
CP-violating Higgs decays to gauge boson plus Higgs or Higgs-to-Higgs decays
cannot be  
identified unless all of these decay channels are observed. Since the
absolute size of CP violation is stringently constrained by
experiment, the identification of CP-violating Higgs
bosons will be a non-trivial task requiring precision measurements,
high luminosities and the
combination of various CP-violating observables. In particular tight
constraints on the CP-violating phases arise from the non-observation
of electric dipole moments (EDMs). Accordingly, we discuss in detail
in the main part of the paper the role of EDMs in constraining the
parameter space of the CP-violating NMSSM. By taking into account 
the latest experimental constraints from the Higgs data and the
measurements of the EDMs, we investigate
the size of the EDMs as a function of the CP-violating
phases. This allows for conclusions on the overall allowed size of CP violation in the
NMSSM in view of the newest experimental results. Subsequently, we
investigate how CP violation can be identified by combining 
Higgs-to-Higgs decays and Higgs decays to gauge boson plus Higgs. This
is complemented 
by the discussion of measuring CP violation in fermionic Higgs
decays. \s

The investigation of the CP-violating NMSSM Higgs sector considered
here takes into account higher order corrections both to the parameters and
the observables. Radiative corrections to the Higgs boson masses are
crucial to lift the SM--like Higgs boson mass to the
observed value of 125~GeV. The connection of the Higgs self-couplings
and masses through the Higgs potential requires the inclusion of
higher order corrections also in the Higgs self-interactions to
consistently describe Higgs-to-Higgs decays, which can become relevant
for spectra with light Higgs bosons and/or sizeable values of the
singlet coupling $\lambda$ \cite{Barbieri:2013hxa,Nhung:2013lpa,Ellwanger:2013ova,Han:2013sga,Munir:2013dya,King:2012is,King:2012tr,King:2014xwa,Cao:2014kya,Wu:2015nba,Buttazzo:2015bka}. In the Higgs boson decays 
into SM particles and coloured SUSY particles in particular QCD
corrections play an important role and have to be included. Also
electroweak corrections can become important. They cannot be adapted
from the SM or MSSM, however, but require an explicit calculation, as
has been done so far in the NMSSM only for the Higgs boson decays into
squarks \cite{Baglio:2015noa}. \s

It is worth recalling the status of higher order corrections to the
NMSSM Higgs boson masses.
In the CP-conserving NMSSM the one-loop mass corrections are available 
\cite{Ellwanger:1993hn,Elliott:1993ex,Elliott:1993uc,Elliott:1993bs,Pandita:1,Pandita:1993hx,Ellwanger:2005fh,Degrassi:2009yq,Staub:2010ty,Ender:2011qh}, 
as well as two-loop results of O($\alpha_t \alpha_s + \alpha_b \alpha_s$) in the
approximation of zero external momentum \cite{Degrassi:2009yq}. First
corrections beyond order O($\alpha_t \alpha_s + \alpha_b \alpha_s$)
have been provided in \cite{Goodsell:2014pla}. The one-loop
corrections to the Higgs masses of the CP-violating NMSSM have been
calculated by 
\cite{Ham:2001kf,Ham:2007mt,Ham:2001wt,Ham:2003jf,Funakubo:2004ka,Graf:2012hh}
and the logarithmically enhanced two-loop effects have been given in
\cite{Cheung:2010ba}.  The two-loop corrections to the Higgs
boson masses of the CP-violating NMSSM in the Feynman diagrammatic
approach with vanishing external momentum at O($\alpha_t \alpha_s$)
have been computed in \cite{Muhlleitner:2014vsa}. The one-loop
corrections to the trilinear Higgs self-couplings for the
CP-conserving NMSSM \cite{Nhung:2013lpa} are available and have 
recently been extended to include the two-loop corrections at
order ${\cal O}(\alpha_t \alpha_s)$ in the approximation of
vanishing external momentum in the CP-violating case \cite{Muhlleitner:2015dua}.
Several public codes are on the market for the computation of the NMSSM mass
spectrum for a given parameter set. These are the stand-alone codes 
 {\tt NMSSMTools}
\cite{Ellwanger:2004xm,Ellwanger:2005dv,Ellwanger:2006rn}, {\tt SOFTSUSY}
\cite{Allanach:2001kg,Allanach:2013kza} and {\tt NMSSMCALC}
\cite{Baglio:2013vya,Baglio:2013iia}, and the NMSSM implementations
of the programs 
{\tt FlexibleSUSY} \cite{Athron:2014yba,Athron:2014wta} and {\tt SPheno}
\cite{Porod:2003um,Porod:2011nf} which are based on {\tt SARAH}
\cite{Goodsell:2014pla,Staub:2010jh,Staub:2012pb,Staub:2013tta,Goodsell:2014bna}.\footnote{For
a comparison of the codes, see \cite{Staub:2015aea}.} An extension of
the program package {\tt NMSSMTools} to include also the CP-violating
NMSSM has recently been announced in \cite{Domingo:2015qaa}. \s 

For the phenomenological analysis in this paper we have 
implemented the EDMs in the Fortran package {\tt NMSSMCALC}. It 
computes for the CP-conserving and CP-violating case the two-loop
NMSSM Higgs boson masses at ${\cal O} (\alpha_t 
\alpha_s)$ and the Higgs boson widths and branching ratios including
the dominant higher order corrections.\footnote{The program package
{\tt NMSSMCALC} including the computation of the EDMs is made publicly
available and can be downloaded from the url:
http://www.itp.kit.edu/$\sim$maggie/NMSSMCALC/.}
\s

The layout of the paper is as follows. In
Section~\ref{sec:cpviolnmssm} we introduce the CP-violating NMSSM
Lagrangian and set our notation. Section~\ref{sec:edms} briefly
recapitulates the computation of the SUSY contributions to the
EDMs. In Section~\ref{sec:edmconstr} we discuss in detail the EDMs
induced by various CP-violating phases present in the 
NMSSM. In subsection~\ref{sec:natural} this investigation is performed in the
subspace of the Natural NMSSM that features a rather light Higgs
spectrum with good discovery prospects for all Higgs bosons of the
NMSSM. In subsection~\ref{sec:large}, we extend the investigation to an enlarged
NMSSM parameter range. Section~\ref{sec:pheno} is devoted to the prospects for
measuring CP violation in the Higgs couplings involving $Z$ bosons
through the combination of signal rates, subsection~\ref{sec:rates},
and in the Higgs couplings to fermions,
subsection~\ref{sec:fermions}. Section~\ref{sec:concl} summarizes and
concludes the paper. 

\section{The Lagrangian of the CP-violating NMSSM
\label{sec:cpviolnmssm}}
We work in the framework of the CP-violating NMSSM with a
scale-invariant superpotential and a discrete $\mathbb{Z}^3$
symmetry. The Higgs potential is obtained from the superpotential, the
soft SUSY breaking Lagrangian and the $D$-term contributions. The
NMSSM superpotential in terms of the two Higgs doublet superfields
$\hat{H}_d$ and $\hat{H}_u$, the singlet superfield $\hat{S}$, the
quark and lepton superfields and their charged conjugates, with the
superscript $c$, $\hat{Q}, \hat{U}^c, \hat{D}^c, \hat{L}, \hat{E}^c$,
is given by
\beq
W_{NMSSM} = \epsilon_{ij} [y_e \hat{H}^i_d \hat{L}^j \hat{E}^c + y_d
\hat{H}_d^i \hat{Q}^j \hat{D}^c - y_u \hat{H}_u^i \hat{Q}^j \hat{U}^c]
- \epsilon_{ij} \lambda \hat{S} \hat{H}^i_d 
\hat{H}^j_u + \frac{1}{3} \kappa \hat{S}^3 \;. 
\eeq
The $i,j=1,2$ are the indices of the $SU(2)_L$ fundamental
representation, and $\epsilon_{ij}$ is the totally antisymmetric
tensor with $\epsilon_{12}=\epsilon^{12}=1$, where we adopt the
convention to sum over equal indices. Colour and generation indices
have been suppressed. As we neglect generation mixing, the Yukawa
couplings $y_e, y_d$ and $y_u$ are diagonal, and complex phases can be
reabsorbed by redefining the quark fields without effect on the
physical meaning \cite{Kobayashi:1973fv}. The dimensionless parameters
$\lambda$ and $\kappa$ are complex in case of CP violation. \s

The soft SUSY breaking Lagrangian in terms of the scalar component
fields $H_u, H_d$ and $S$ reads
\begin{align}
{\cal L}_{\text{soft},\text{ NMSSM}} =& -m_{H_d}^2 H_d^\dagger H_d - m_{H_u}^2
H_u^\dagger H_u -
m_{\tilde{Q}}^2 \tilde{Q}^\dagger \tilde{Q} - m_{\tilde{L}}^2 \tilde{L}^\dagger \tilde{L}
- m_{\tilde{u}_R}^2 \tilde{u}_R^* 
\tilde{u}_R - m_{\tilde{d}_R}^2 \tilde{d}_R^* \tilde{d}_R 
\nonumber \\\nonumber
& - m_{\tilde{e}_R}^2 \tilde{e}_R^* \tilde{e}_R - (\epsilon_{ij} [y_e A_e H_d^i
\tilde{L}^j \tilde{e}_R^* + y_d
A_d H_d^i \tilde{Q}^j \tilde{d}_R^* - y_u A_u H_u^i \tilde{Q}^j
\tilde{u}_R^*] + \mathrm{h.c.}) \\
& -\frac{1}{2}(M_1 \tilde{B}\tilde{B} + M_2
\tilde{W}_i\tilde{W}_i + M_3 \tilde{G}\tilde{G} + \mathrm{h.c.})\\ \nonumber
&- m_S^2 |S|^2 +
(\epsilon_{ij} \lambda 
A_\lambda S H_d^i H_u^j - \frac{1}{3} \kappa
A_\kappa S^3 + \mathrm{h.c.}) \;,
\label{eq:softnmssm}
\end{align}
where a sum over all three quark and lepton generations is implicit. 
The $\tilde{Q}$ and $\tilde{L}$ denote the complex scalar components
of the corresponding quark and lepton superfields and, {\it e.g.}~for
the first generation are $\tilde{Q} = (\tilde{u}_L, \tilde{d}_L)^T$
and $\tilde{L}=(\tilde{\nu}_L,\tilde{e}_L)^T$. In the CP-violating
NMSSM the soft SUSY breaking trilinear couplings $A_x$
($x=\lambda,\kappa,d,u,e$) and the gaugino mass parameters $M_k$
($k=1,2,3$) of the bino, wino and gluino fields
$\tilde{B},\tilde{W}_i$ ($i=1,2,3$) and $\tilde{G}$, are
complex. Exploiting the $R$-symmetry either $M_1$ or $M_2$ can be
chosen to be real. On the other hand the soft SUSY breaking mass
parameters of the scalar fields, $m_X^2$ ($X=S,H_d,H_u, \tilde{Q},
\tilde{u}_R, \tilde{d}_R, \tilde{L}, \tilde{e}_R$) are real.  \s

The Higgs potential finally is obtained as
\beq
V_{H}  &=& (|\lambda S|^2 + m_{H_d}^2)H_d^\dagger H_d+ (|\lambda S|^2
+ m_{H_u}^2)H_u^\dagger H_u +m_S^2 |S|^2 \nonumber \\
&& + \frac{1}{8} (g_2^2+g_1^{2})(H_d^\dagger H_d-H_u^\dagger H_u )^2
+\frac{1}{2} g_2^2|H_d^\dagger H_u|^2 \label{eq:higgspotential} \\ 
&&   + |-\epsilon^{ij} \lambda  H_{d,i}  H_{u,j} + \kappa S^2 |^2+
\big[-\epsilon^{ij}\lambda A_\lambda S   H_{d,i}  H_{u,j}  +\frac{1}{3} \kappa
A_{\kappa} S^3+\mathrm{h.c.} \big] \;,
\nonumber
\eeq
with $g_1$ and $g_2$ denoting the $U(1)_Y$ and $SU(2)_L$ gauge
couplings, respectively. Expanding the two Higgs doublets and the
singlet field about their VEVs, $v_d, v_u$ and
$v_s$, two more CP-violating phase, $\varphi_u$ and $\varphi_s$, are
introduced,
\beq
H_d =
 \bpmatrix \frac{1}{\sqrt{2}}(v_d + h_d +i a_d)\\ h_d^- \epmatrix,\quad
H_u = e^{i\varphi_u}\bpmatrix
h_u^+ \\ \frac{1}{\sqrt 2}(v_u + h_u +i a_u)\epmatrix,\quad
S= \frac{e^{i\varphi_s}}{\sqrt{2}}(v_s + h_s +ia_s)
\;.~\label{eq:Higgs_decomposition}  
\eeq
The VEVs $v_u$ and $v_d$ are related to $v\approx 246$~GeV through
$v^2=v_d^2+ v_u^2$ and their ratio is parametrized by
$\tan\beta=v_u/v_d$. 
The phase $\varphi_u$ affects the top quark mass. We absorb this phase
into the left-handed and right-handed top fields through
\beq
t_L \to e^{-i\varphi_u/2}\,t_L \quad \mbox{and} \quad t_R \to
e^{i\varphi_u/2}\,t_R \;, \label{eq:topfieldrephase}
\eeq
so that the top Yukawa coupling is kept real. This alters all
couplings with one top quark. Replacing
Eq.~(\ref{eq:Higgs_decomposition}) in Eq.~(\ref{eq:higgspotential})
yields the Higgs potential
\beq
V_H & = & V_H^{\mbox{\scriptsize const}}  +  t_{h_d} h_d + t_{h_u} h_u +
t_{h_s} h_s  +  t_{a_{d}} a_d+  t_{a_{u}} a_u 
+  t_{a_{s}} a_s  \\ \non
&& + 
 \frac{1}{2} \phi^{0,T}  {\mathcal{M}_{\phi\phi}} \, \phi^0 +
 \phi^{c,\dagger} {\mathcal{M}_{h^+h^-}} \, \phi^c 
+V_H^{\phi^3,\phi^4} \;,
\eeq
with $\phi^0 \equiv (h_d, h_u, h_s, a_d, a_u, a_s)^T$ and $\phi^c \equiv
((h_d^-)^*,h_u^+)^T$. The tadpole coefficients are denoted by $t_\phi$
($\phi=h_d, h_u, h_s, a_d, a_u, a_s$), ${\cal M}_{\phi\phi}$ is the
$6\times 6$ mass matrix for the neutral Higgs bosons and $M_{h^+ h^-}$
the $2\times 2$ mass matrix for the charged Higgs states. Constant
terms are summarized in $V_H^{\mbox{\scriptsize const}}$ and the
trilinear and quartic interactions in $V_H^{\phi^3,\phi^4}$. A few
remarks on the tadpoles and mass matrices are in order, without
repeating their explicit expressions here, which can be found in
Ref.~\cite{Graf:2012hh}. The tadpole coefficients vanish at tree level
due to the minimization conditions of the Higgs potential. 
Rewriting the complex parameters $\lambda, \kappa, A_\lambda$ and
$A_\kappa$ as
\beq
\lambda = |\lambda| e^{i\varphi_\lambda} \;, \; \kappa = |\kappa|
e^{i\varphi_\kappa} \;, \; A_{\lambda} = |A_{\lambda}|
e^{i\varphi_{A_\lambda}} \quad \mbox{and} \quad
A_{\kappa} = |A_{\kappa}|
e^{i\varphi_{A_\kappa}} \;,
\eeq
three phase combinations appear at tree level in the tadpoles and the
mass matrices, 
\bea
\varphi_x &=& \varphi_{A_\lambda}+
\varphi_1 \,, \label{eq:phasecomb1}\\ 
\varphi_y &=& \varphi_2 - \varphi_1 \,, \label{eq:phasecomb2} \\ 
\varphi_z &=& \varphi_{A_\kappa} + \varphi_2 \,,
\label{eq:phasecomb3}
\eea
where we have introduced
\beq
\varphi_1 &=& \varphi_\lambda + \varphi_s + \varphi_u
\,, \label{eq:phase1} \\
\varphi_2 &=& \varphi_\kappa + 3 \varphi_s \;. \label{eq:phase2}
\eeq
Two of the three combinations
Eqs.~(\ref{eq:phasecomb1})-(\ref{eq:phasecomb3}) can be eliminated at
lowest order by applying the 
minimization conditions $t_{a_d}=t_{a_s}=0$. We express $\varphi_x$
and $\varphi_z$ in terms of $\varphi_y$. All mass matrix elements
mixing the CP-even and CP-odd interaction states, ${\cal M}_{h_i
  a_j}$, are then proportional to $\sin \varphi_y$. At tree level,
this is the only CP-violating phase in the Higgs sector. The rotation
from the interaction to the mass eigenstates $h_i$ ($i=1,...,5$) is
performed by applying two consecutive rotations. The first rotation
with matrix ${\cal R}^G$ separates the would-be Goldstone bosons. The
second one with the matrix ${\cal R}$ performs the rotation to the
mass eigenstates, {\it i.e.}
\beq
(h_d,h_u,h_s,a,a_s, G)^T &=&  \mathcal{R}^G~(
h_d,h_u,h_s,a_d,a_u,a_s)^T\,, \non \\ 
(h_1,h_2,h_3,h_4,h_5, G)^T &=& \mathcal{R} ~(h_d,h_u,h_s,a,a_s,
G)^T\,, \label{eq:rotationtreelevel}
\eeq
with the diagonal mass matrix
\beq
\diag(m_{h_1}^2,m_{h_2}^2,m_{h_3}^2,m_{h_4}^2,m_{h_5}^2,0)&=&
\mathcal{R} \mathcal{M}_{hh} \mathcal{R}^T\,, \quad \mathcal{M}_{hh}=
\mathcal{R}^G\mathcal{M}_{\phi\phi}(\mathcal{R}^G)^T. 
\label{eq:massmatrix}
\eeq
The mass eigenstates $h_i$ are ordered by ascending mass, where the
lightest mass is given by $m_{h_1}$. \s

The tree-level Higgs potential can be parametrized by the following
set of independent parameters
\be 
t_{h_d},t_{h_u},t_{h_s},t_{a_d},t_{a_s},M_{H^\pm}^2,v,\sin \theta_W,
e,\tan\beta,|\lambda|,v_s,|\kappa|,\text{Re}A_{\kappa},\sin \varphi_y\,.
\ee
We have chosen to use $v$ and $\sin\theta_W$, where $\theta_W$ denotes
the weak mixing angle, instead of $M_W$ and $M_Z$. This is more
convenient in view of the inclusion of the two-loop corrections to the
Higgs boson masses in the gaugeless limit.\footnote{See
  \cite{Muhlleitner:2014vsa} for details.} Furthermore, in accordance
with the SUSY Les Houches Accord (SLHA)
\cite{Skands:2003cj,Allanach:2008qq} the real part of $A_\kappa$ is 
used as input parameter. The imaginary part is eliminated by the
tadpole conditions. This distinction is not necessary for $\lambda$
and $\kappa$, as both the real and imaginary parts are given in the
SLHA convention and can be related to the respective absolute values
and phases. Note finally, that the effective higgsino mixing
parameter is given by
\beq
\mu_{\text{eff}}= \frac{|\lambda| v_s
  e^{i(\varphi_s+\varphi_\lambda)}}{\sqrt{2}} \;. \label{eq:mueff}
\eeq 

At higher order in the Higgs mass corrections (and self-couplings) the
CP-violating phases entering the Higgs sector at tree level are not
related any more. The phases $\varphi_1$ and $\varphi_2$ appear
independently in the neutralino sector, while the chargino and
up-type squark sector depend on the phase $\varphi_1$. We therefore
have two independent CP-violating phases that appear, if we choose to
determine the phases $\varphi_{A_\kappa}$ and $\varphi_{A_\lambda}$
from the tadpole conditions. Of course non-vanishing $\varphi_1$ or
  $\varphi_2$ will automatically imply non-vanishing
  $\varphi_{A_\lambda}$ and $\varphi_{A_\kappa}$. However, since we
  consider the latter two as derived quantities, which are fixed via
  the tadpole conditions, we will not mention them explicitly in the
  following discussion. 
The higher order corrections introduce further complex phases stemming
from the complex soft SUSY breaking trilinear couplings and gaugino mass
parameters, that enter the couplings and SUSY particle masses involved
in the loop corrections. 

\section{Electric Dipole Moments \label{sec:edms}}
\begin{figure}[t]
\begin{center}
\vspace*{-0.2cm}
\includegraphics[width=1.025\textwidth]{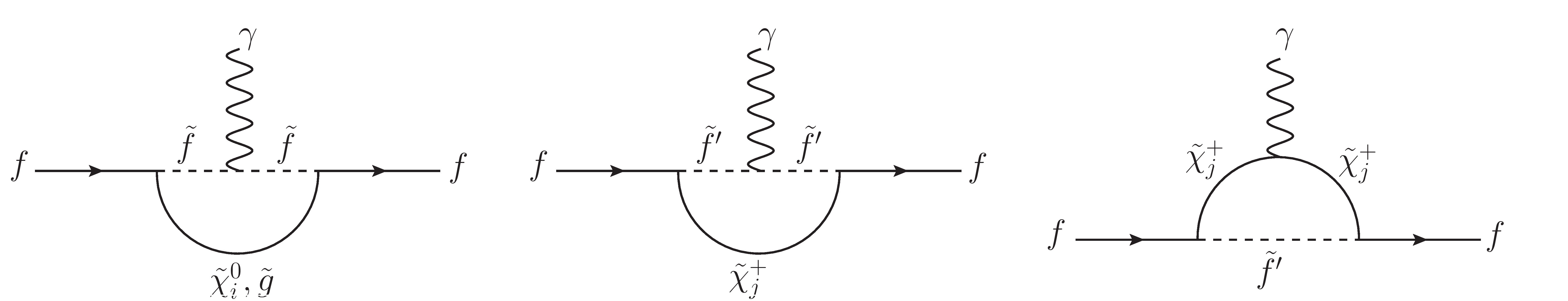} 
\caption{Generic one-loop diagrams contributing to the EDMs of the
  electron and light quarks ($f=e,u,d,s$). \label{fig:oneloopdiags}} 
\end{center}
\end{figure}
\begin{figure}[t]
\begin{center}
\vspace*{-0.2cm}
\includegraphics[width=1.025\textwidth]{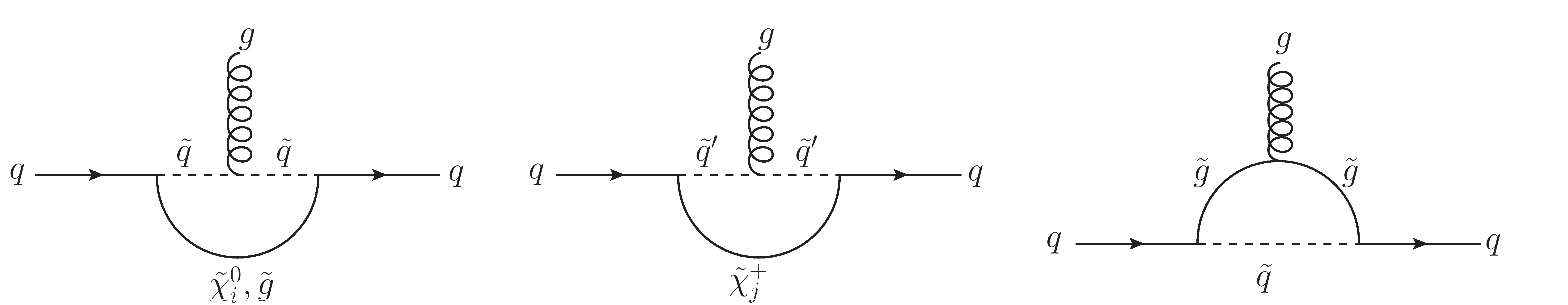} 
\caption{Generic one-loop diagrams contributing to the CEDMs of the
  light quarks ($q=u,d,s$).  \label{fig:oneloopdiagscedm}}
\end{center}
\end{figure}
\begin{figure}[t]
\begin{center}
\vspace*{-0.2cm}
\includegraphics[width=1.025\textwidth]{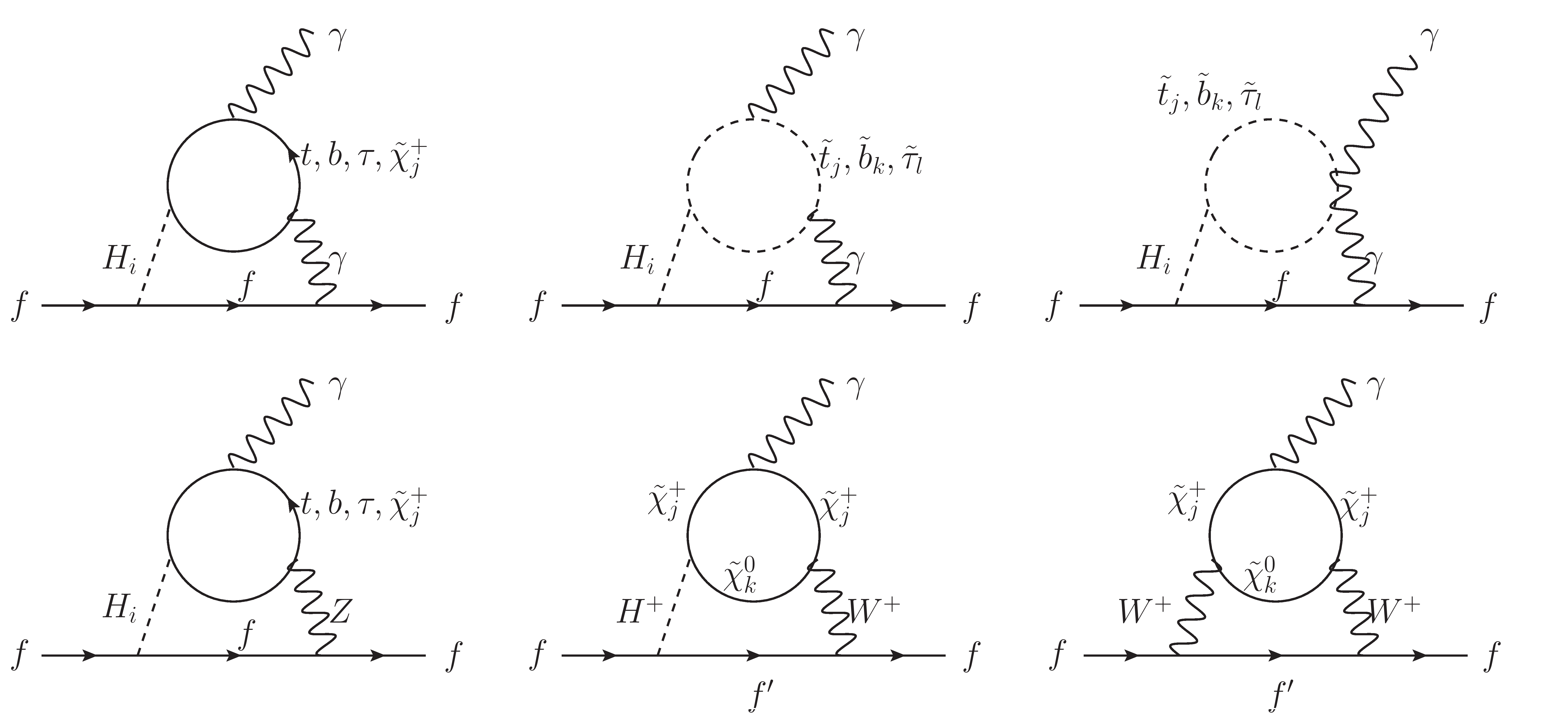} 
\caption{Generic two-loop Barr-Zee type diagrams contributing to the
  EDMs of the electron and light quarks
  ($f=e,u,d,s$). Upper: $\gamma H$, and lower from
    left to right: $ZH$, $WH$ and $WW$.\label{fig:twoloopedm}}
\end{center} 
\end{figure}
\begin{figure}[t]
\begin{center}
\vspace*{-0.2cm}
\includegraphics[width=1.025\textwidth]{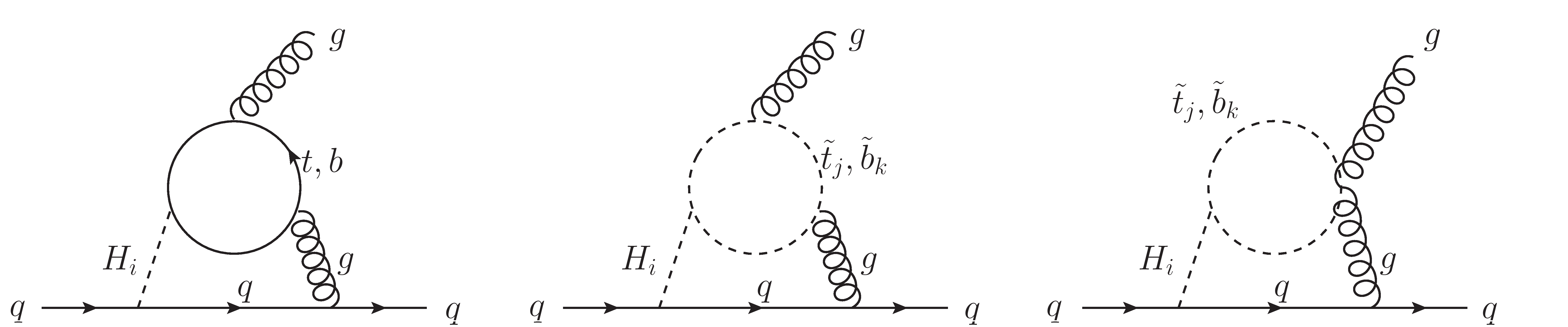} 
\caption{Generic two-loop Barr-Zee type diagrams contributing to the
  CEDMs of the light quarks
  ($q=u,d,s$).\label{fig:twoloopcedm}}
\end{center} 
\end{figure}

CP violation would manifest itself in the generation of EDMs. The
non-observation of any EDMs so far poses stringent constraints on
CP-violating phases. These have to be taken into account, when
discussing possible CP-violating effects in the Higgs sector. The
constraints may become weaker in case of accidental cancellations of
the various contributions to the EDMs, even for lighter SUSY particles
with masses below ${\cal O}(1\mbox{ TeV})$ \cite{cancellations,elliseal}. In our analysis we
take into account all relevant CP-induced contributions to the
observable EDMs. In particular we consider the compatibility with the
experimental upper bounds on the EDMs, which are 
\beq
\begin{array}{lll}
\mbox{Electron EDM \cite{eedm}} &: \qquad & \sim 1\cdot 10^{-28} e \,\mbox{cm} \\[0.1cm]
\mbox{Thallium EDM \cite{thaledm}} &: \qquad & \sim 9 \cdot 10^{-25} e
\,\mbox{cm} \\[0.1cm]
\mbox{Neutron EDM \cite{neutedm}} &: \qquad & \sim 3 \cdot 10^{-26} e
\, \mbox{cm} \\[0.1cm]
\mbox{Mercury EDM \cite{mercedm}} &: \quad & \sim 3.1 \cdot 10^{-29} e
\, \mbox{cm} \;,
\end{array}
\label{eq:edmexp}
\eeq
where the electron EDM is estimated from the thorium monoxide experiment. 
These observable EDMs receive form factor contributions from the
electric dipole moment, the 
chromo-electric dipole moment (CEDM), the two-loop Weinberg
three-gluon operator and 
the Higgs-exchange four-fermion operators. All of these contain contributions that are generated by
CP-violating Higgs mixing at tree level. In the EDM and the CEDM we
consider one- and two-loop contributions. The two-loop contributions
stemming from CP violation in the Higgs sector at tree level, which is specific to the NMSSM, can become
important when the CP phases of the MSSM parameters are set to zero
\cite{cheungeal}. Such a configuration can  be achieved by choosing $\varphi_2\neq 0$, while keeping
$\varphi_1=0$. Therefore we will refer to $\varphi_2$ as the NMSSM-specific phase in the following.
The one-loop EDMs of the electron and the light quarks $u,d,s$ are induced by chargino,
$\tilde{\chi}_j^{\pm}$ ($j=1,2$), and
neutralino, $\tilde{\chi}_i^0$ ($i=1,...,5$), exchange diagrams, {\it
  cf.}~Fig.~\ref{fig:oneloopdiags}. For the quarks also gluino,
$\tilde{g}$, exchange diagrams contribute. The light quarks
furthermore  have CEDMs which are also generated by chargino,
neutralino and gluino loops as shown in Fig.~\ref{fig:oneloopdiagscedm}. The
one-loop EDMs have been computed before and the formulae can be found in
\cite{cheungeal,elliseal}. At two-loop level
the Higgs mediated Barr-Zee type diagrams contribute significantly to
the EDMs. They are mediated by neutral Higgs couplings to two photons,
$\gamma\gamma H^0_i$ \cite{elliseal}, the charged Higgs coupling to the
charged $W$ boson and a photon, $\gamma H^\pm W^\mp$, the $\gamma
W^\pm W^\mp$coupling \cite{ellis2010}, and the couplings between a neutral
Higgs boson, a photon and a $Z$ boson, $\gamma H_i^0 Z$ 
\cite{gamhz}. We will denote these contributions in the following as 
$\gamma H$, $WH$, $WW$ and $ZH$, respectively. The diagrams are
displayed in Fig.~\ref{fig:twoloopedm}.
Additionally, CEDMs of the light quarks $u$, $d$ and $s$ are 
generated by two-loop Higgs-mediated Barr-Zee graphs
\cite{cheungeal,elliseal}, {\it cf.}~Fig.~\ref{fig:twoloopcedm}. For
the Weinberg operator we take into 
account the contributions from the Higgs-mediated two-loop diagrams
\cite{weinbergop} and additionally the contribution from the quark-squark-gluino
exchange contribution \cite{daieal}. The coefficients of the four-fermion operators,
finally, are generated from the $t$-channel exchanges of the CP-violating neutral
Higgs bosons \cite{elliseal}. \s

We briefly describe how the observable EDMs are obtained from the
various contributions introduced above. For explicit formulae and more
details we refer the reader to \cite{cheungeal,elliseal} and
references therein. The Thallium EDM receives contributions from the
electron EDM and additionally from the CP-odd electron-nucleon
interaction \cite{thal1,thal2}. For the neutron EDM three different
hadronic approaches are considered. These are the Chiral Quark Model
(CQM), the Parton Quark Model (PQM) and the QCD
sum rule technique. In the non-relativistic CQM the
quark EDMs are estimated via naive dimensional analysis \cite{cqm}. The
PQM uses low-energy data related to the constituent-quark contribution
to the proton spin combined with isospin symmetry \cite{pqm}. In the
third approach finally QCD sum rule techniques
\cite{sumrule1,sumrule2,sumrule3} are applied to determine the
EDM. Depending on the approach, the neutron EDM is composed of the
contributions from the EDMs and CEDMs of the light quarks, the Weinberg
operator and the CP-odd four-fermion operators, see \cite{elliseal} for
details. By using QCD sum rules \cite{sumrule1,sumrule3} the Mercury
EDM is estimated from the EDMs induced by the 
Schiff moment\footnote{For the Schiff moment several approximations
  exist \cite{Ellis:2011hp}.}, the electron EDM, the
contribution due to the CP-odd electron-nucleon 
interaction and the contributions from the couplings of
electron-nucleon interactions. Details are given in \cite{cheungeal}.\s

\section{EDM constraints \label{sec:edmconstr}}
In this section we investigate the influence of the various
CP-violating phases on the EDMs and the resulting constraints on
possible CP-violating scenarios. We will present results for the
phases $\varphi_1, \varphi_2$ and the phase $\varphi_{A_t}$, which
arises from the top/stop sector, and the phases $\varphi_{M_i}$ of the
gaugino mass parameters $M_i$ ($i=1,2,3$). The EDMs also
depend on the phases $\varphi_{A_j}$ of the soft SUSY
breaking trilinear couplings $A_j$ ($j=b,u,c,d,s$). 
The influence of the phases $\varphi_{A_j}$ of the sbottom and the
first and second generation squark sector is by far subleading
compared to the effects of the other phases. The reason is that the
trilinear couplings and hence their phases come in combination with the
quark masses, which are small in this case. \s

In order to find viable parameter
points we perform a scan in the NMSSM parameter space and keep only
those points that are in accordance with the LHC Higgs data.
That this is the case has been checked with the help of
 the programs {\tt HiggsBounds}
 \cite{Bechtle:2008jh,Bechtle:2011sb,Bechtle:2013wla} and {\tt
   HiggsSignals} \cite{Bechtle:2013xfa}.  
For the computation of the Higgs boson masses, the effective
couplings, the decay widths and branching ratios of the SM and
NMSSM Higgs bosons, that are needed as inputs for {\tt HiggsBounds}
and {\tt HiggsSignals}, the Fortran code {\tt NMSSMCALC}
\cite{Baglio:2013vya,Baglio:2013iia} has been used.
Besides the masses at two-loop level, it provides the SM and NMSSM
decay widths and branching 
ratios including the state-of-the-art higher order corrections. We
demanded that the valid scenarios feature a Higgs boson with mass 
around 125~GeV. With {\tt
  HiggsBounds} we have checked whether or not the Higgs spectrum is
excluded at the 95\% confidence level (CL) with respect to the LEP,
Tevatron and LHC measurements. The package {\tt HiggsSignals} tests
for the compatibility of the SM-like Higgs boson with the Higgs
observation data. We required the $p$-value, that is given out, to be at
least 0.05, corresponding to a non-exclusion at 95\% CL.

\subsection{Natural NMSSM \label{sec:natural}}
We performed a parameter scan in the subspace of the Natural NMSSM as
defined in \cite{King:2014xwa}. It features a rather light overall Higgs mass
spectrum and gives good discovery prospects for all NMSSM Higgs
scalars. It is characterized by
\beq
0.6 \le |\lambda| \le 0.7 \,, \; |\kappa| \le 0.3 \,, \; 1.5 \le \tan\beta
 \le 2.5\, ,\; 100\mbox{ GeV} \le |\mu_{\text{eff}}| \le 185 \mbox{
   GeV} \, .
\label{eq:cond1}
\eeq
The small $\kappa$ values lead to an approximate Peccei-Quinn
symmetry. Note, that in this parameter region the second lightest
of the Higgs bosons that are dominantly CP-even, is
SM-like\footnote{The Higgs boson with mass close to 125~GeV is forced
  to be mostly CP-even due to the requirement to be compatible
  with the LHC Higgs data.}. The soft SUSY
breaking trilinear NMSSM couplings are varied in the interval
\beq
|A_\lambda| \le 2 \mbox { TeV} \qquad \mbox{and} \qquad 
|A_\kappa| \le 2 \mbox { TeV} \;.
\label{eq:cond2}
\eeq
The remaining soft SUSY breaking trilinear couplings and masses have
been chosen as
\beq
|A_U|, |A_D|, |A_L| \le 2 \mbox { TeV} \qquad
\mbox{with} \qquad 
U \equiv u,c,t, \;, D\equiv d,s,b \;, L=e, \mu, \tau,
\label{eq:cond3}
\eeq
\vspace*{-0.7cm}
\beq
M_{\tilde{u}_R,\tilde{c}_R} =
M_{\tilde{d}_R,\tilde{s}_R}=M_{\tilde{Q}_{1,2}}=M_{\tilde{e}_R,\tilde{\mu}_R}=
M_{\tilde{L}_{1,2}} = 3 \mbox{ TeV} \;,
\label{eq:cond4}
\eeq
\vspace*{-0.7cm}
\beq
600 \mbox{ GeV} \le M_{\tilde{t}_R} = M_{\tilde{Q}_3} \le 3 \mbox{
 TeV} \; , \;
600 \mbox{ GeV} \le M_{\tilde{\tau}_R} = M_{\tilde{L}_3} \le 3 \mbox{
 TeV} \; , \;
M_{\tilde{b}_R} = 3 \mbox{ TeV} \;,
\label{eq:cond5}
\eeq
\vspace*{-0.7cm}
\beq
100 \mbox{ GeV} \le |M_1| \le 1 \mbox{ TeV} \;, \;
200 \mbox{ GeV} \le |M_2| \le 1 \mbox{ TeV} \;, \;
1.3 \mbox{ TeV} \le |M_3| \le 3 \mbox{ TeV} \;.
\label{eq:cond6}
\eeq
All scenarios have been checked for compatibility with the lower bound on the
charged Higgs mass \cite{chargedhiggs} and respect the 
exclusion limits on the SUSY particle masses
\cite{susyexclusion1,susyexclusion1a,susyexclusion2}.
Note also, that the signs of the gaugino masses $M_1$ and $M_2$ 
have only a marginal effect on the features of the NMSSM
Higgs sector. The NMSSM-specific input parameters
$\lambda,\kappa,A_\lambda$ and $A_\kappa$ as well as all other soft
SUSY breaking masses and trilinear couplings, according to the SLHA
format, are understood as $\overline{\mbox{DR}}$ parameters taken at
the SUSY scale $M_S=\sqrt{M_{\tilde{t}_R}M_{\tilde{Q}_3}}$. In
{\tt NMSSMCALC} also $\tan\beta$ is assumed to be given at the SUSY scale. \s

\begin{figure}[h!]
\begin{center}
\vspace*{-0.2cm}
\includegraphics[height=0.45\textwidth,angle=-90]{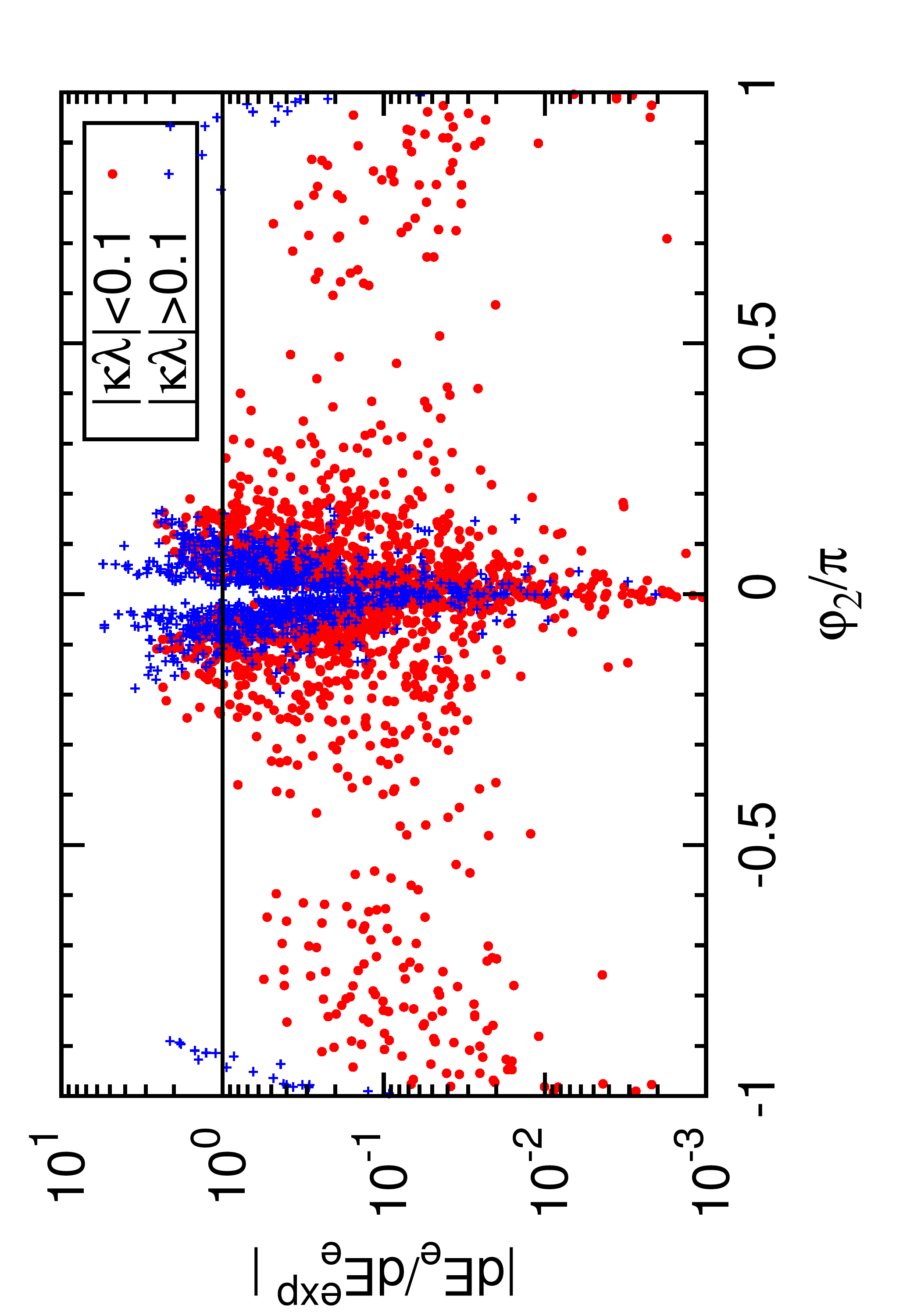}\includegraphics[height=0.45\textwidth,angle=-90]{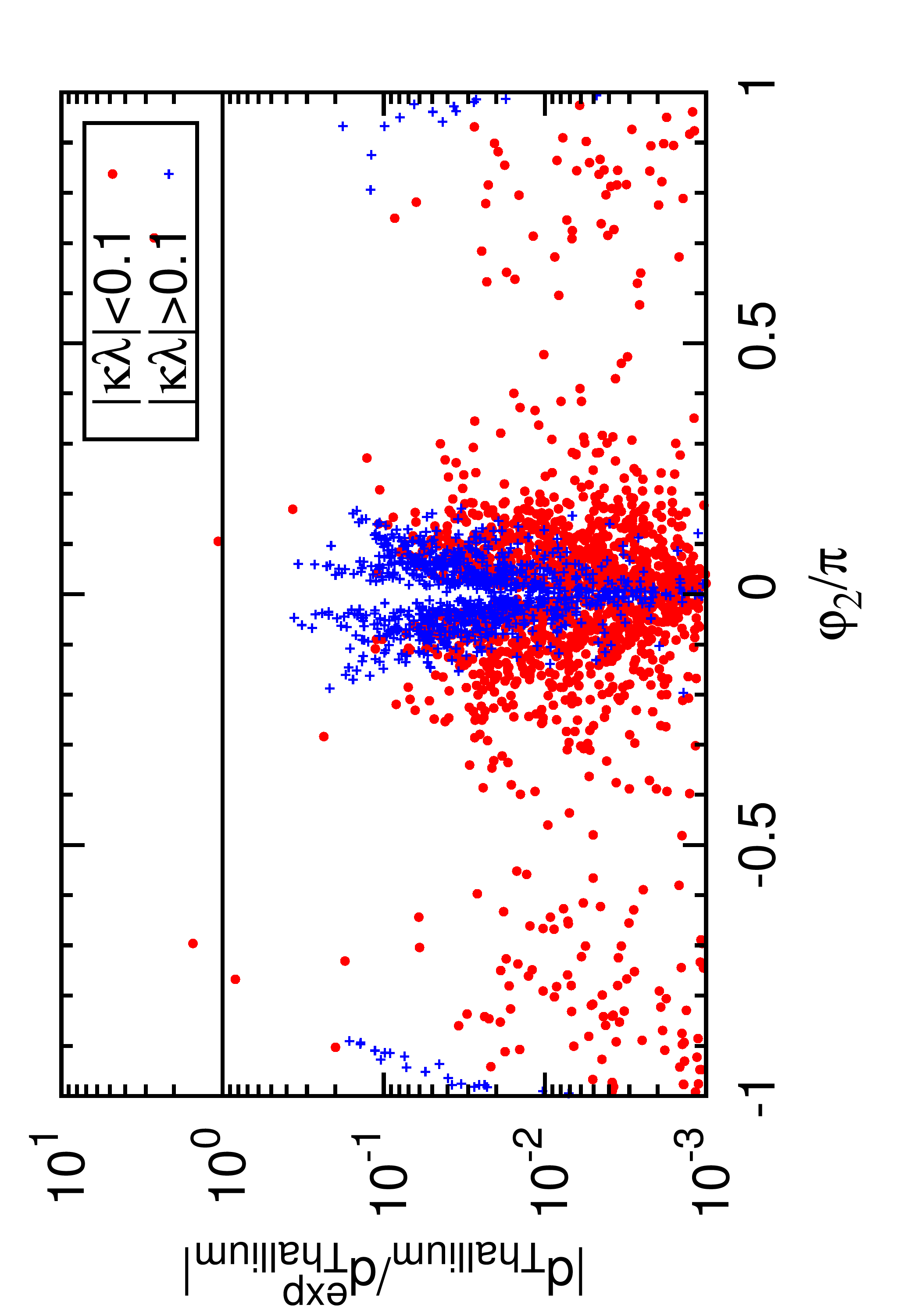} \\[-2mm]
\includegraphics[height=0.45\textwidth,angle=-90]{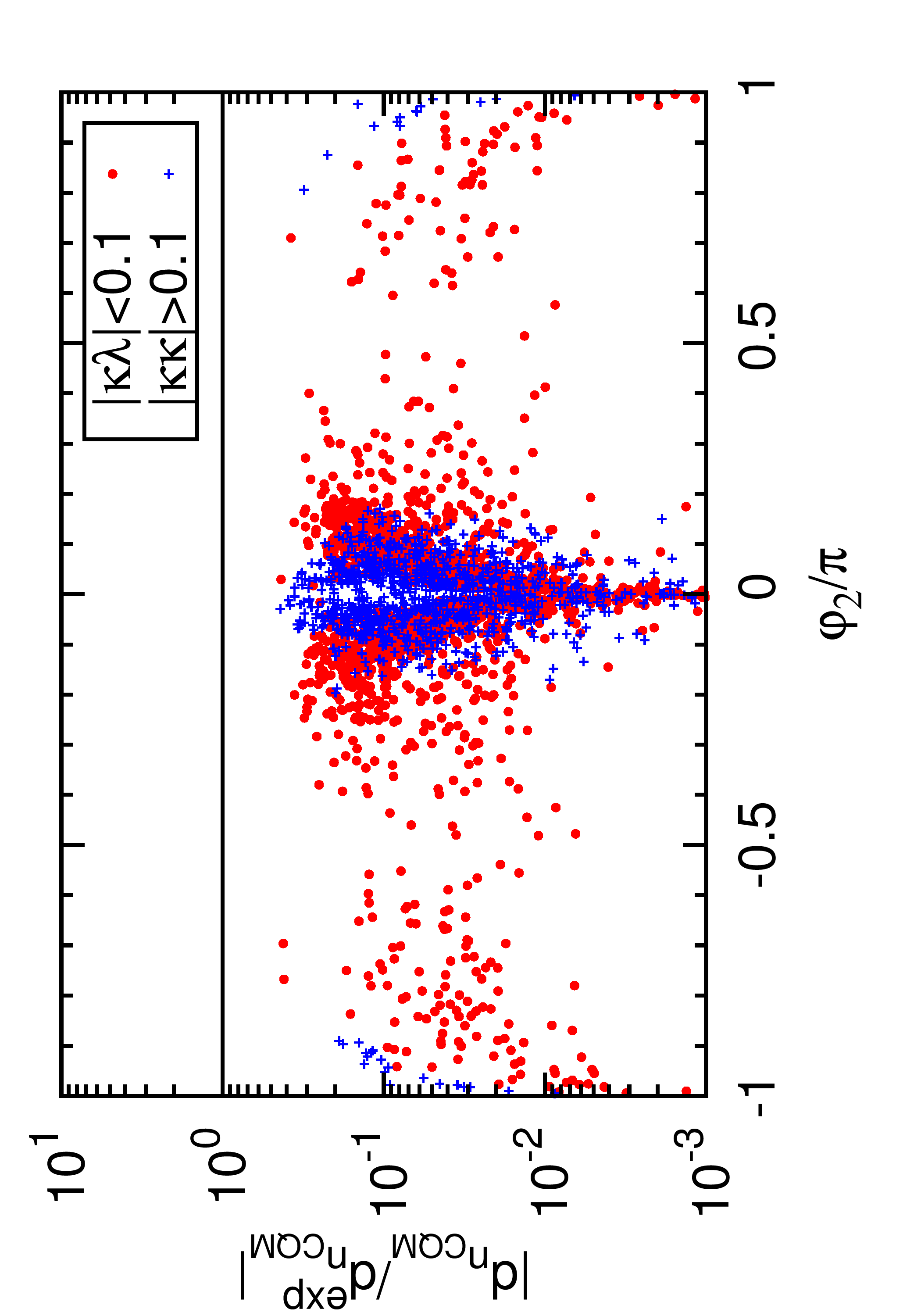}\includegraphics[height=0.45\textwidth,angle=-90]{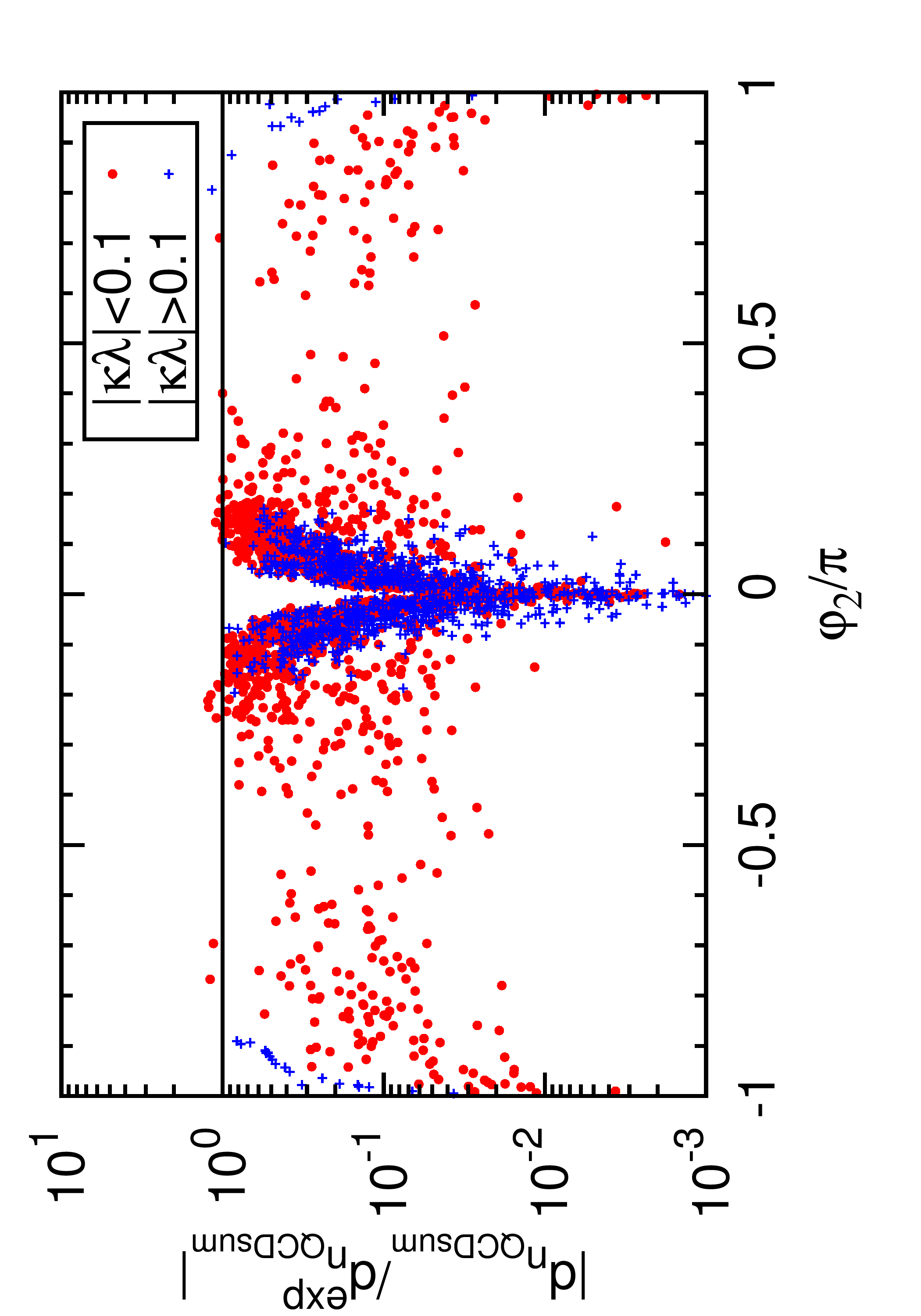} \\[-2mm]
\includegraphics[height=0.45\textwidth,angle=-90]{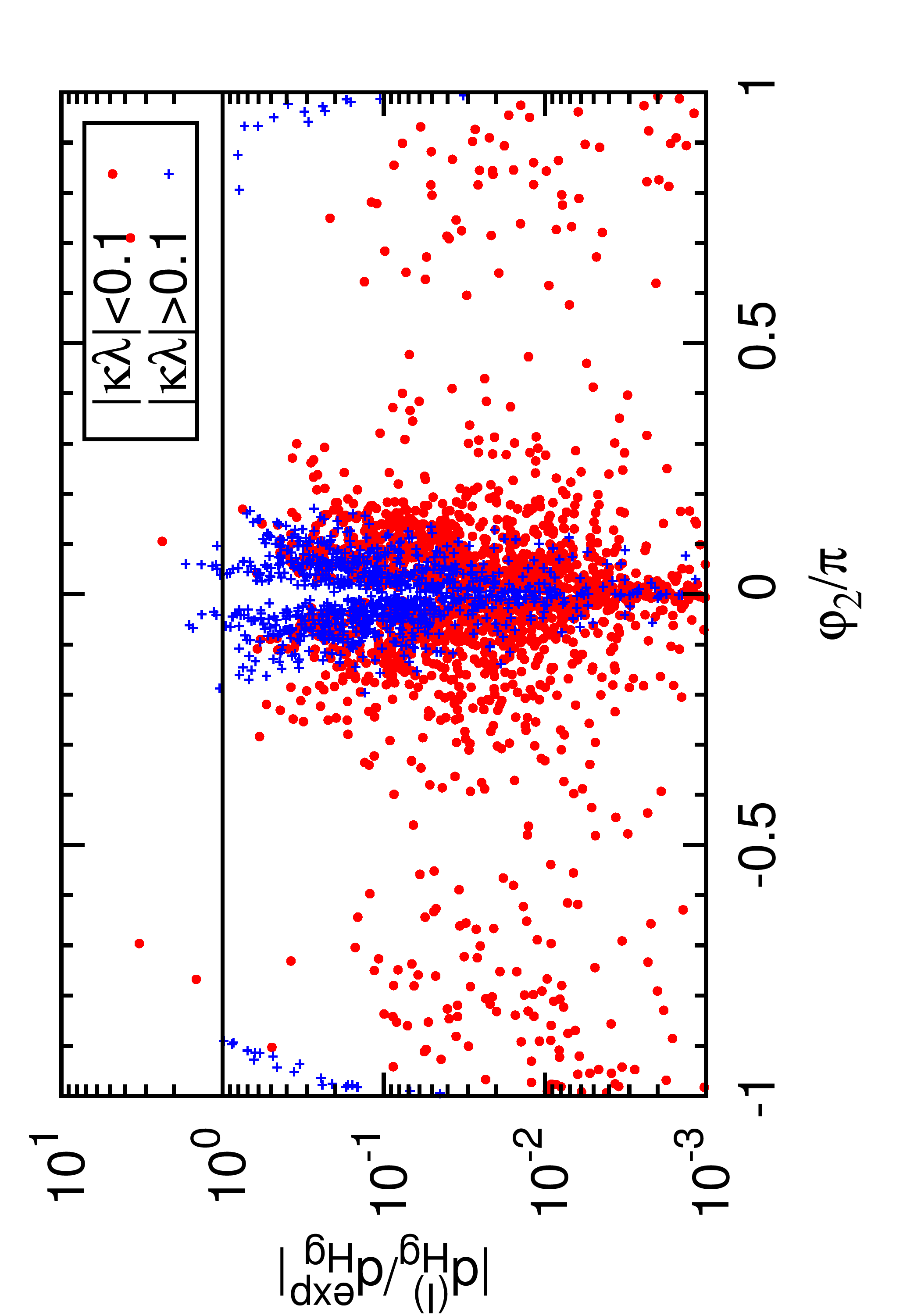}\includegraphics[height=0.45\textwidth,angle=-90]{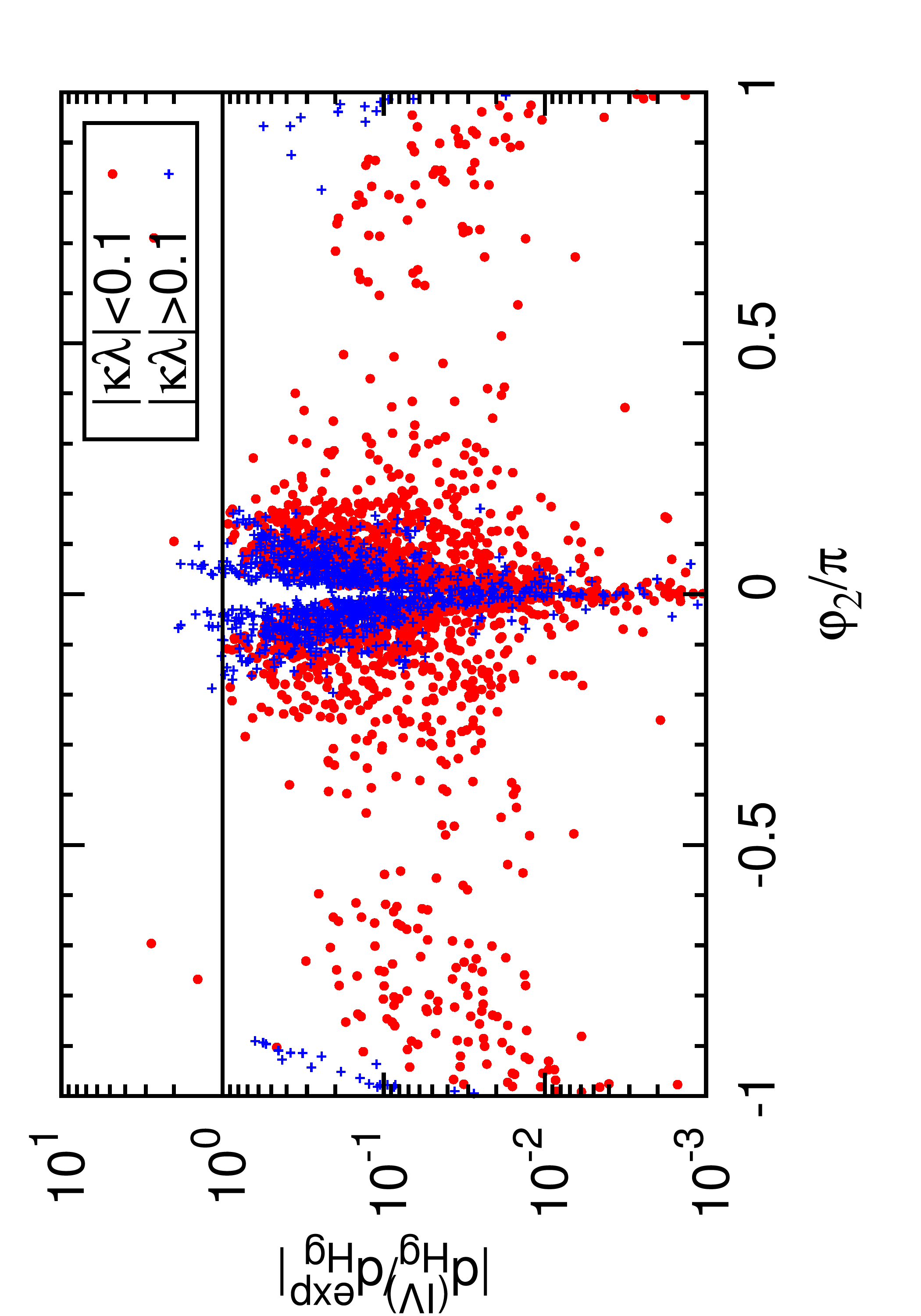} \\
\caption{Absolute values of the electron (upper
  left), Thallium (upper right), neutron 
  (middle) and Mercury (lower) EDMs as a function of
  $\varphi_2$, normalized to the respective experimental upper bound;
  red: $|\kappa \lambda| <0.1$, blue: $|\kappa \lambda| > 0.1$. 
\label{fig:edmcomplete}}
\end{center}
\end{figure}
\underline{\it Variation of $\varphi_2$:}
For the scenarios of our scan that are compatible with all above
described constraints we computed the various EDMs and checked for
their compatibility with the experimental values. The results are shown
in Fig.~\ref{fig:edmcomplete} for the electron, the Thallium, the neutron and
the Mercury EDM. In these plots we have only varied the NMSSM specific
phase $\varphi_2$. All other complex phases have been set to
zero. We hence investigate here solely tree-level CP violation, as it can 
occur in the NMSSM. As the phase
$\varphi_2$ only appears in the NMSSM we 
call this in the following {\it NMSSM-type CP violation}. 
For each EDM, the absolute values of the
computed EDMs in the valid scenarios are shown, normalized to the
respective experimental upper bound, as given in
Eq.~(\ref{eq:edmexp}). All points above 1 are hence  
in conflict with the EDM data. For the neutron EDM
results are given for two different hadronic approaches, the chiral
quark model and the one applying QCD sum rule
techniques, Fig.~\ref{fig:edmcomplete} (middle
row).\footnote{In {\tt 
  NMSSMCALC} we also implemented the Parton Quark Model, but do
not show explicit results here.} In the Mercury EDM,
Fig.~\ref{fig:edmcomplete} (lower row), 
we have taken into account the uncertainties in the calculation of the
contribution from the Schiff moment by showing results for two of the
four values given in the literature, $d_{\text{Hg}}^{\mbox{I,II,III,IV}} [S]$
\cite{cheungeal,Ellis:2011hp}. The superscripts in $d_{\text{Hg}}$
indicate which value has been applied. As can be inferred from the
plots the most stringent constraints arise from the electron EDM,
Fig.~\ref{fig:edmcomplete} (upper left). The
constraints from the Thallium EDM, Fig.~\ref{fig:edmcomplete} (upper
right), which mostly depends on the 
electron EDM $dE_e$, are not as stringent. The results
for the neutron EDM based on the CQM and the QCD sum rule technique
differ by about 
a factor 2-3 in accordance with previous results in the literature
\cite{cheungeal,Domingo:2015qaa}. The pictures for the
Mercury EDM based on d$_{\text{Hg}}^{\mbox{I}}$ and
d$_{\text{Hg}}^{\mbox{IV}}$ are almost the 
same and similar to the ones based on d$_{\text{Hg}}^{\mbox{II}}$ and
d$_{\text{Hg}}^{\mbox{III}}$, which are  
not shown here. The electron EDM is larger for larger values of
$|\kappa \lambda|$, {\it i.e.} $|\kappa \lambda| > 0.1$. Note, that
CP violation in the tree-level Higgs sector is induced by terms
proportional to $|\kappa \lambda| \sin (\varphi_1 - \varphi_2)$ so
that for larger absolute 
values of $|\kappa \lambda|$ the CP-violating phase plays a more important role in
the EDM. The figures show, that despite the exclusion of some points
by the tight electron EDM constraints, scenarios are viable that
feature a large CP-violating phase, including possible maximum
CP violation $\varphi_2 = \pm \pi/2$. Finally let us remark that 
the asymmetry in the scattering of the points at $\varphi_2= 0$
and $\pm \pi$ is due to the fact, that our scan only extends over
positive values of $\lambda$ and that $\kappa$ appears always in the
product $\kappa \lambda$.  \s 

\begin{figure}[h!]
\begin{center}
\vspace*{-0.2cm}
\includegraphics[height=0.45\textwidth,angle=-90]{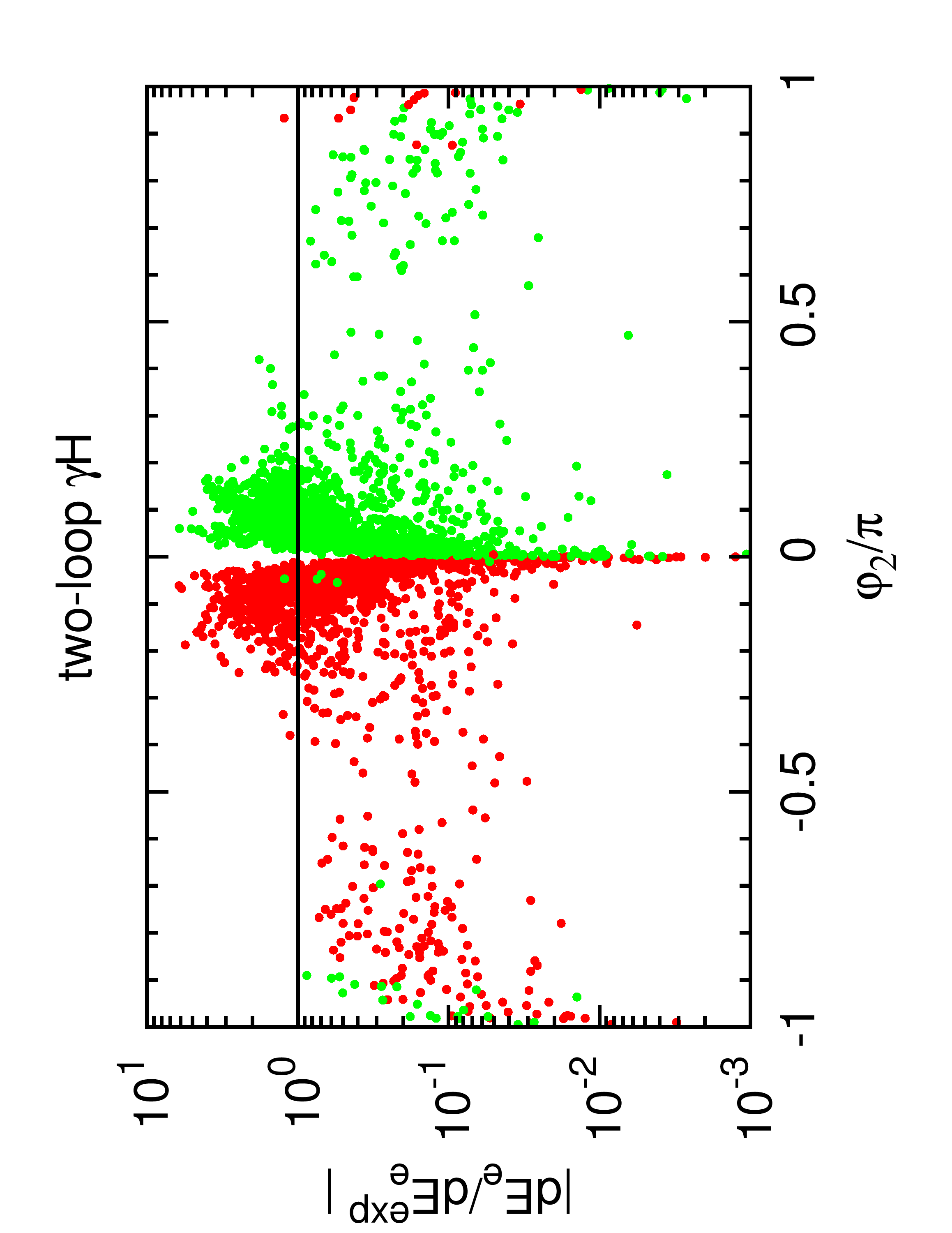}
 \includegraphics[height=0.45\textwidth,angle=-90]{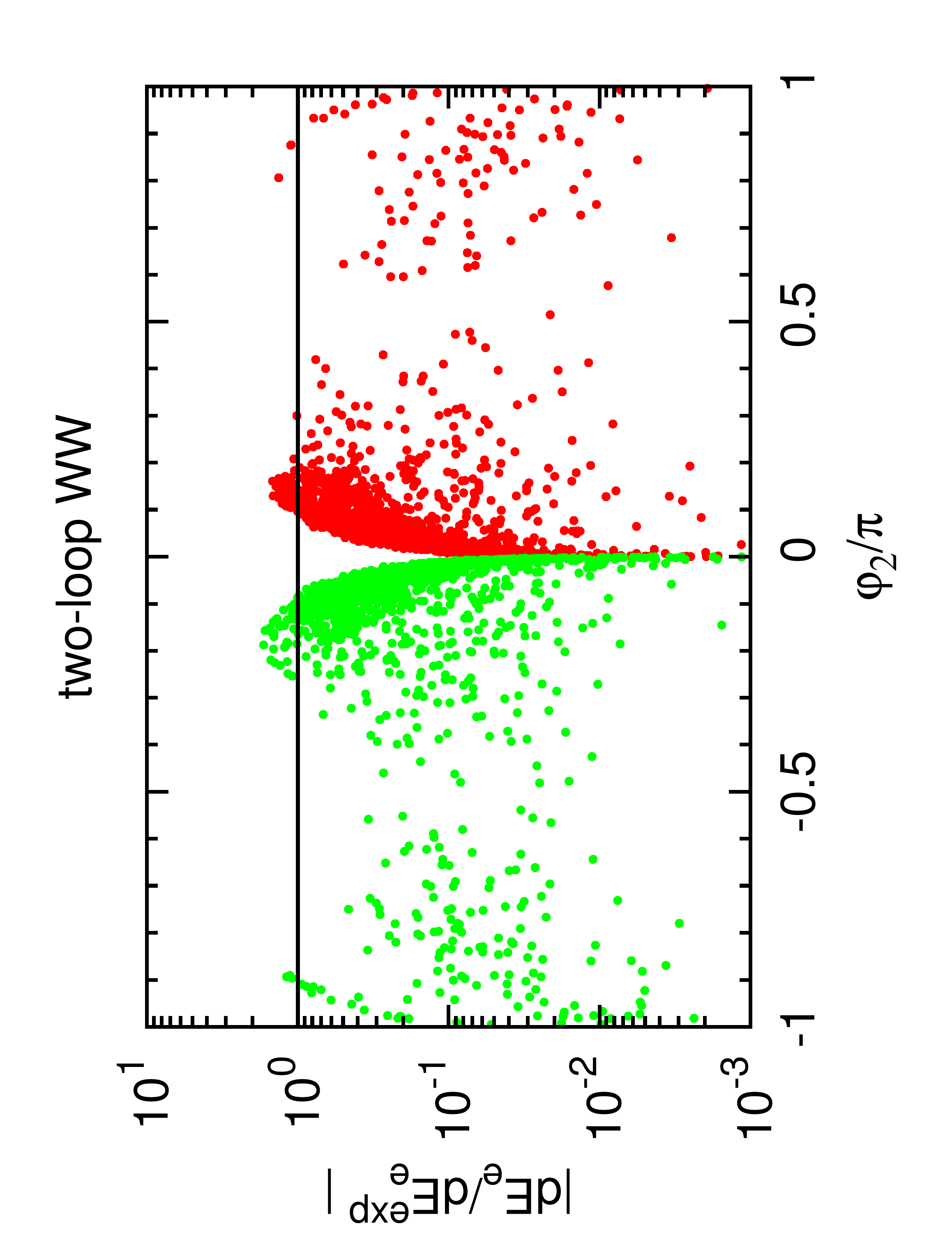}
\caption{Absolute values of the two-loop $\gamma H$
  (left) and $WW$ (right) contributions to the 
  electron EDM normalized to the experimental upper bound. Red points
  represent contributions with negative sign, green points those with 
  positive sign. \label{fig:edmbarrzee}} 
\end{center}
\end{figure}
Investigating the various contributions in detail, we find that at
one-loop level only the neutralino exchange diagrams contribute to the
electron EDM. All other one-loop contributions do not depend on
$\varphi_2$. The one-loop contributions turn out to be well below the
exclusion bounds. The dominant part comes from the two-loop diagrams
where the $\gamma H$ contribution is the most relevant one, but also
$WW$ is significant and comes with a different sign, see
Fig.~\ref{fig:edmbarrzee}. The $WH$ and $ZH$ contributions are about
an order of magnitude smaller. \s

The analysis of the Thallium EDM shows that the one-loop contributions
are tiny. The two-loop and the four-fermion operator contributions are
roughly of the same size and come with opposite sign, so that they
partly cancel each other. Concerning the neutron EDM both in the CQM
and in the QCD sum rule approximation the one-loop contributions are
irrelevant and the two-loop part is dominated by the Weinberg operator
contribution and the Barr-Zee type contributions to the CEDM of the
quarks. In the QCD sum rule approach the latter two come with opposite
signs. The four-fermion operator part that also contributes here, is
small. The largest contributions to the Mercury EDM originate from the
electron EDM. Also the contributions induced by the Schiff moment are
of comparable order, but come with a different sign. \s

\underline{\it Variation of $\varphi_{A_t}$:} We now turn to the
discussion of the effects of a non-vanishing phase of the stop
trilinear soft SUSY breaking coupling, $\varphi_{A_t}$. This is a
phase that also appears in the MSSM and we call this in the following
{\it MSSM-type CP violation}. 
\begin{figure}[h!]
\begin{center}
\vspace*{-0.2cm}
\includegraphics[height=0.45\textwidth,angle=-90]{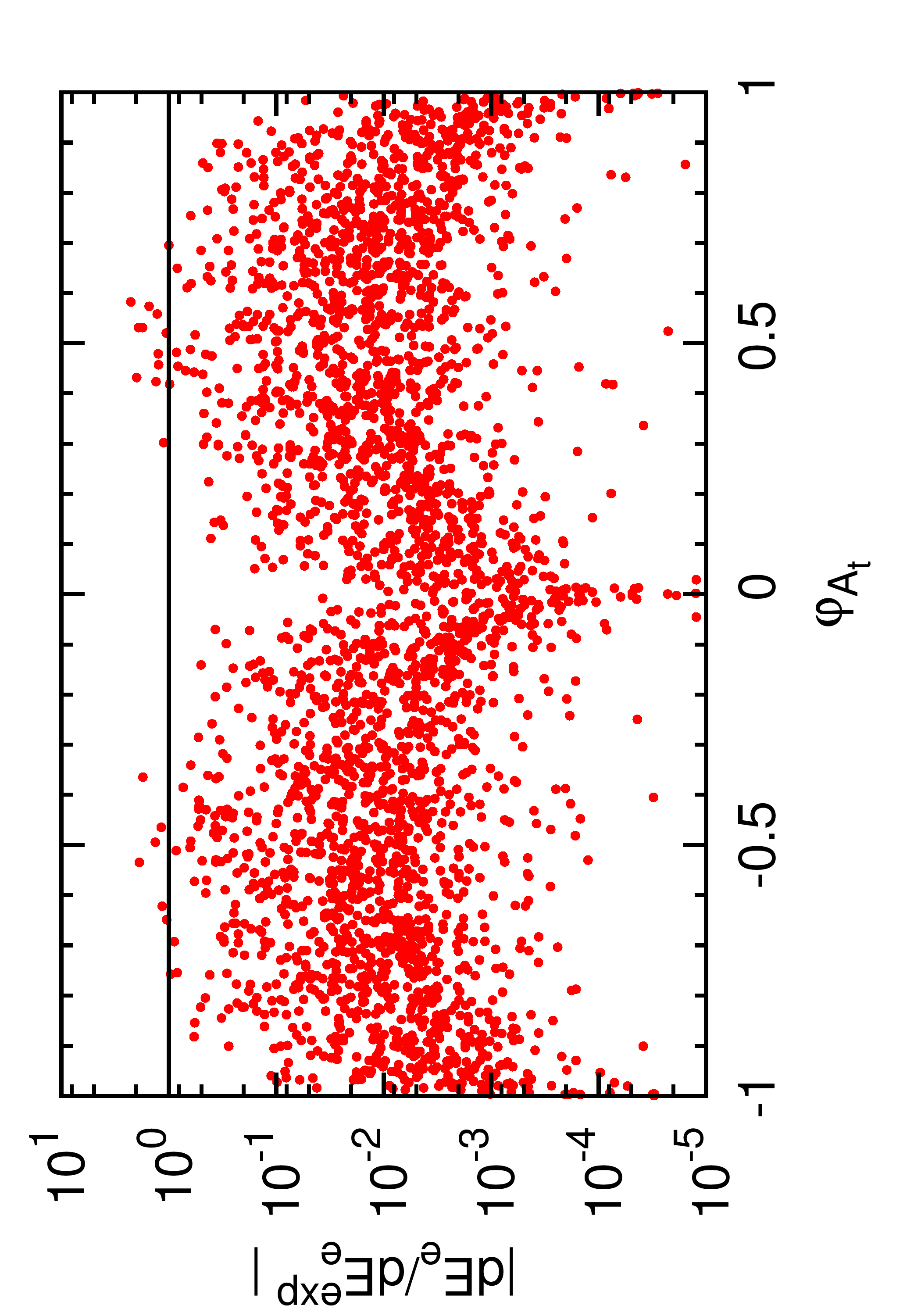}\includegraphics[height=0.45\textwidth,angle=-90]{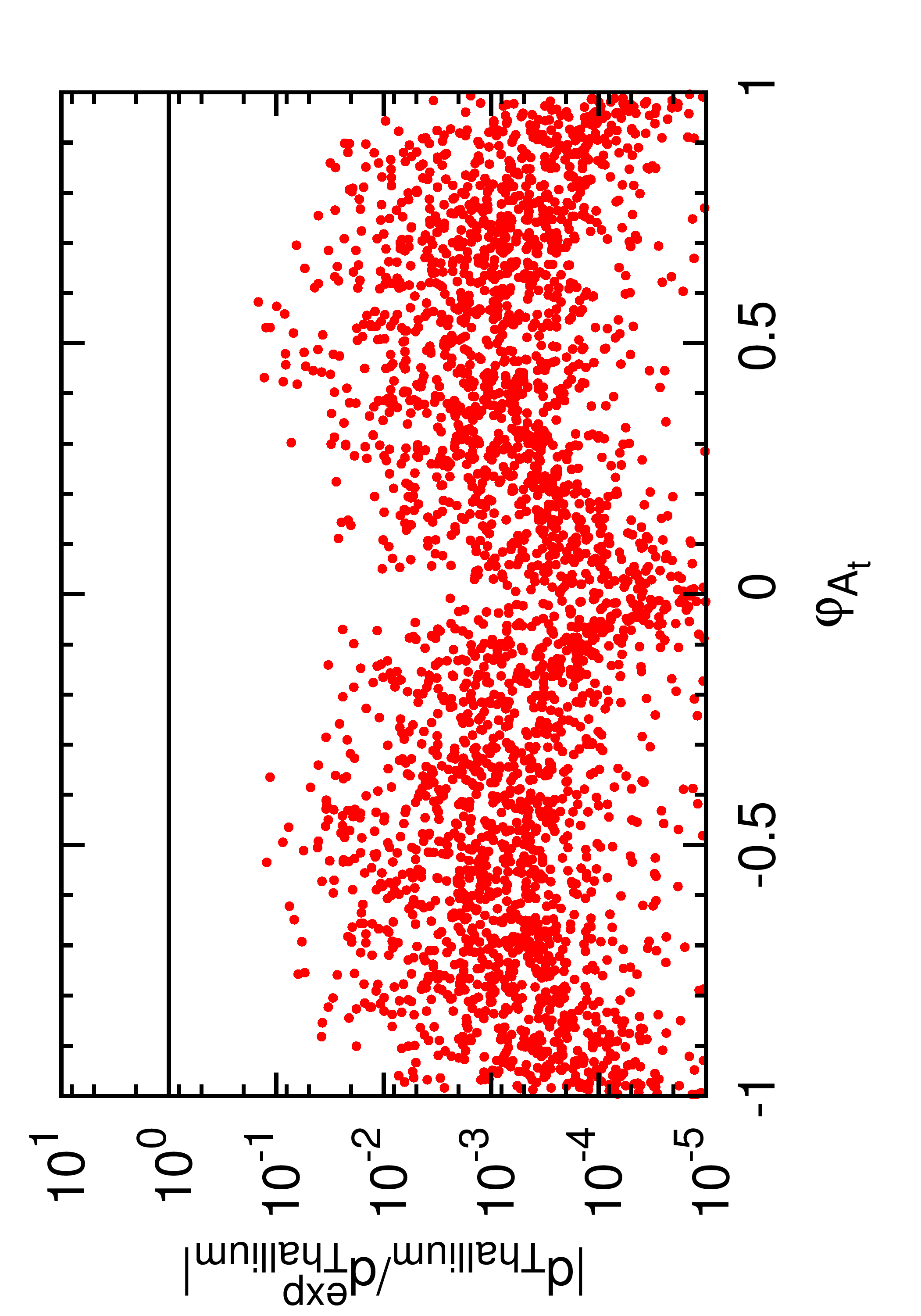} \\[-2mm]
\includegraphics[height=0.45\textwidth,angle=-90]{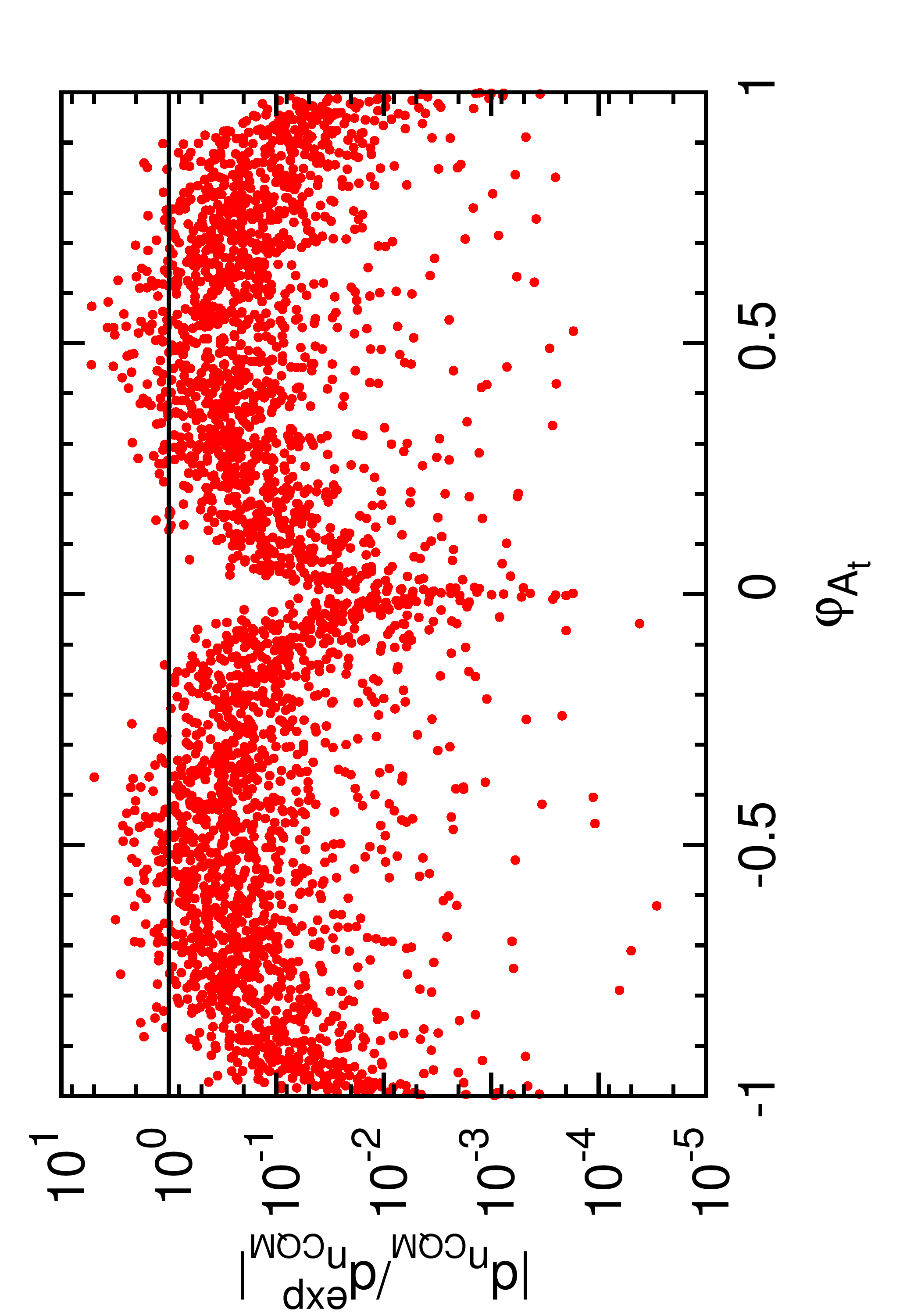}\includegraphics[height=0.45\textwidth,angle=-90]{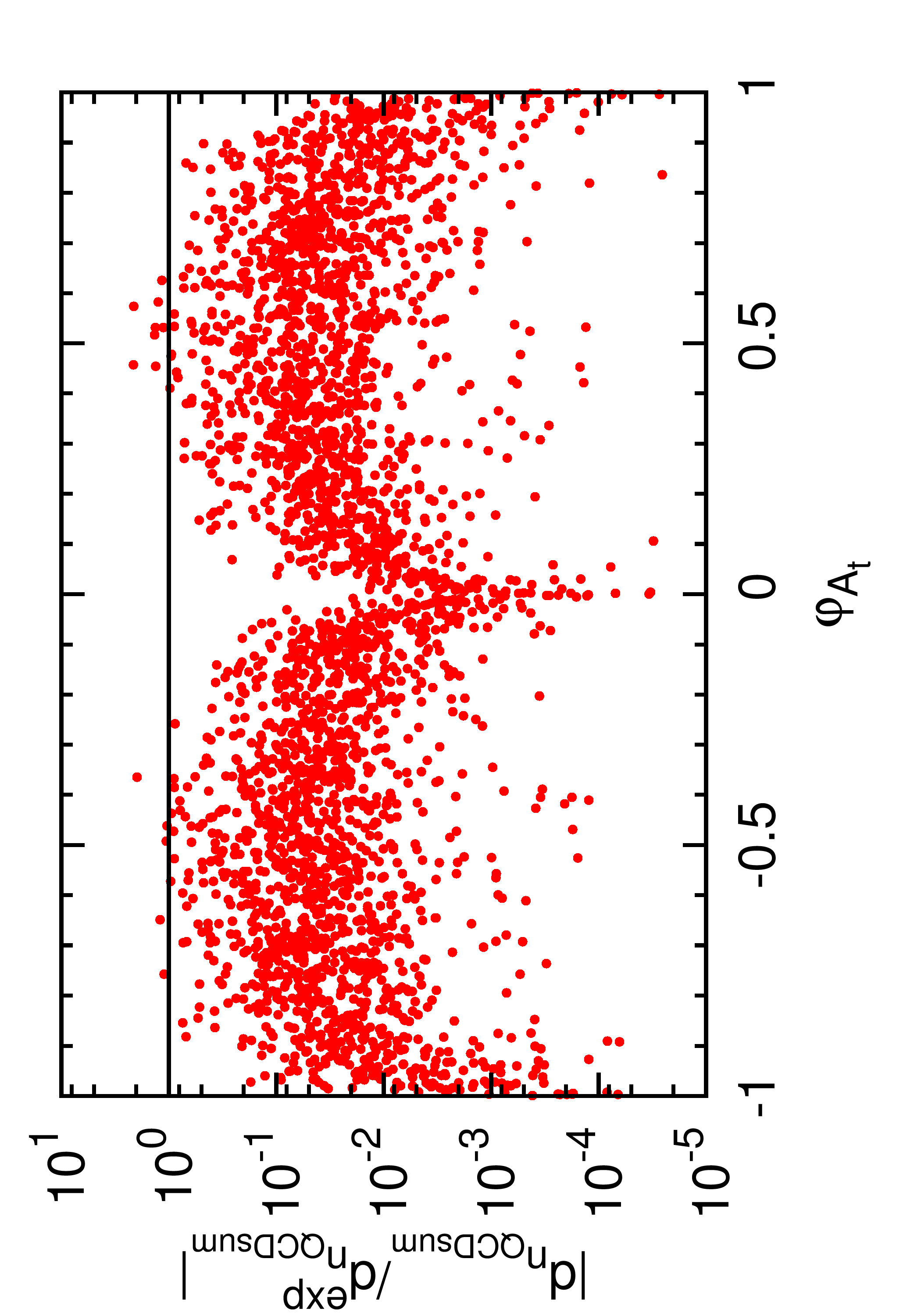} \\[-2mm]
\includegraphics[height=0.45\textwidth,angle=-90]{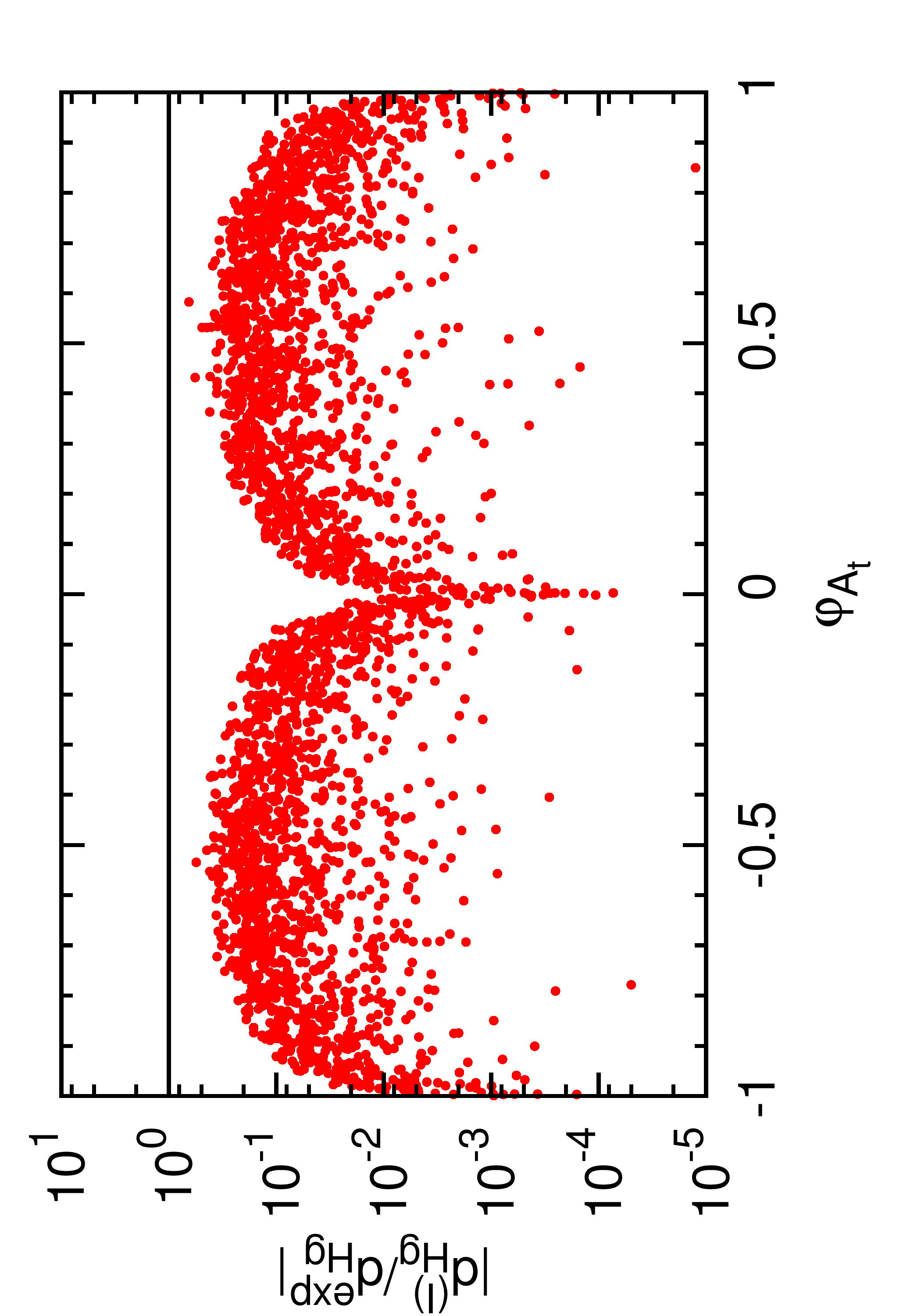}\includegraphics[height=0.45\textwidth,angle=-90]{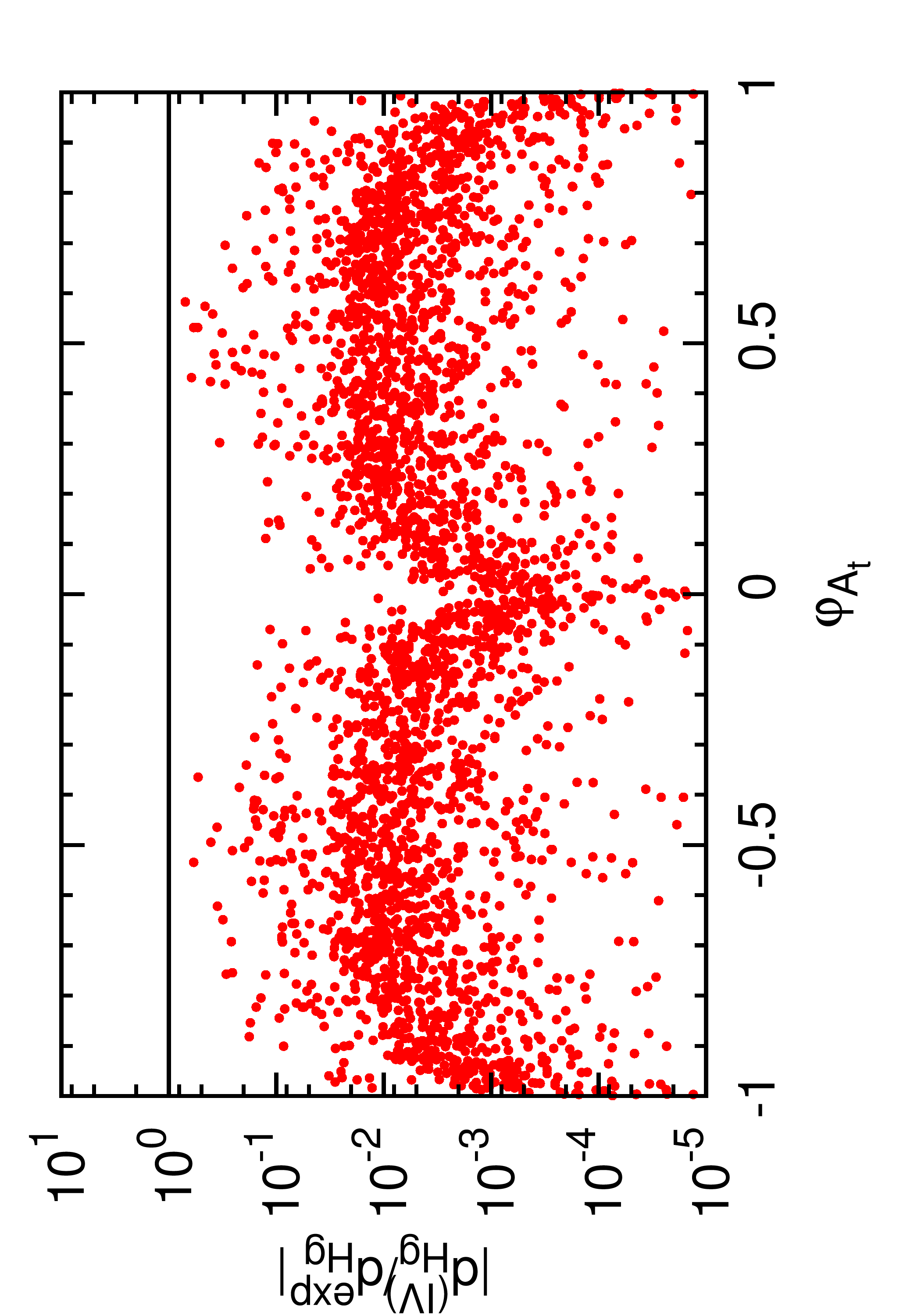} \\
\caption{Same as Fig.~\ref{fig:edmcomplete}, but for the variation of
  $\varphi_{A_t}$. All other CP-violating phases are set to zero. \label{fig:atedm}}
\end{center}
\end{figure}
Figure~\ref{fig:atedm} shows the absolute values of
the observable EDMs computed for the NMSSM scenarios from our scan
normalized to the experimental upper bounds, now as a function of
$\varphi_{A_t}$. All other possible CP-violating phases have been set
to zero. As can be inferred from the figure, the most important
observable EDMs induced by $\varphi_{A_t}$ are the electron EDM
$dE_e$, Fig.~\ref{fig:atedm} (upper left), and the neutron EDM $dn$,
Fig.~\ref{fig:atedm} (middle row), both, however, being well below the
experimental values for most of the parameter points.  
In the Mercury EDMs differences in the application of different
Schiff moment contributions are now well visible. They are not of
relevance though, as the obtained values remain below the experimental
limit. \s

\begin{figure}[ht!]
\begin{center}
\vspace*{-0.5cm}
\includegraphics[height=0.45\textwidth,angle=-90]{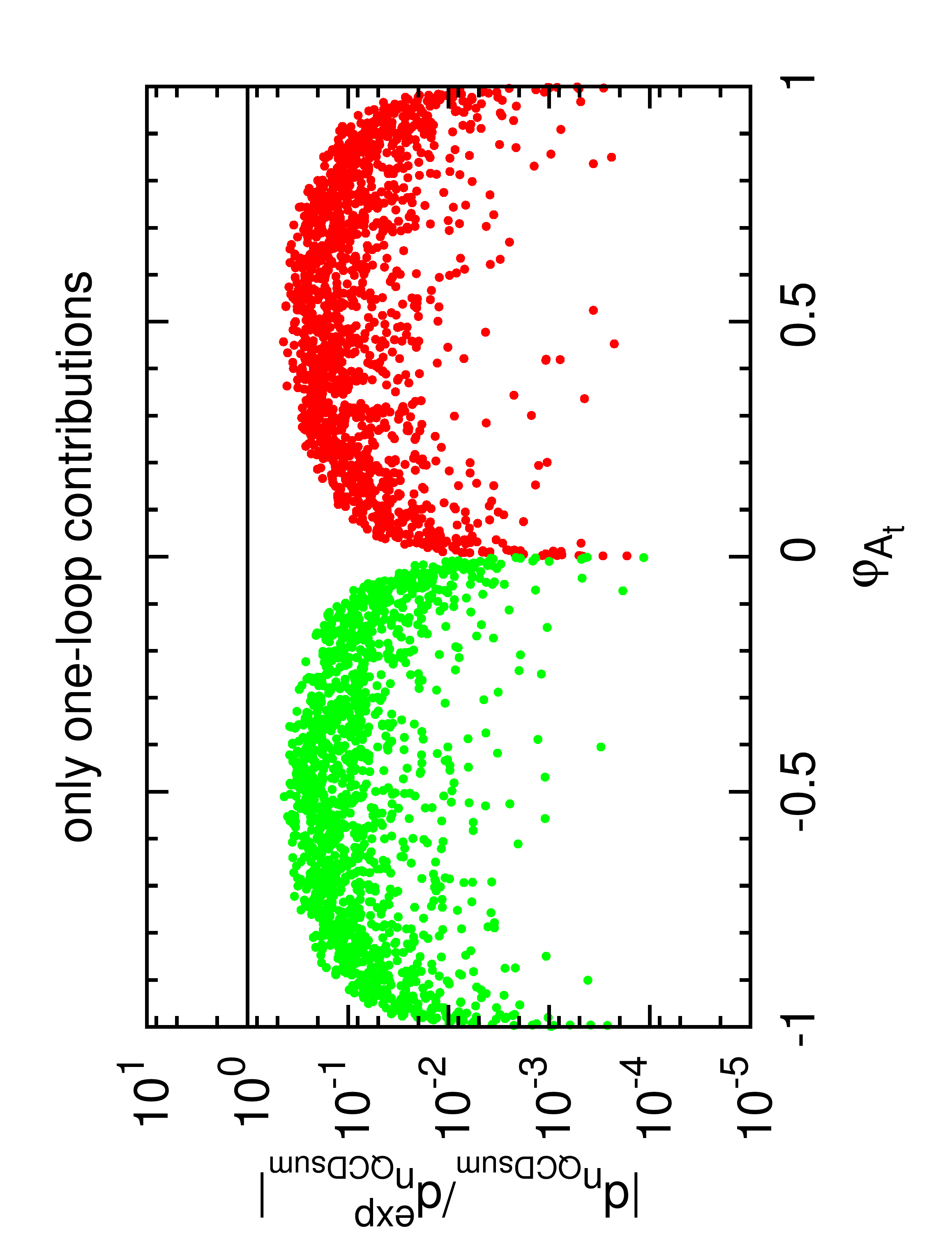}\includegraphics[height=0.45\textwidth,angle=-90]{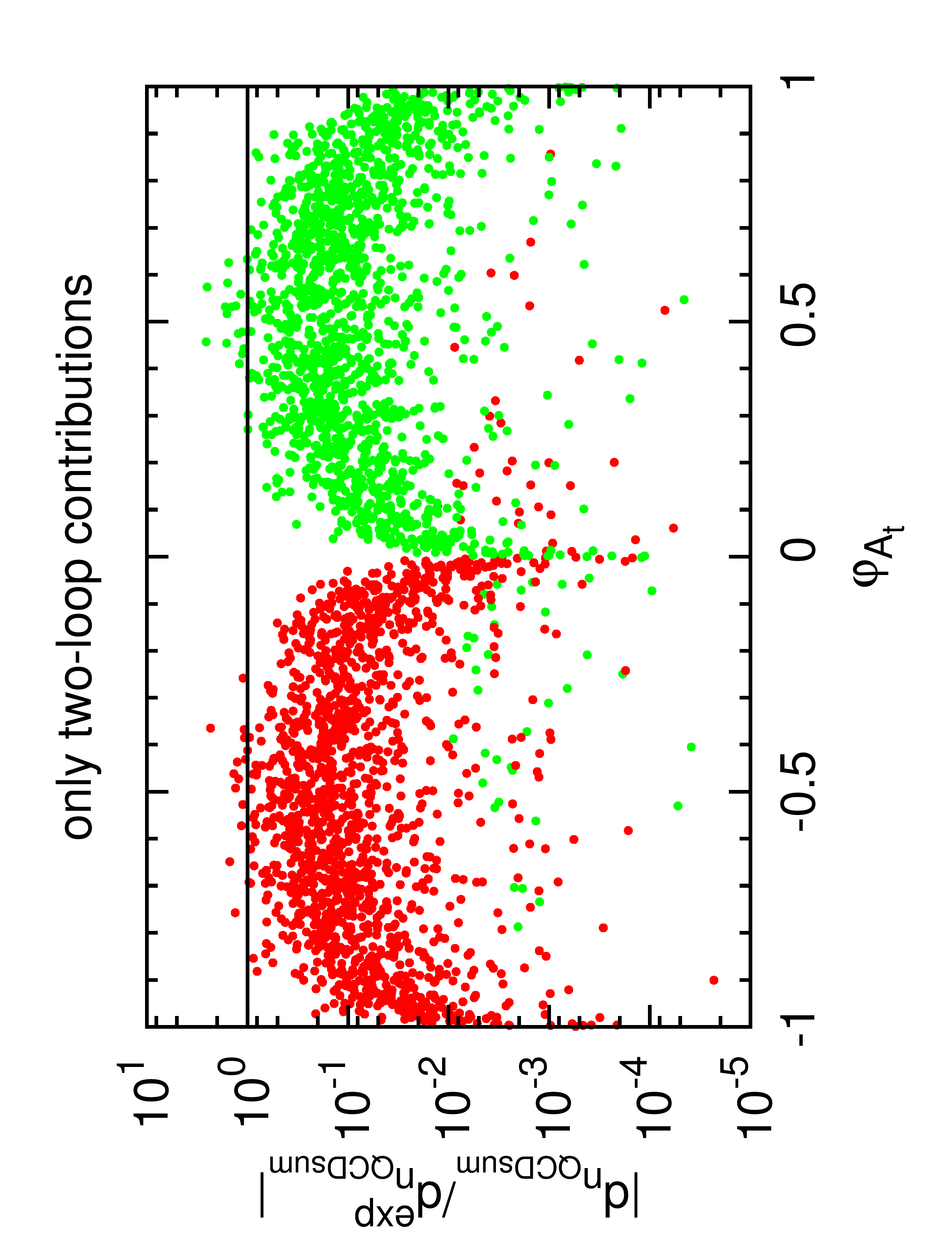}
\vspace*{-0.5cm}
\caption{Absolute values of the one- (left) and two-loop contributions
to the neutron EDM in the QCD sum rule approach, normalized to the
measured upper bound. Red points represent contributions with negative sign,
green points those with positive sign. \label{fig:1to2comp}} 
\end{center}
\end{figure}
\begin{figure}[h!]
\begin{center}
\vspace*{0cm}
\includegraphics[height=0.45\textwidth,angle=-90]{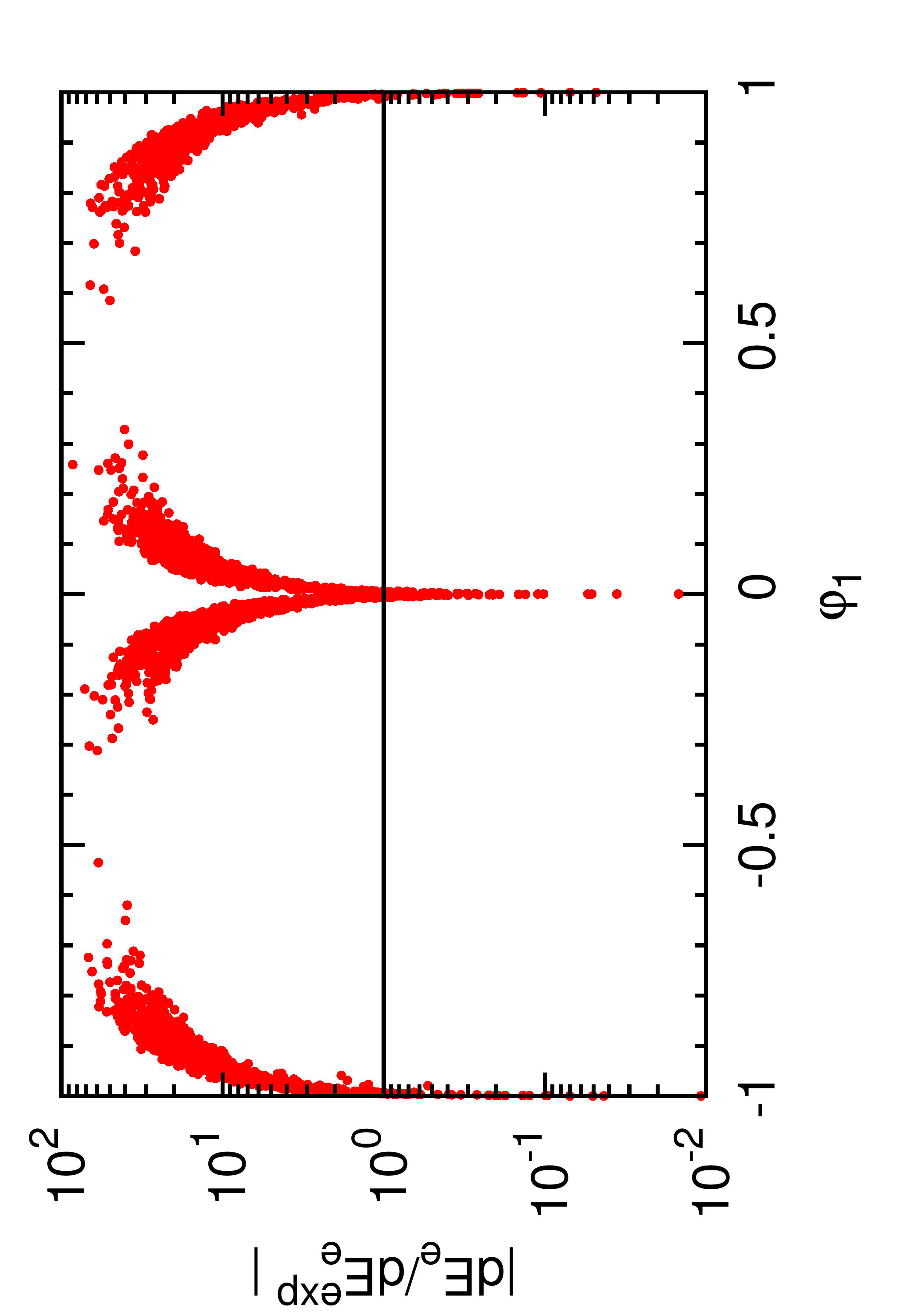}\includegraphics[height=0.45\textwidth,angle=-90]{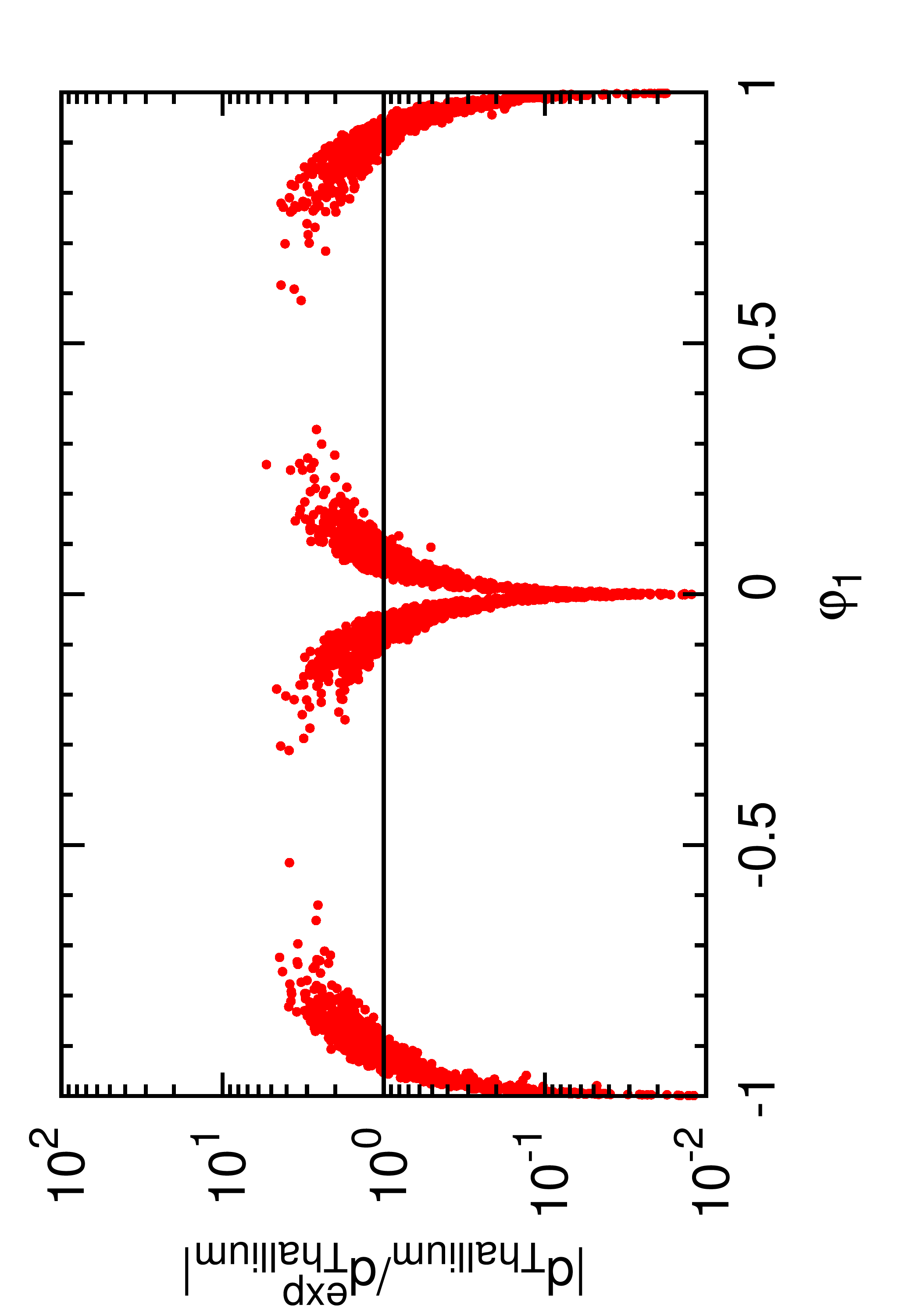} \\[-3mm]
\includegraphics[height=0.45\textwidth,angle=-90]{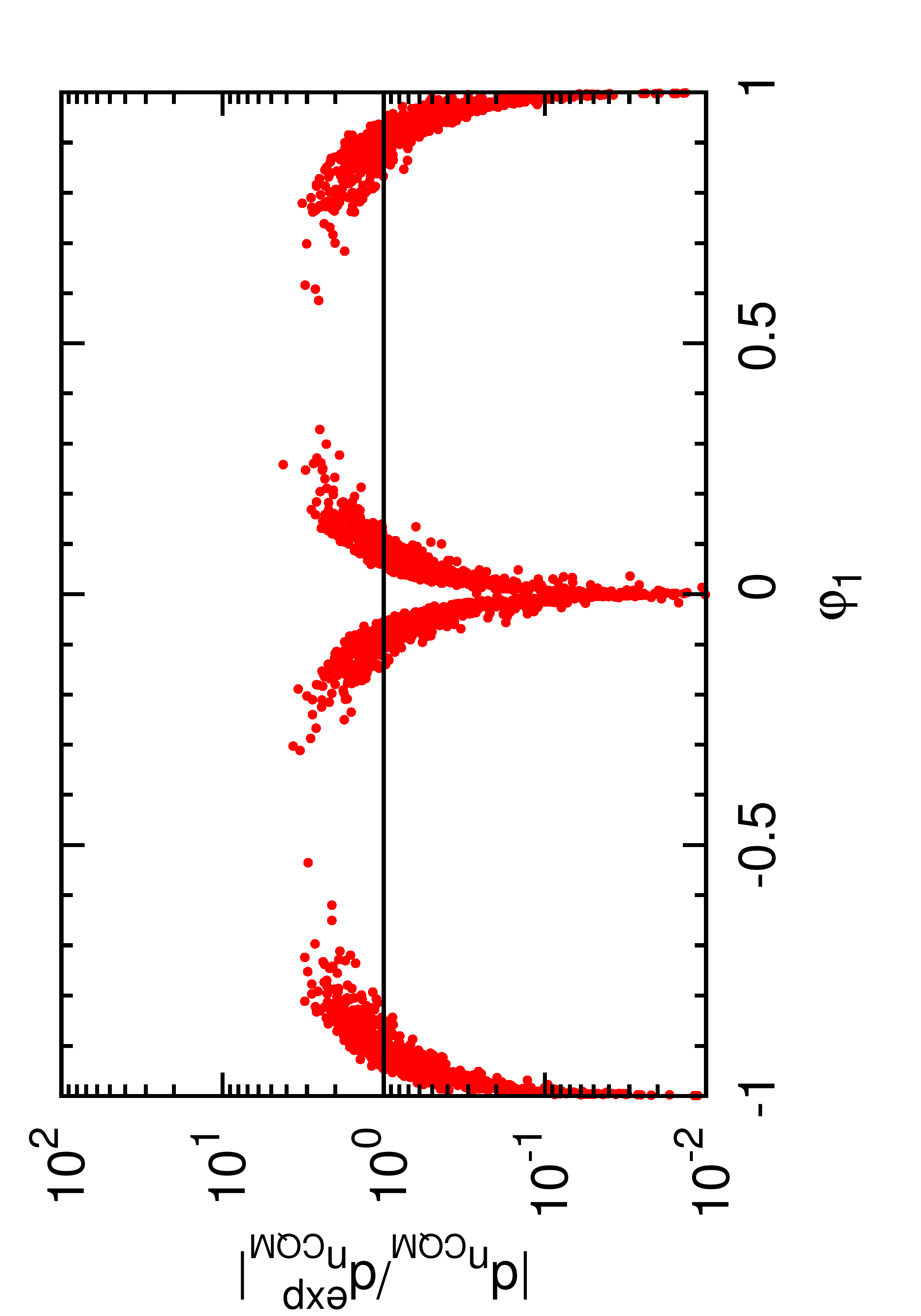}\includegraphics[height=0.45\textwidth,angle=-90]{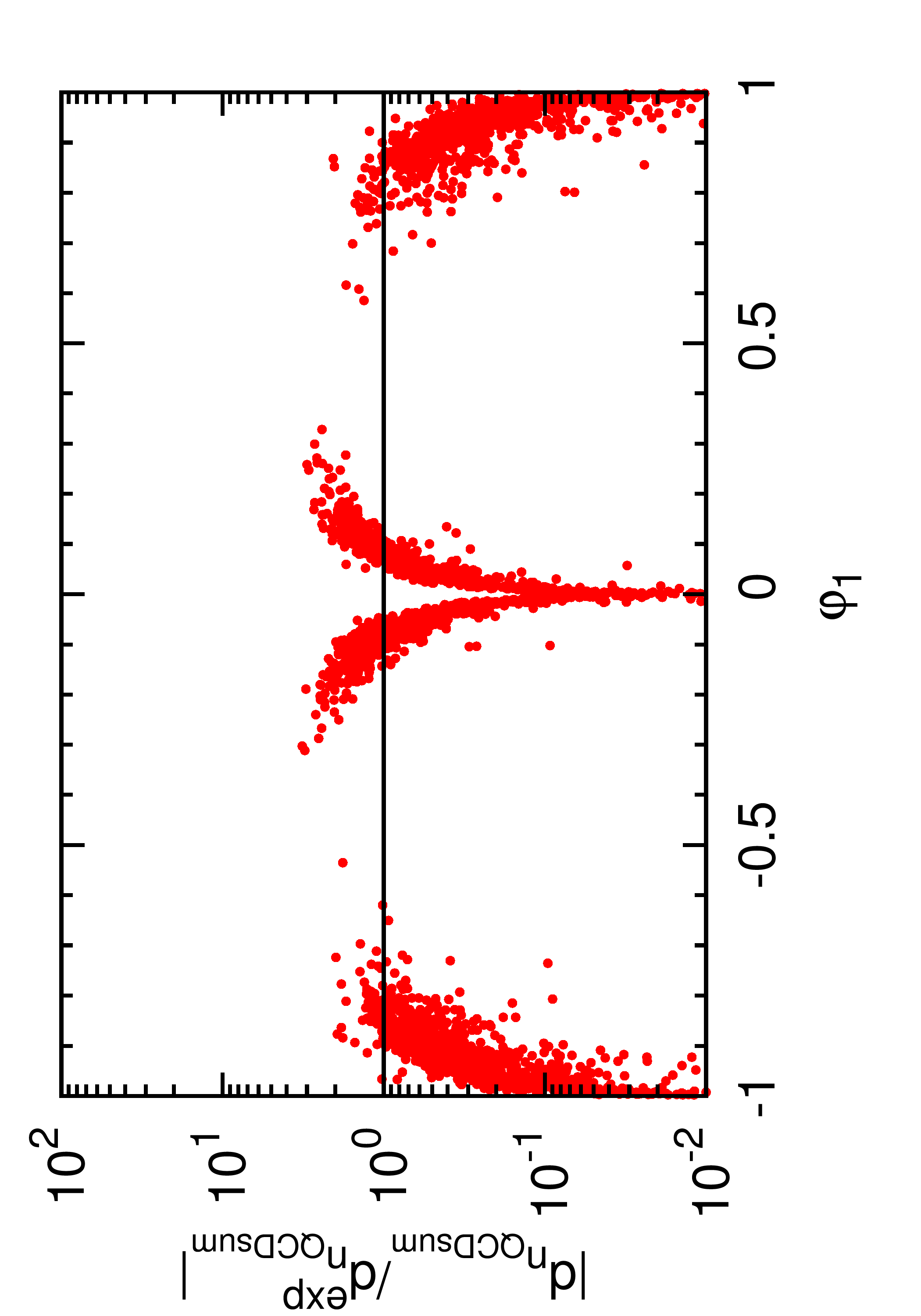} \\[-3mm]
\includegraphics[height=0.45\textwidth,angle=-90]{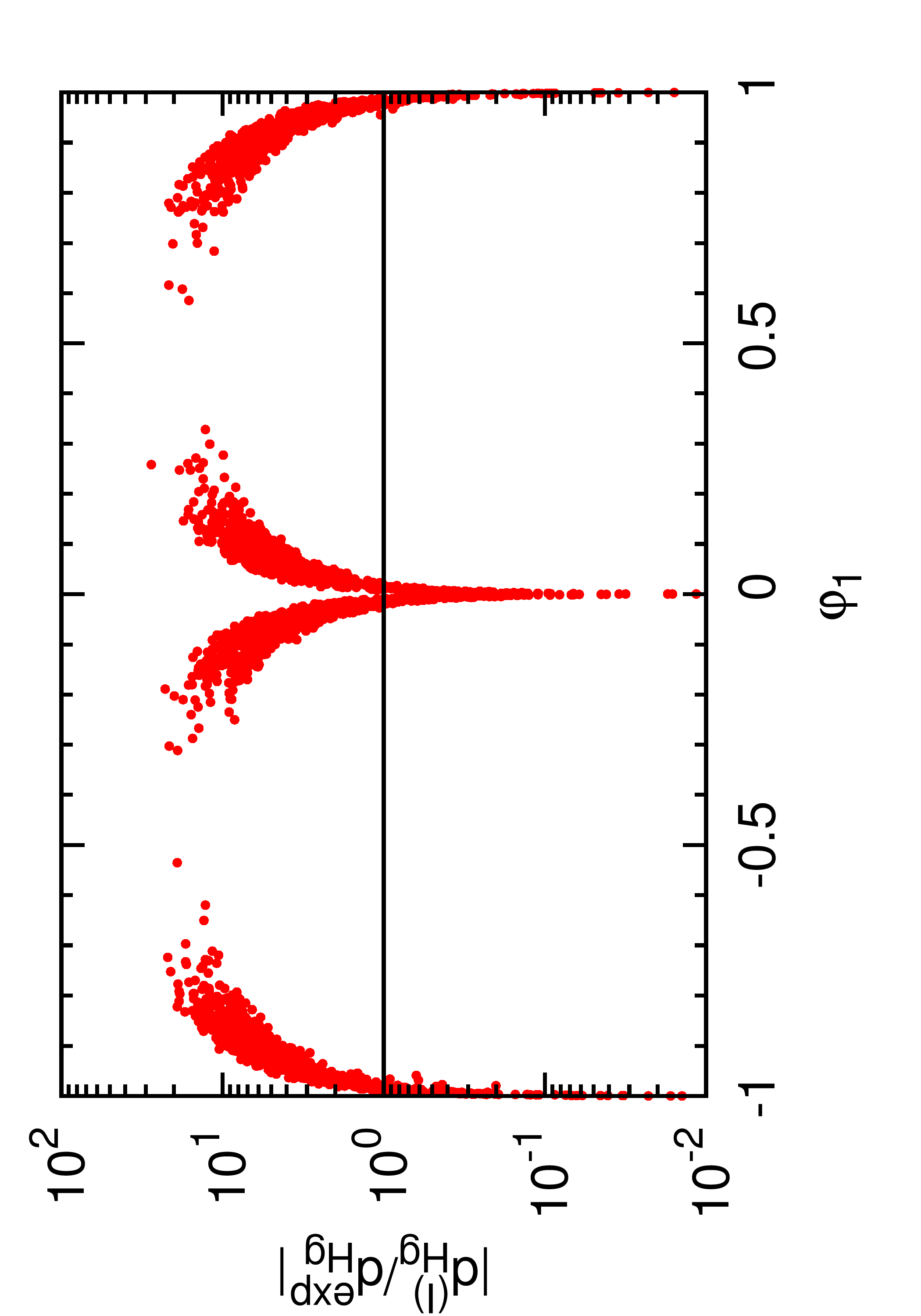}\includegraphics[height=0.45\textwidth,angle=-90]{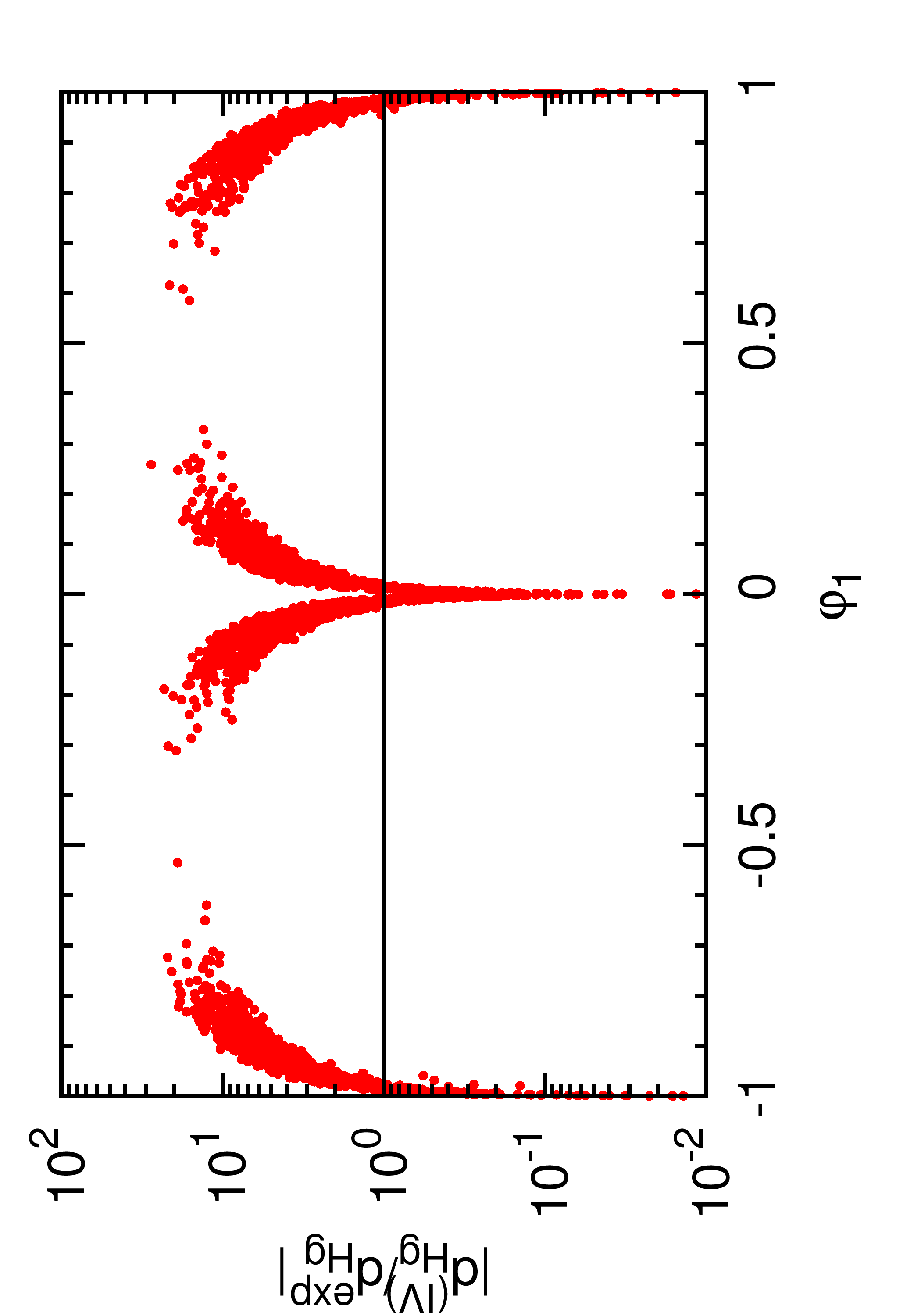}
\\[-0.5cm]
\caption{Absolute values of the electron (upper left), Thallium (upper
  right), neutron (middle) and Mercury (lower) EDMs as a function of
  $\varphi_1$, normalized to the respective experimental upper
  bound. \label{fig:lambdaedm}}
\end{center}
\vspace*{-0.8cm}
\end{figure}

Note that a non-zero phase $\varphi_{A_t}$ can
induce at one-loop level only for the up-quark an EDM and a CEDM through the
neutralino and gluino exchange diagrams, but solely when stops are involved, {\it
  cf.}~Figs.~\ref{fig:oneloopdiags} and \ref{fig:oneloopdiagscedm}. At
two-loop level, besides the Weinberg operator, only the Barr-Zee type
diagrams $\gamma H$ lead to a non-vanishing contribution, {\it
  cf.}~Figs.~\ref{fig:twoloopedm} and \ref{fig:twoloopcedm}. 
All other diagrams involve no couplings which contain $\varphi_{A_t}$. 
Therefore the electron and the Thallium EDM only receive two-loop
$\gamma H$ contributions, {\it cf.}~Fig.~\ref{fig:twoloopedm} (upper),
which change sign with the phase $\varphi_{A_t}$. In the neutron
EDM the two-loop contributions, which also include those from the
Weinberg operator, are about one to two orders of magnitude larger
than the one-loop contributions, depending on the approximation. Both
come with opposite sign, {\it cf.}~Fig.~\ref{fig:1to2comp}. The
Weinberg operator provides the 
dominant part at two-loop level, while the EDM and CEDM contributions
to the quarks, which enter at one- and two-loop level, are of about the
same size and come with opposite sign. \s

\underline{\it Variation of $\varphi_1$:} A non-vanishing phase
$\varphi_1$ leads to CP violation in the Higgs sector already at
tree level and hence generates an {\it NMSSM-type CP violation}. As
the phase $\varphi_1$ also enters the effective higgsino
parameter $\mu_{\text{eff}}$, {\it cf.}~Eq.~(\ref{eq:mueff}), it also
generates CP violation in the doublet higgsino and in the sfermion
sector as it occurs in the MSSM, and therefore leads to {\it MSSM-type
CP violation}. In Fig.~\ref{fig:lambdaedm} we show the effect of a
complex phase $\varphi_1$ on the values of the EDMs. All other
phases have been set to zero. As can be inferred from the plots, CP
violation induced by $\varphi_1$ is strongly constrained by the
EDMs. In particular the induced electron EDM is by a factor up to 100
times larger than the experimental upper bound. At one-loop level the
chargino contributions, that are MSSM-like, are important. The
Barr-Zee type two-loop contributions are an order of magnitude larger
than the ones induced by $\varphi_2 \ne 0$. In the Thallium EDM
the one- and two-loop contributions are of same size and come with the
same sign. They dominate over the four-fermion operator
part. For the neutron EDM the one-loop contributions dominate,
 while the contributions originating from the two-loop diagrams and the 
Weinberg operator are about one order of magnitude smaller and cancel
each other partly. In the Mercury EDM, the dominant part is built up by the
electron EDM. \s

In order to disentangle how much of the observed effect originates
from NMSSM-type or, respectively, MSSM-type CP violation, we varied the
phases $\varphi_1$ and $\varphi_2$ at the same time. Setting
$\varphi_1=\varphi_2$ allows to turn off tree-level CP
violation in the NMSSM Higgs sector, since this yields a vanishing
CP-violating phase in the 
tree-level Higgs sector. Hence only the CP-violating MSSM-type
contributions induced by $\varphi_1$ remain. The resulting plots,
which we do not show explicitly here, look strikingly similar to
Fig.~\ref{fig:lambdaedm}. The major difference is that in
Fig.~\ref{fig:lambdaedm} there are hardly any  points around the
maximally CP-violating phases $\varphi_1=\pm\pi/2$, whereas for the
$\varphi_1=\varphi_2$ variation these points exist. The gap in
Fig.~\ref{fig:lambdaedm} around $\varphi_1=\pm\pi/2$
can be attributed
to the fact that we demand compatibility with the Higgs data. However,
if the CP violation in the tree-level Higgs sector is too strong, {\it
  i.e.}~if the CP-odd admixture of the SM-like boson becomes too
large, the signals of the SM-like Higgs boson are not compatible with
the experimental values any more. Regarding the EDMs we conclude that in the
parameter space of the Natural NMSSM the MSSM-type CP-violating
contributions induced by $\varphi_1$ dominate by far over the
NMSSM-ones generated by $\varphi_1$. \s
\begin{figure}[hb!]
\begin{center}
\vspace*{-0.2cm}
\includegraphics[height=0.45\textwidth,angle=-90]{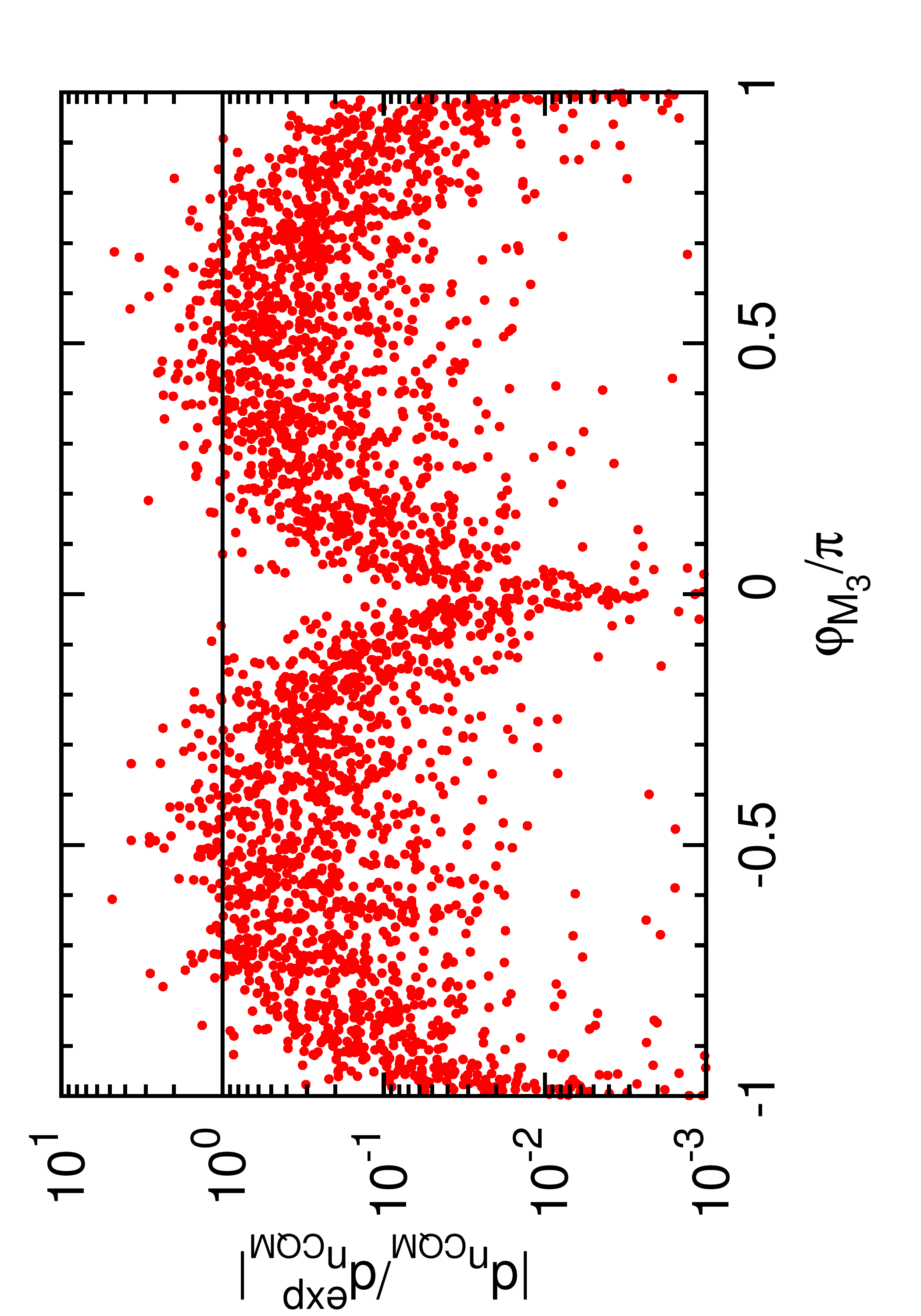}\includegraphics[height=0.45\textwidth,angle=-90]{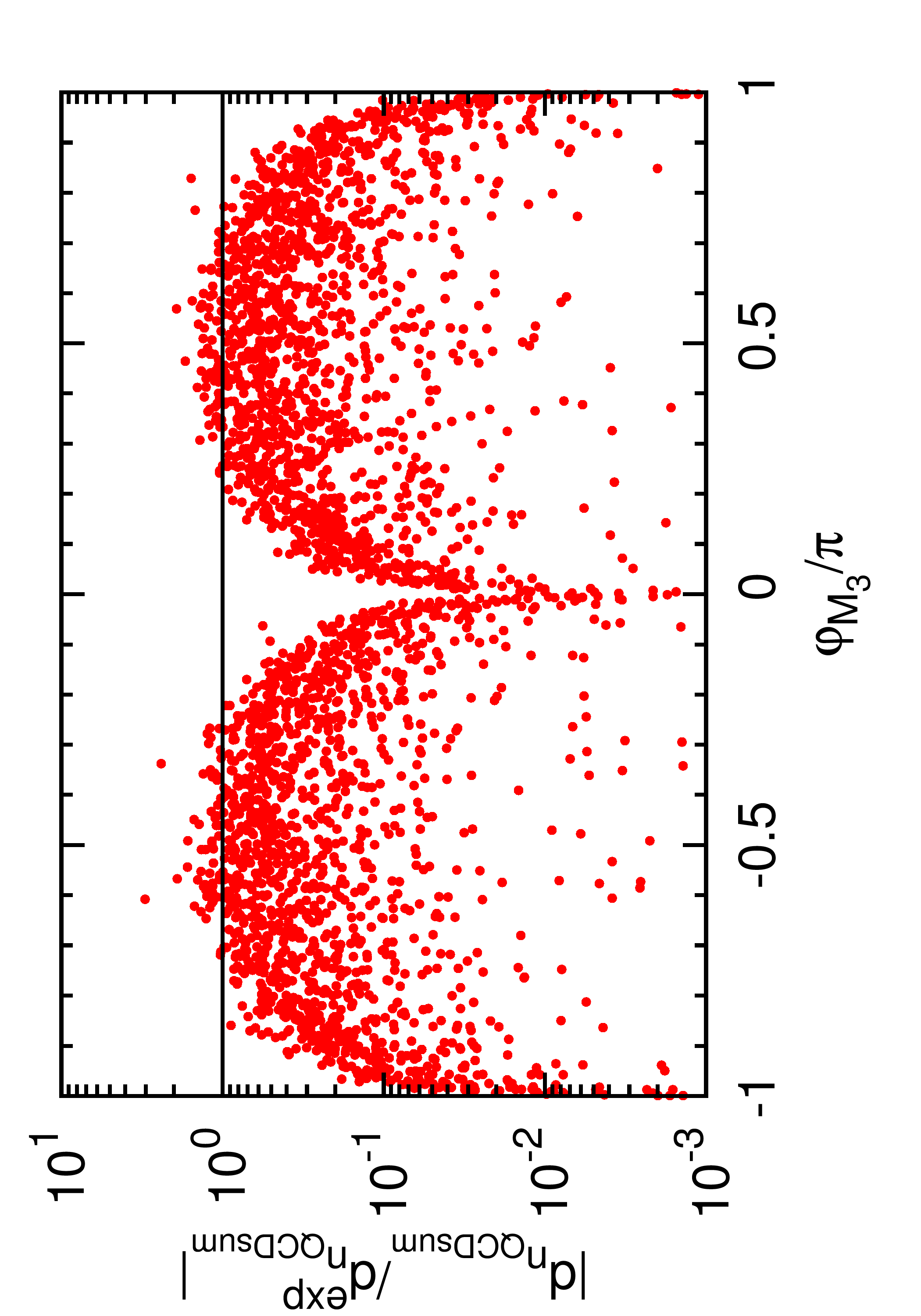} 
\\[-2mm]
\includegraphics[height=0.45\textwidth,angle=-90]{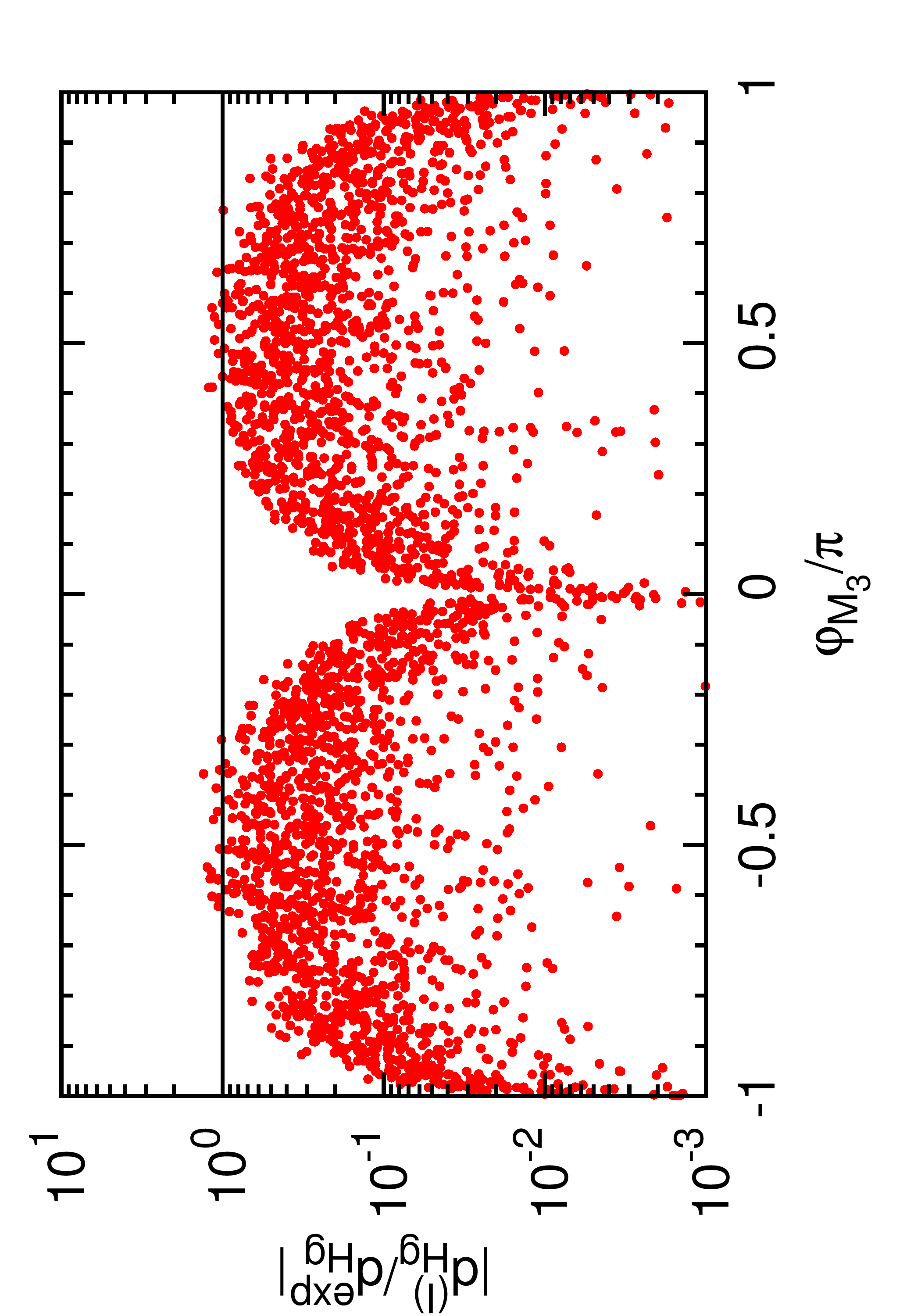}\includegraphics[height=0.45\textwidth,angle=-90]{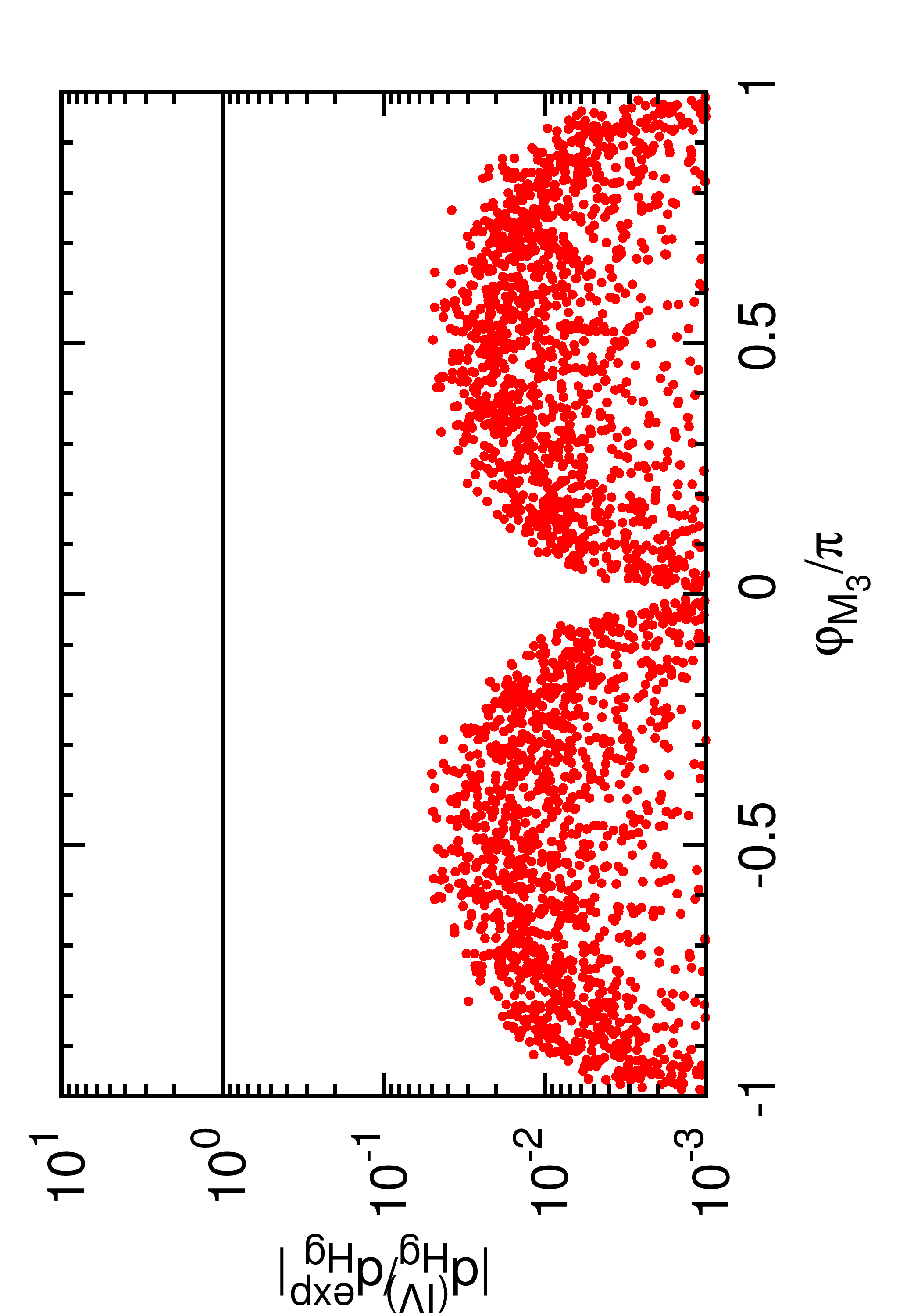} 
\\
\caption{Absolute values of the neutron 
  (upper) and Mercury (lower) EDMs as a function of
  $\varphi_{M_3}$, normalized to the respective experimental upper bound. 
\label{fig:m3edm}}
\end{center}
\end{figure}

\underline{\it Variation of the phases $\varphi_{M_i} (i=1,2,3$):}
We comment here on the influence of the phases $\varphi_{M_i}$ of the
gaugino mass parameters $M_i$ on the EDMs. They are other examples for
phases, that appear in the MSSM, too. Compared to the
results from the previously discussed phases $\varphi_1$,  $\varphi_2$
and $\varphi_{A_t}$, the gaugino phases do not lead to more stringent
constraints nor add new effects on the EDMs. \s

\begin{figure}[b!]
\begin{center}
\vspace*{-0.2cm}
\includegraphics[height=0.45\textwidth,angle=-90]{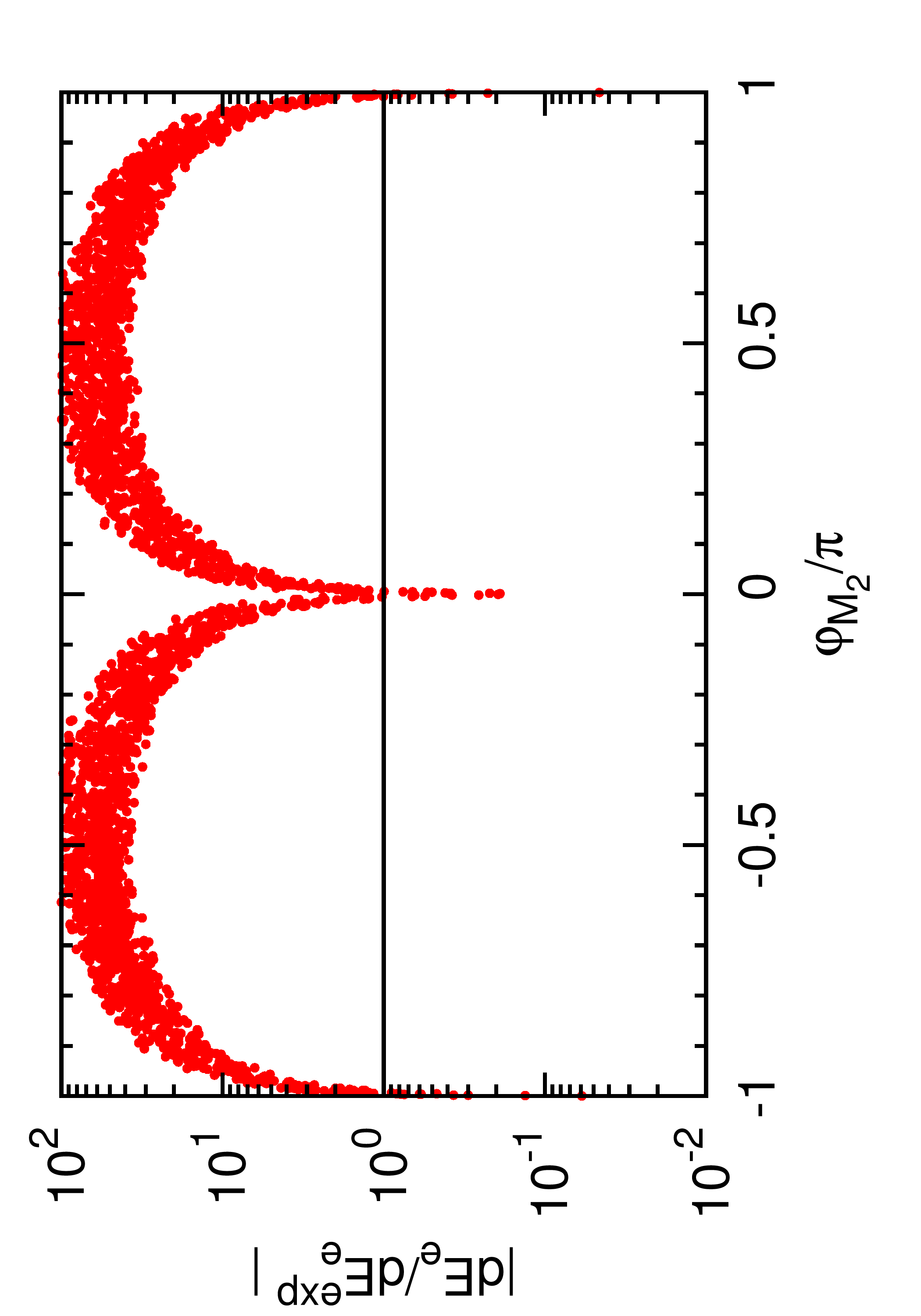}\includegraphics[height=0.45\textwidth,angle=-90]{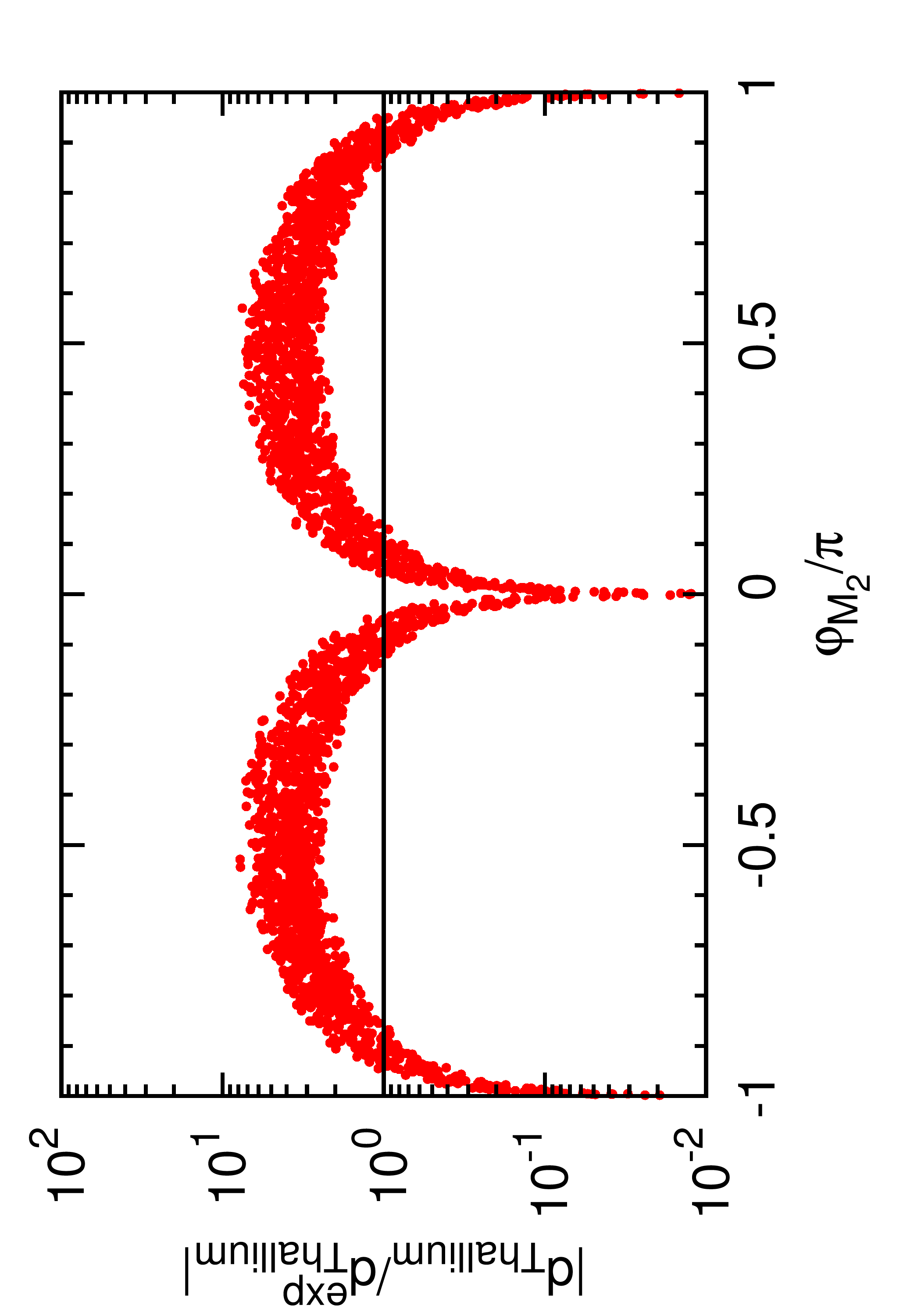} 
\\[-2mm]
\includegraphics[height=0.45\textwidth,angle=-90]{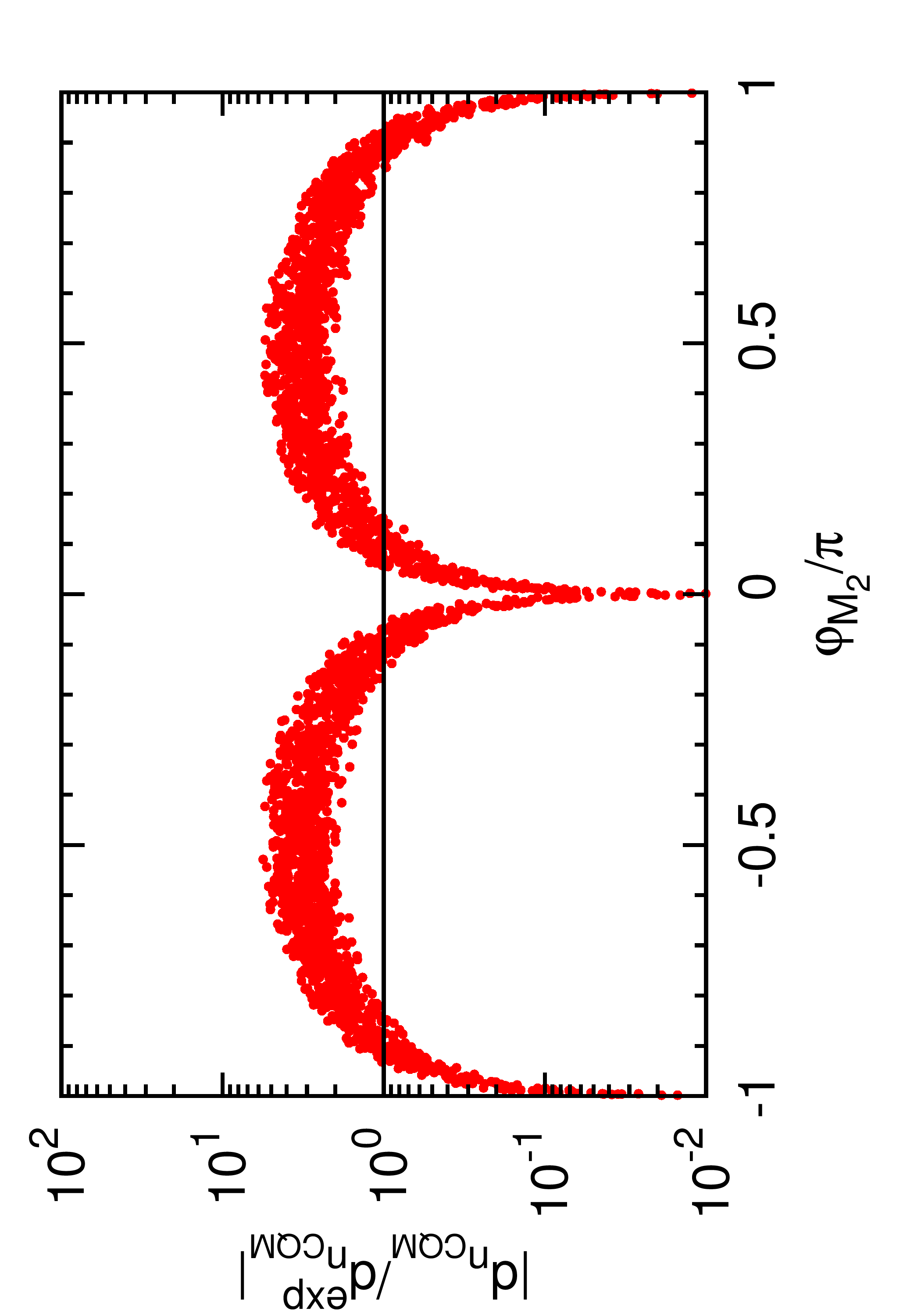}\includegraphics[height=0.45\textwidth,angle=-90]{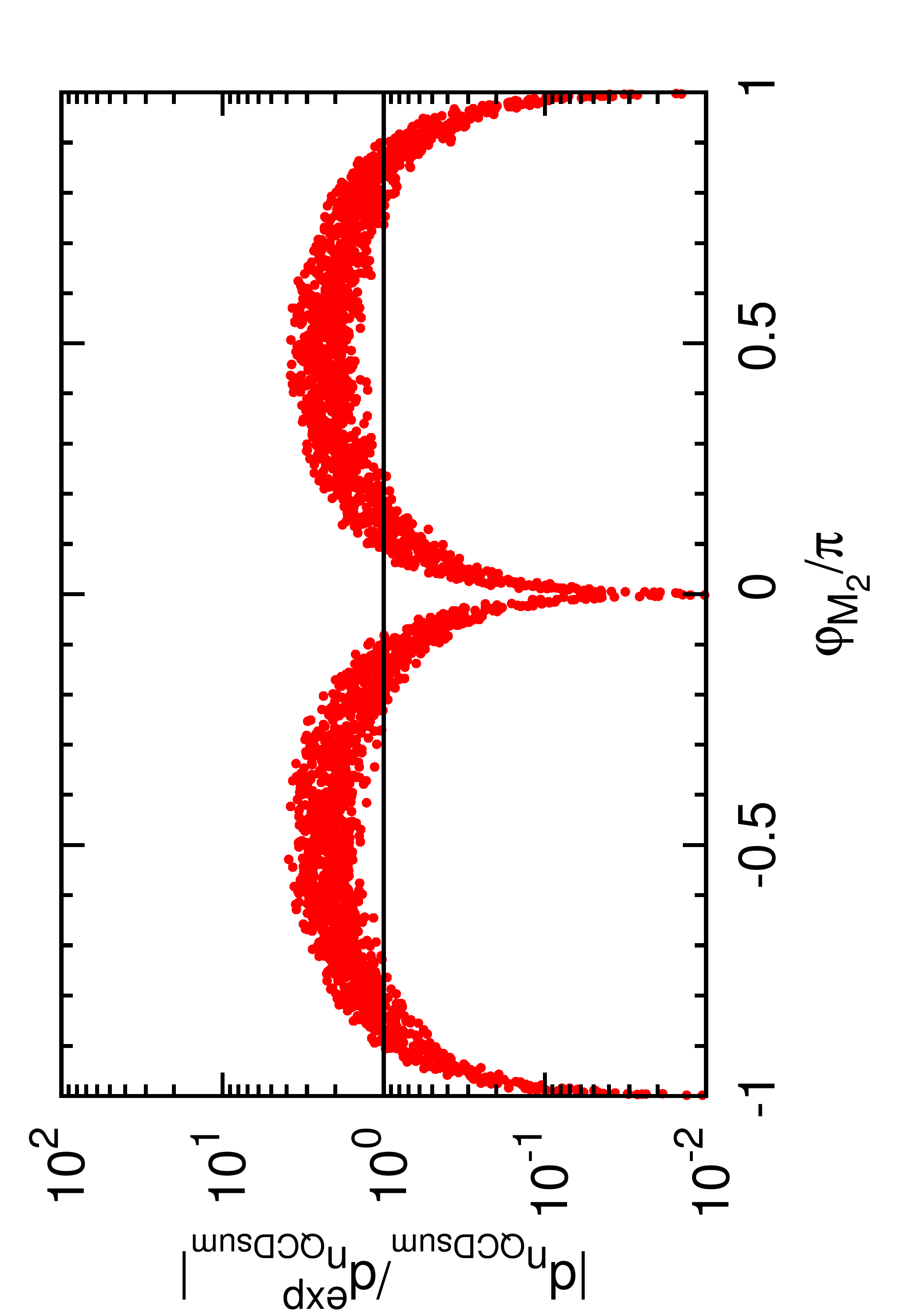} 
\\[-2mm]
\includegraphics[height=0.45\textwidth,angle=-90]{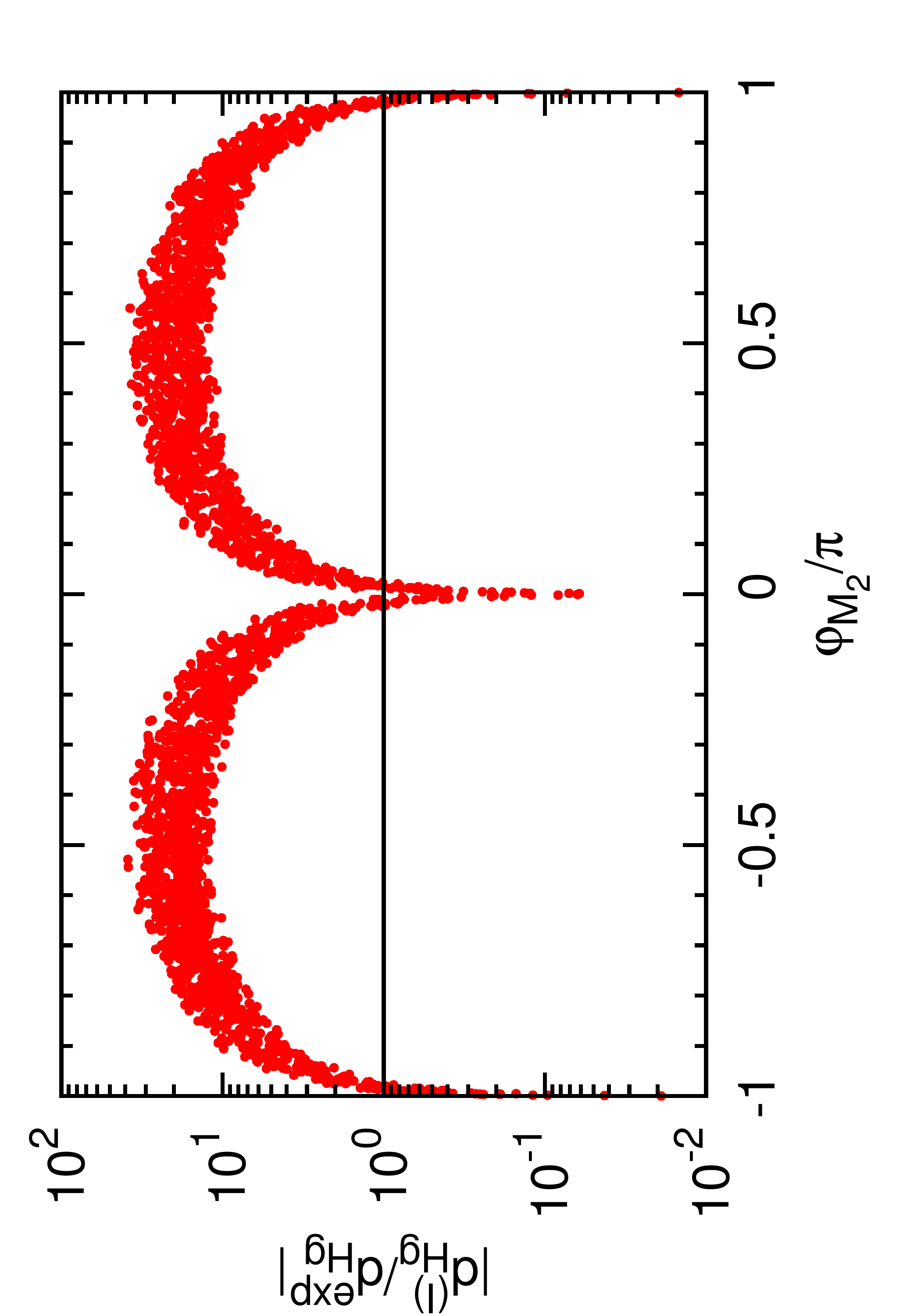}\includegraphics[height=0.45\textwidth,angle=-90]{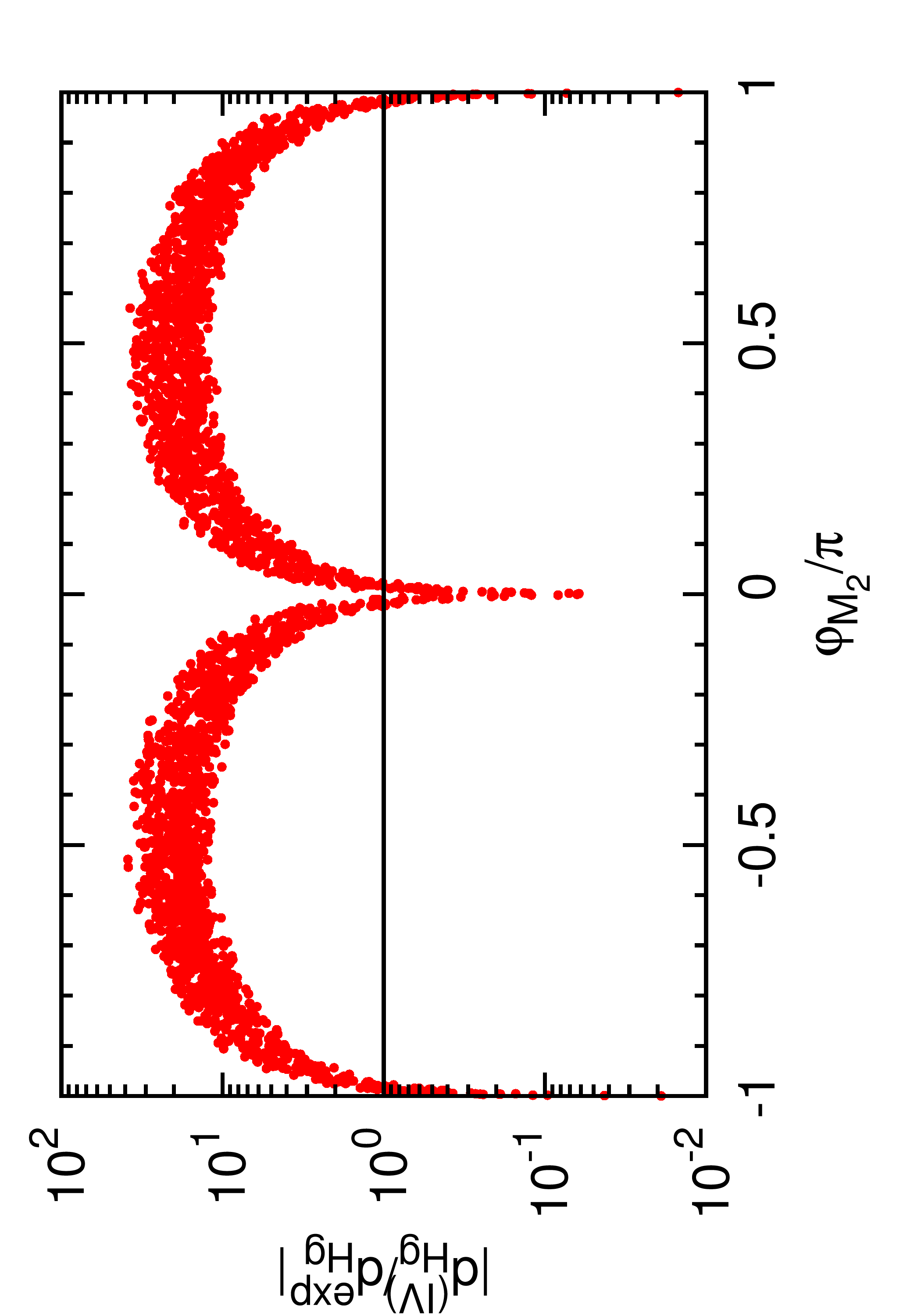} 
\\
\caption{Absolute values of the electron (upper
  left), Thallium (upper right), neutron 
  (middle) and Mercury (lower) EDMs as a function of
  $\varphi_{M_2}$, normalized to the respective experimental upper
  bound. \label{fig:m2edm}} 
\end{center}
\end{figure}
The phase $\varphi_{M_3}$, that arises in the stop and gluino sector,
does not contribute to the electron EDM. Its contributions to the
neutron and Mercury EDM are of the same size as those arising from
$\varphi_{A_t}$, as can be inferred from Fig.~\ref{fig:m3edm}. It shows the absolute
values of the neutron and the Mercury EDMs as a function of
$\varphi_{M_3}$, normalized to the respective experimental upper
bounds.  Here and in the following plots, for the neutron 
EDM results are given in the chiral quark model approach and in the one
based on QCD sum rule techniques. For the Mercury EDM the presented
results are obtained for two different values of the Schiff moment,
$d_{\text{Hg}}^{\mbox{I}} [S]$ and $d_{\text{Hg}}^{\mbox{IV}} [S]$, as
defined in \cite{cheungeal,Ellis:2011hp}. \s

The phase $\varphi_{M_2}$ plays a role in the chargino sector, just as
the phase of the effective $\mu$ parameter, encoded in $\varphi_1$. The
sizes of the EDMs due to $\varphi_{M_2}$ can therefore be expected to
be of the same order as those due to $\varphi_1$. This is confirmed by
the plots given in Fig.~\ref{fig:m2edm}, which show the absolute
values of the electron, Thallium, neutron and Mercury EDMs as a
function of $\varphi_{M_2}$, normalized to the respective 
experimental upper bounds. Since the phase $\varphi_{M_2}$ has only a
marginal effect on the Higgs sector, the compatibility of the Higgs
data is not affected by $\varphi_{M_2}$ and no gap around $\pm \pi/2$
arises as in the results for non-zero $\varphi_1$, {\it
  cf.}~Fig.~\ref{fig:lambdaedm}. \s

\begin{figure}[b!]
\begin{center}
\vspace*{-0.2cm}
\includegraphics[height=0.45\textwidth,angle=-90]{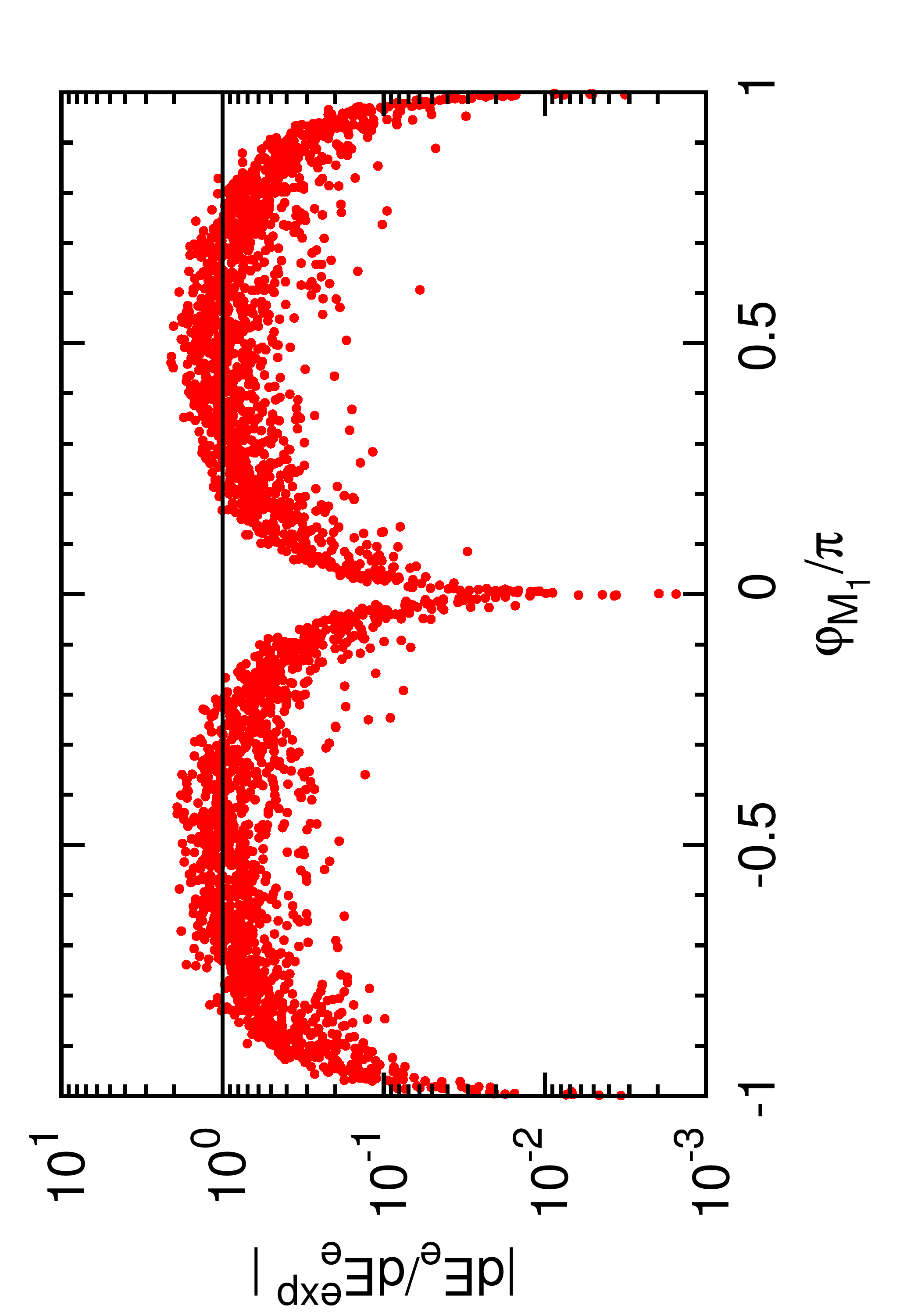}\includegraphics[height=0.45\textwidth,angle=-90]{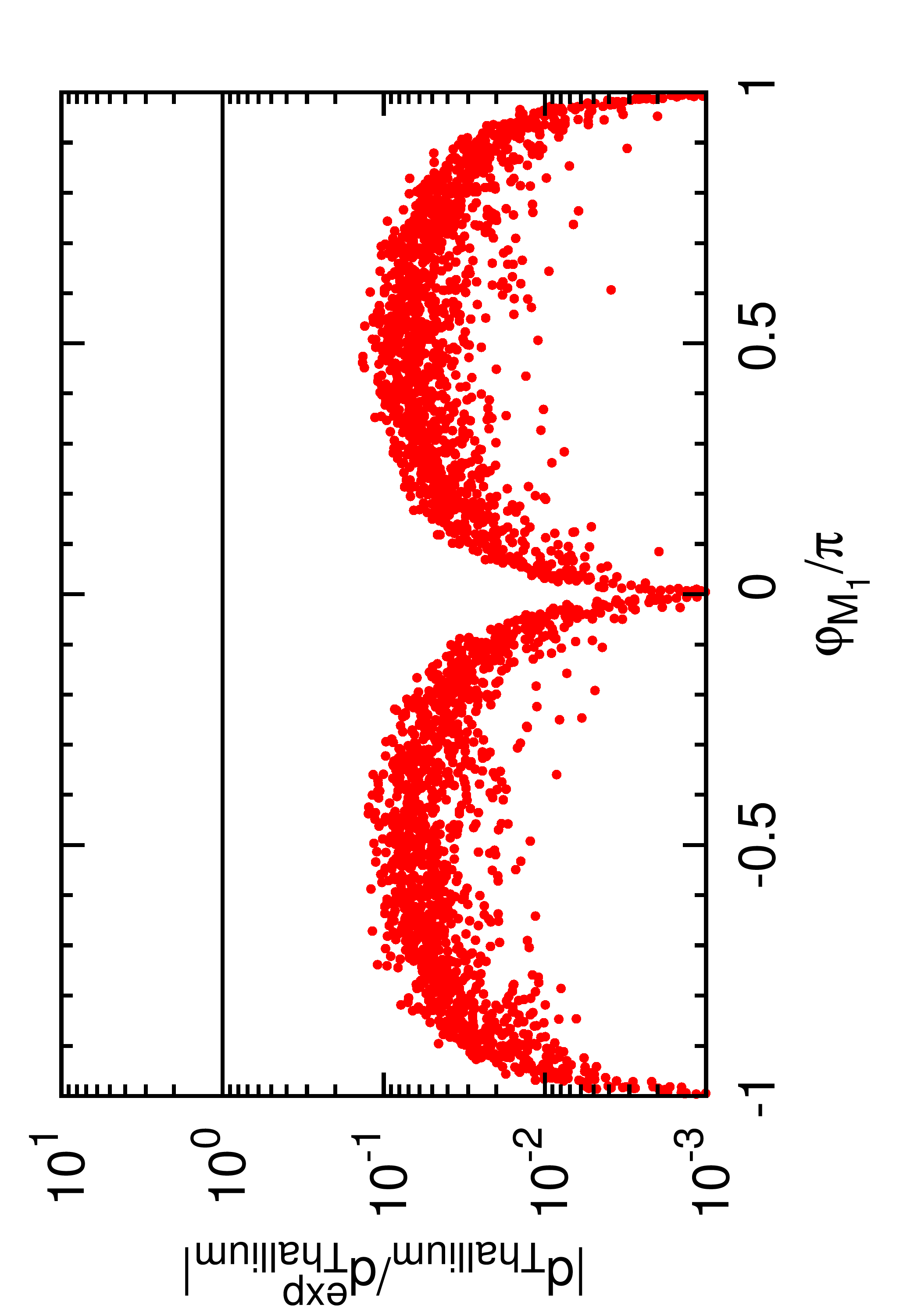} 
\\[-2mm]
\includegraphics[height=0.45\textwidth,angle=-90]{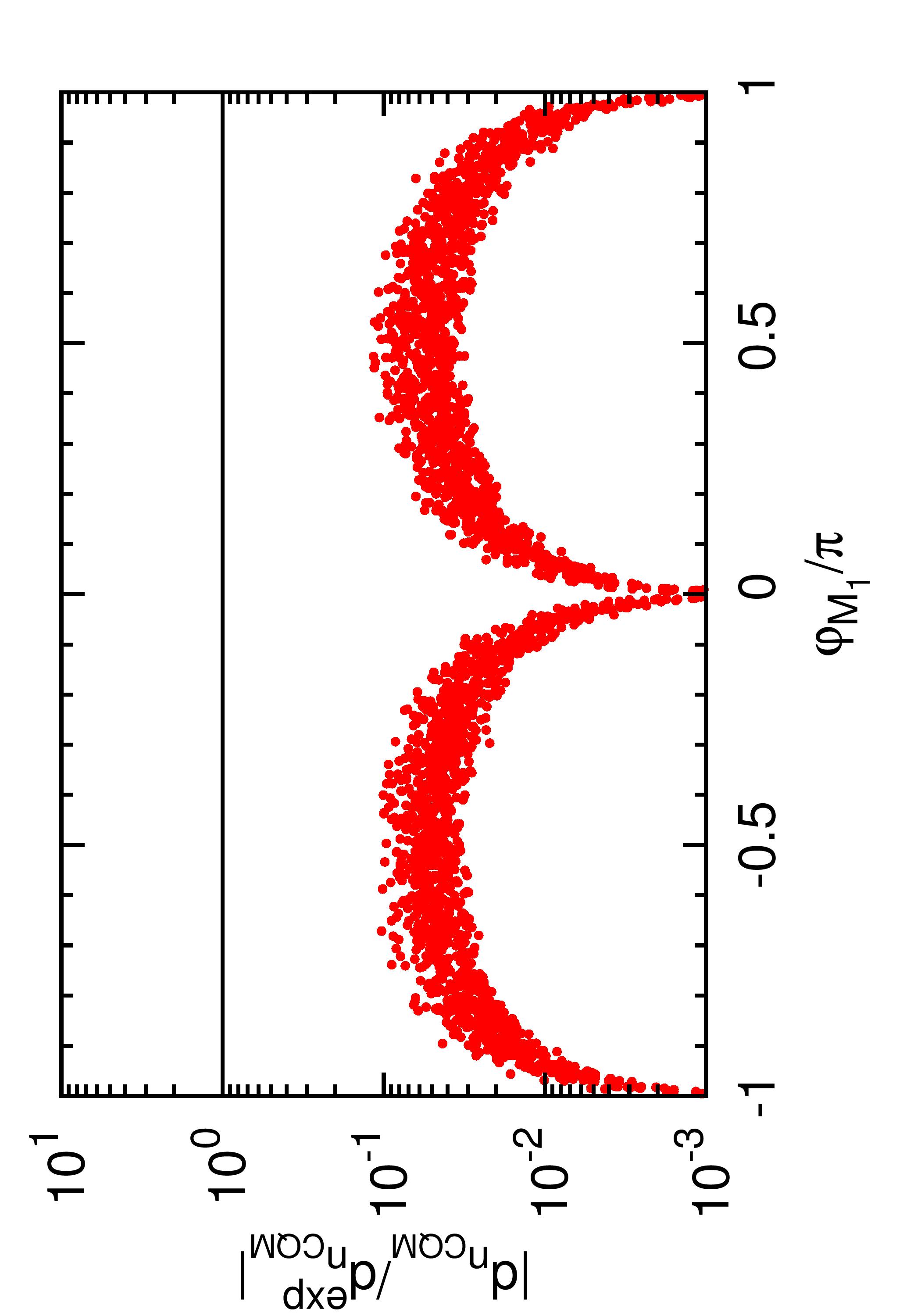}\includegraphics[height=0.45\textwidth,angle=-90]{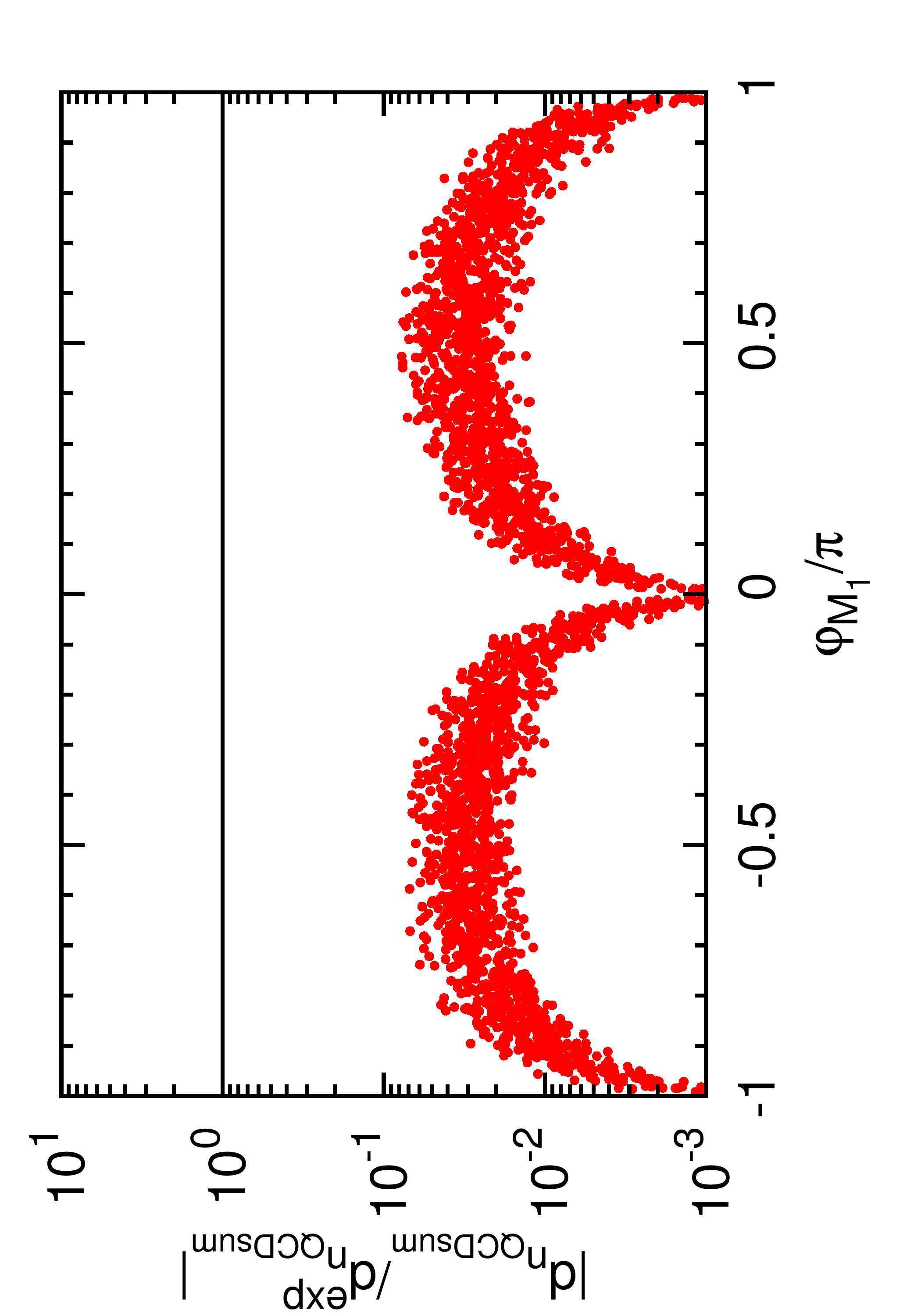} 
\\[-2mm]
\includegraphics[height=0.45\textwidth,angle=-90]{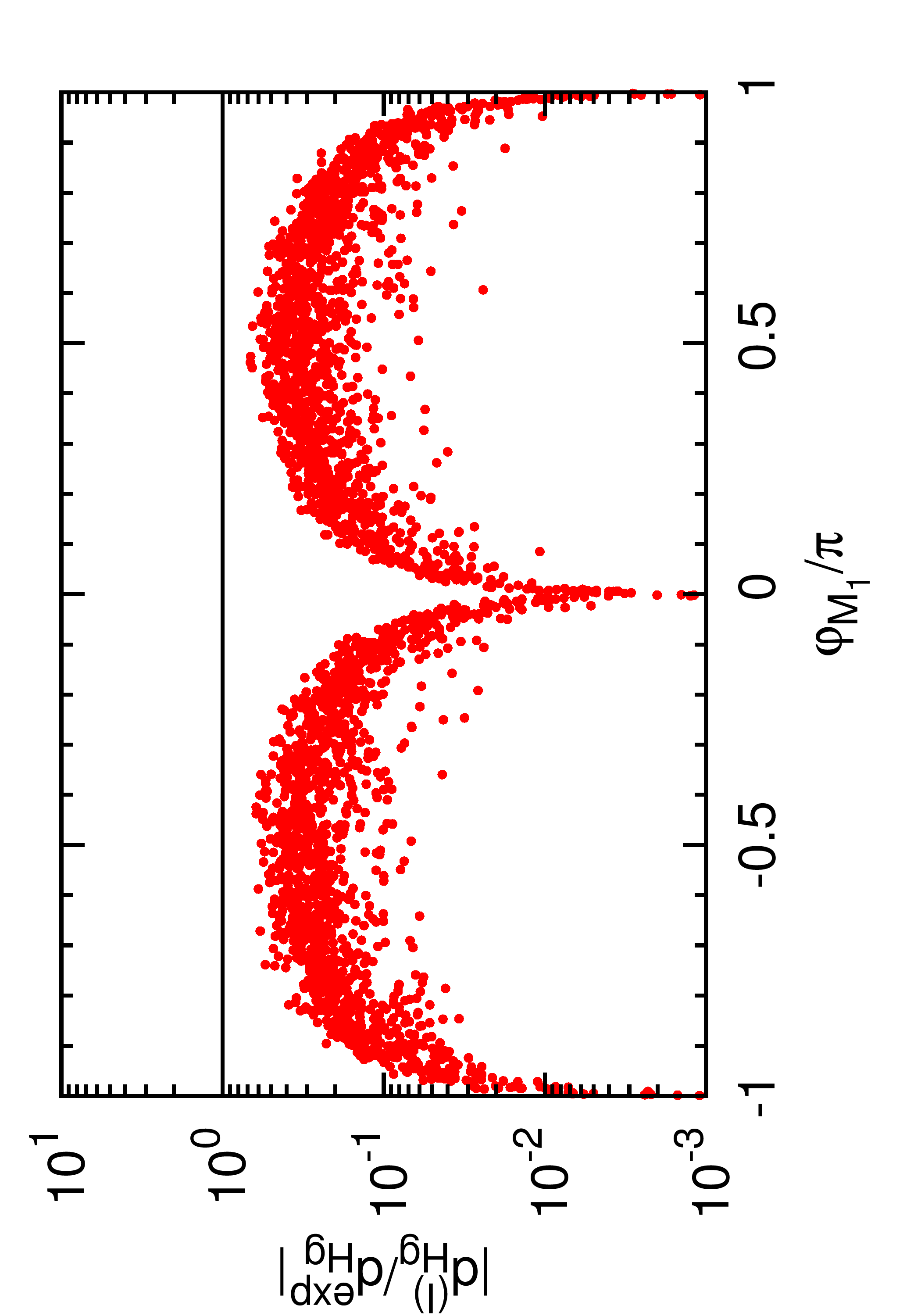}\includegraphics[height=0.45\textwidth,angle=-90]{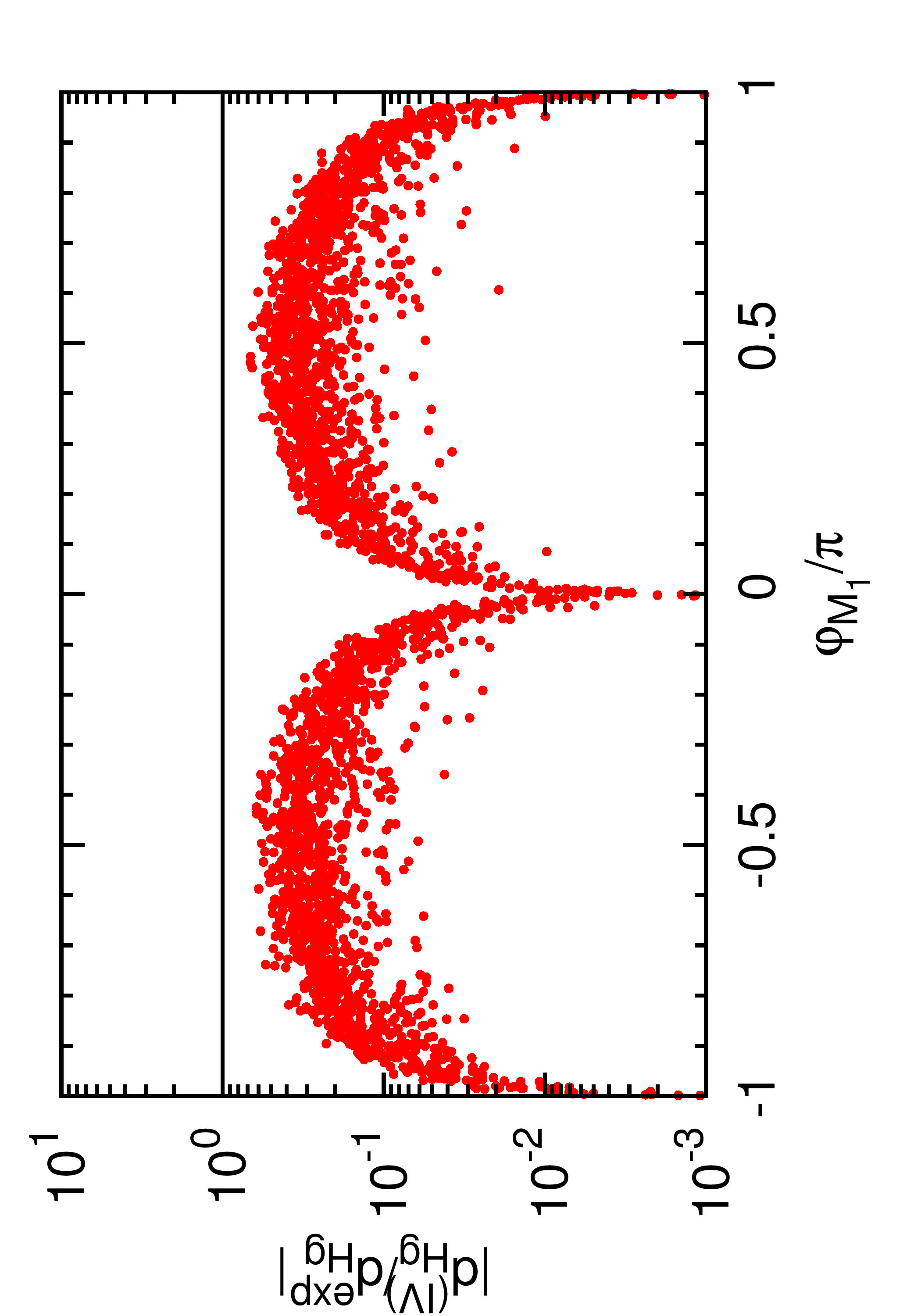} 
\\
\caption{Absolute values of the electron (upper
  left), Thallium (upper right), neutron 
  (middle) and Mercury (lower) EDMs as a function of
  $\varphi_{M_1}$, normalized to the respective experimental upper
  bound. \label{fig:m1edm}} 
\end{center}
\end{figure}
Finally, the phase $\varphi_{M_1}$ arises in the neutralino sector and
therefore only gives contributions to diagrams that involve neutralinos
in the loops. The generated EDMs are smaller than the ones for a
non-vanishing $\varphi_{M_2}$, {\it cf.}~Fig.~\ref{fig:m1edm}. The
phase $\varphi_{M_1}$ therefore does not lead to more stringent
constraints on the CP-violating NMSSM as the ones that have already
been discussed above.   

\subsection{Enlarged NMSSM Parameter Space \label{sec:large}}
We now turn to the discussion of the EDMs in an enlarged NMSSM
parameter space. While the previous scan focused on NMSSM regions that
lead to an overall light Higgs mass spectrum with good discovery
prospects for all Higgs bosons, we here cover a large part of the
NMSSM parameter space. In particular we also allow now for large
values of $\tan\beta$ 
and cover the whole allowed space of $\lambda$ and $\kappa$ while
taking care of the perturbativity constraint
\beq
\sqrt{|\lambda|^2+|\kappa|^2} < 0.7 \;.
\eeq
Additionally, we vary the effective $\mu$ parameter in a large
range. In summary, our scan covers
\beq
1 \le \tan\beta \le 30 \;, \qquad |\lambda| \le 0.7 \;, \qquad
|\kappa| \le 0.7 \;, \qquad |\mu_{\text{eff}}|
\le 1 \mbox{ TeV} \;. \label{eq:cond7}
\eeq 
The ranges of the remaining parameters are the same as given in
Eqs.~(\ref{eq:cond2})-(\ref{eq:cond6}). And we again checked for the
compatibility with the lower bound on the charged Higgs mass
\cite{chargedhiggs} and the exclusion limits on the SUSY particle masses
\cite{chargedhiggs,susyexclusion1,susyexclusion1a,susyexclusion2}. We
now also have scenarios where the lightest of the mostly CP-even-like Higgs
bosons can be SM-like.  \s

\begin{figure}[h!]
\begin{center}
\vspace*{-0.2cm}
\includegraphics[height=0.45\textwidth,angle=-90]{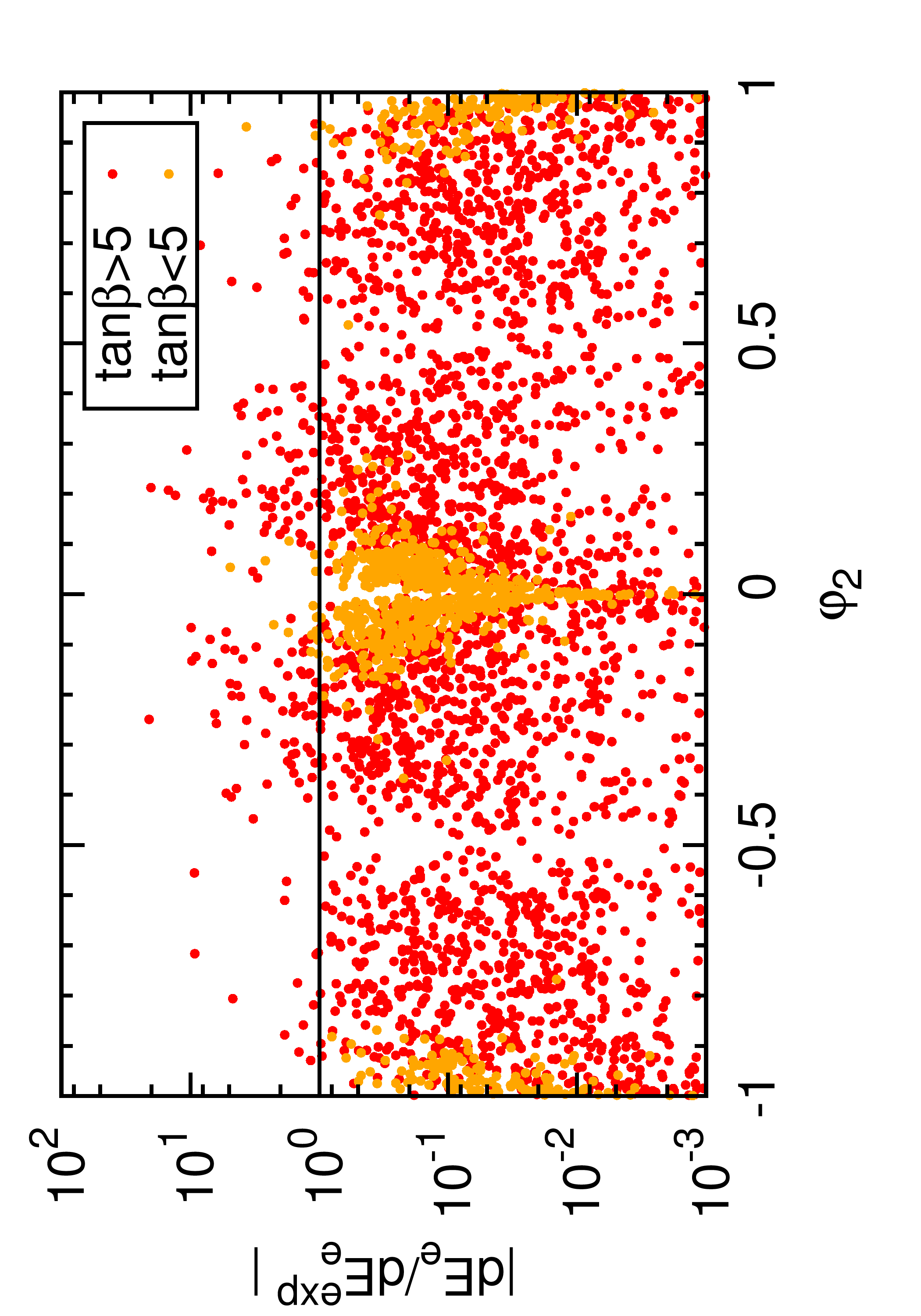}\includegraphics[height=0.45\textwidth,angle=-90]{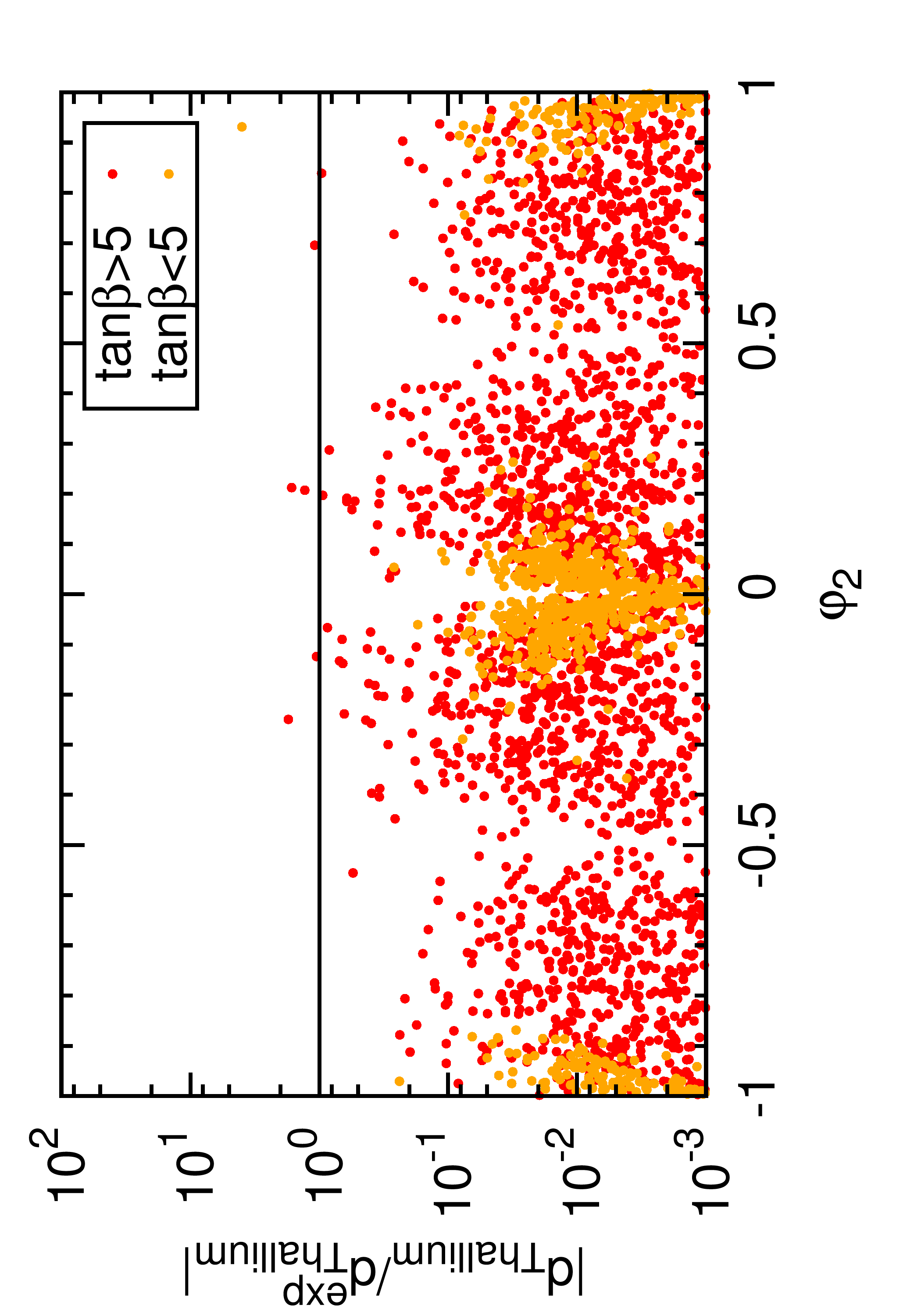} \\[-2mm]
\includegraphics[height=0.45\textwidth,angle=-90]{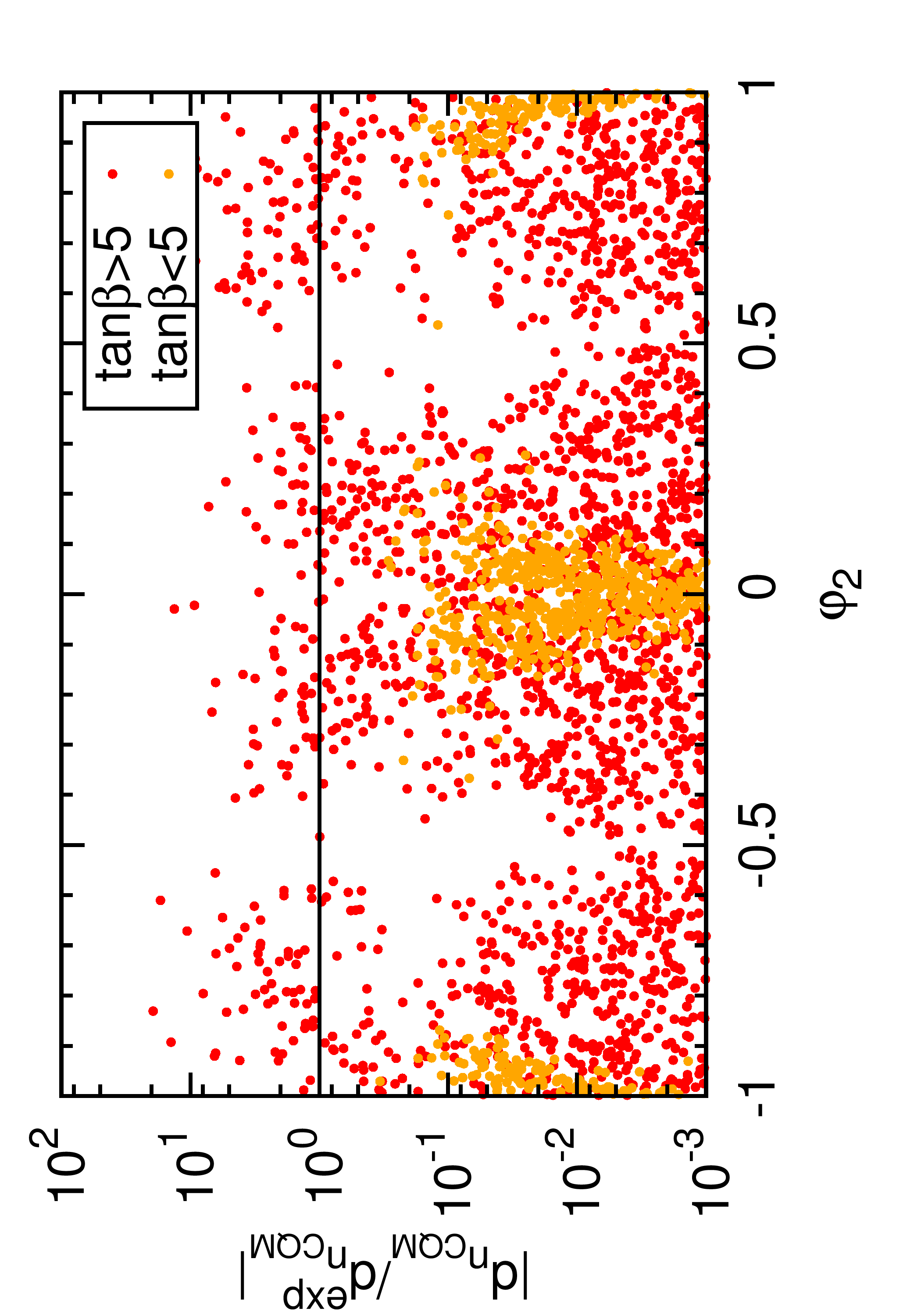}\includegraphics[height=0.45\textwidth,angle=-90]{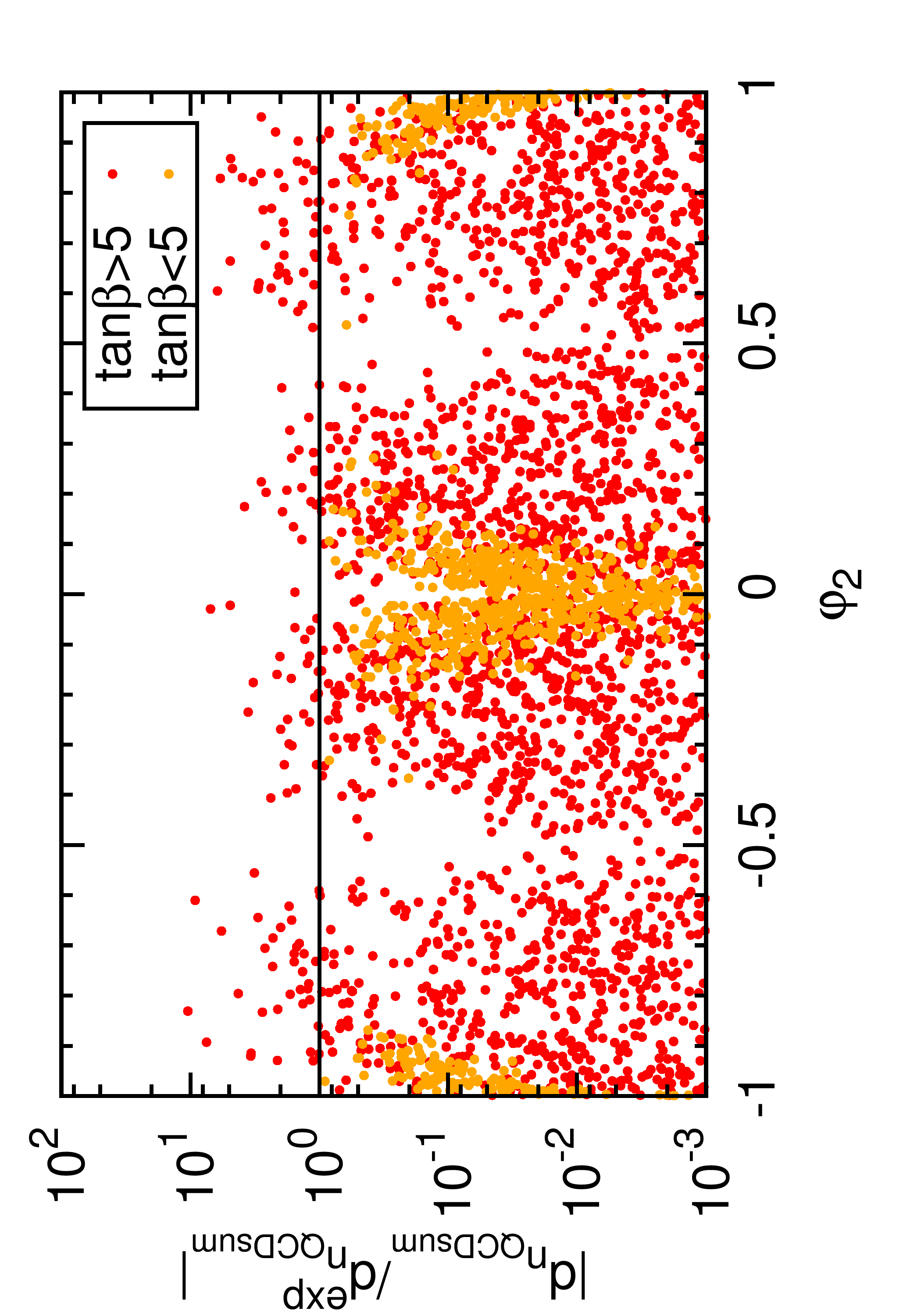} \\[-2mm]
\includegraphics[height=0.45\textwidth,angle=-90]{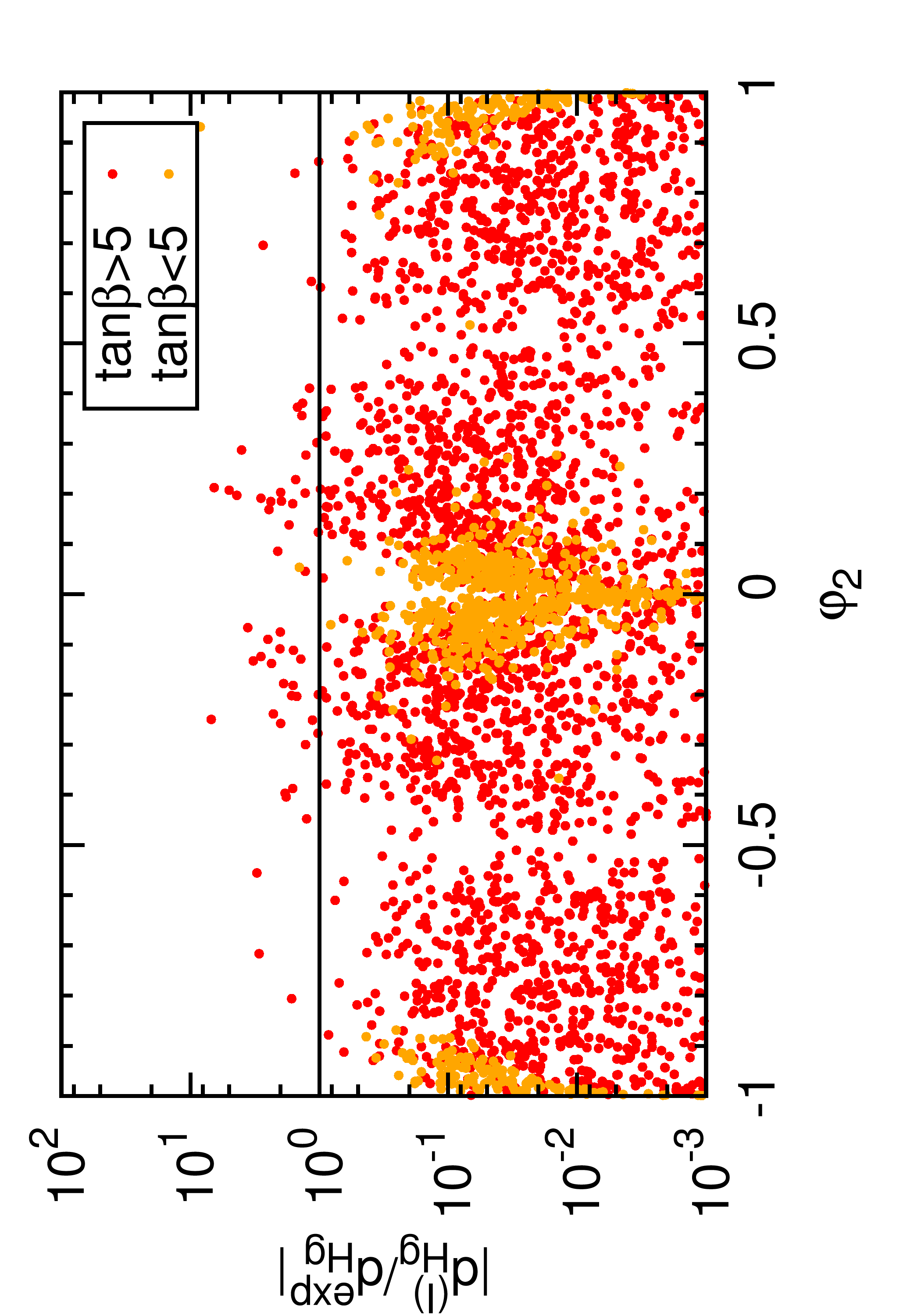}\includegraphics[height=0.45\textwidth,angle=-90]{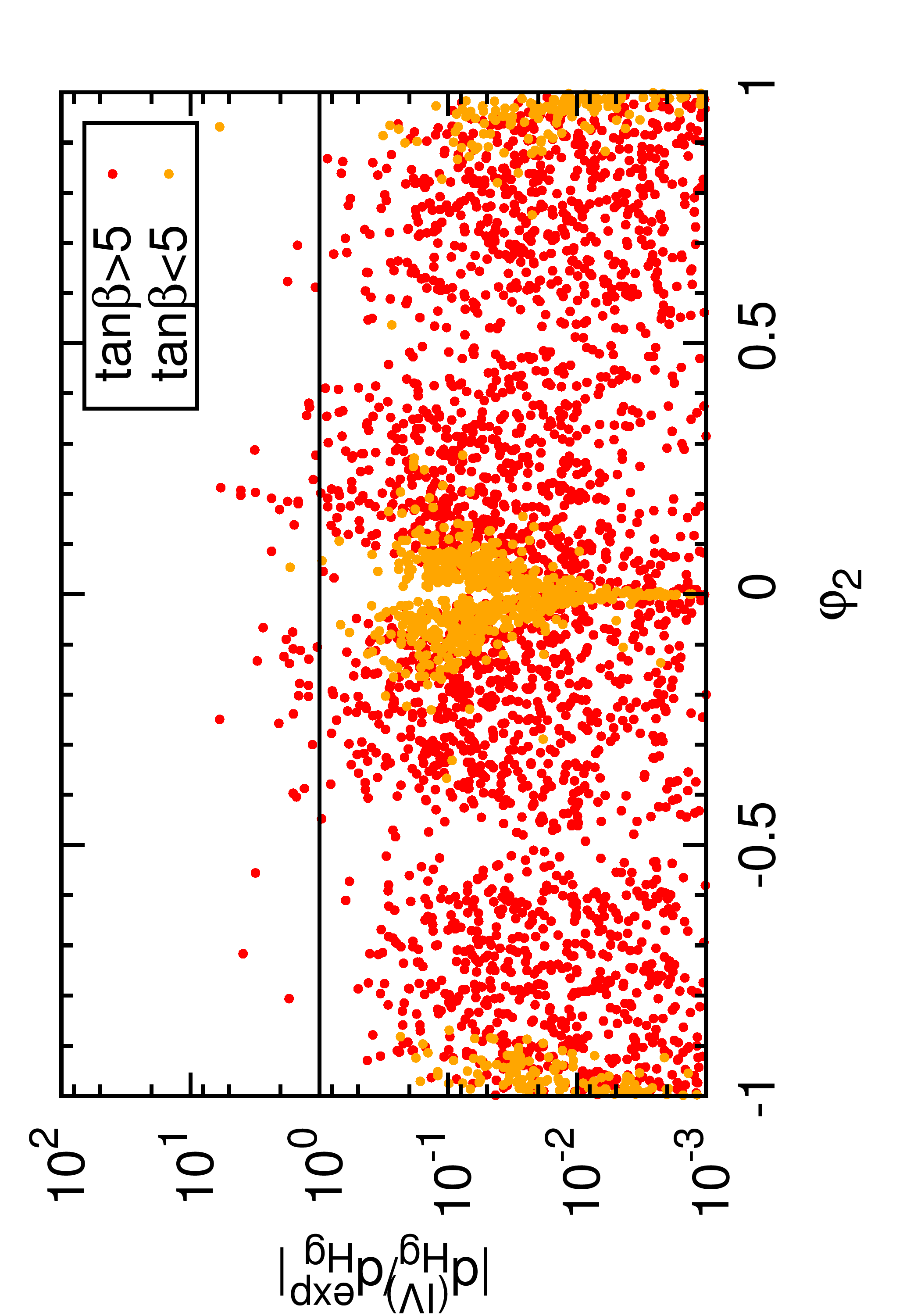} \\
\caption{Large parameter scan: 
  Absolute values of the electron (upper left),
  Thallium (upper right), neutron 
  (middle) and Mercury (lower) EDMs as a function of
  $\varphi_2$, normalized to the respective experimental
  upper bound; orange: $\tan\beta <5$, red: $\tan\beta > 5$. 
  \label{fig:edmcompletelarge}}
\end{center}
\end{figure}
\underline{\it Variation of $\varphi_2$:}
For the scenarios of the large scan we show the results for the EDMs
compared to the experimental values in Fig.~\ref{fig:edmcompletelarge}.
They are plotted for the electron, the Thallium, the neutron and
the Mercury EDM. We have only varied the NMSSM specific
phase $\varphi_2$ and set all other complex phases to
zero.  The comparison with the results of the
Natural NMSSM in Fig.~\ref{fig:edmcomplete} shows, that more parameter
sets now lead to EDMs that exceed the experimental limits. This is
particularly striking for the neutron EDM, where the EDMs can be up to
a factor of 20 larger. At the same time we have more EDMs with very small
values, in particular in the case of large CP-violating phases. The
orange points indicate the results for $\tan\beta$ values below 5,
while red ones refer to larger $\tan\beta$ values. This shows the
strong influence of the $\tan\beta$ parameter on the size of the
EDMs. Larger $\tan\beta$ values lead to larger EDMs and vice
versa. This is in accordance with the findings of Refs.~\cite{elliseal,Abel:2001vy}. 
Note, finally that Fig.~\ref{fig:edmcompletelarge} and all subsequent
figures of this subsection of course contain
the subspace of the Natural NMSSM. As we scan here, however, over a
larger parameter space the distribution of the points may not be
exactly the same as in the corresponding figures for the Natural NMSSM. \s

In the enlarged scan, not only the
electron EDM, but now also the neutron and Mercury EDMs can
lead to stringent constraints on the parameter space. Otherwise, the 
investigation of the individual contributions shows the same pattern
as in the Natural NMSSM subspace. Only the sign of the various contributions
is not related to the sign of the phases as clearly any more as {\it
  e.g.}~in Fig.~\ref{fig:edmbarrzee}.  \s

\begin{figure}[h!]
\begin{center}
\vspace*{-0.2cm}
\includegraphics[height=0.45\textwidth,angle=-90]{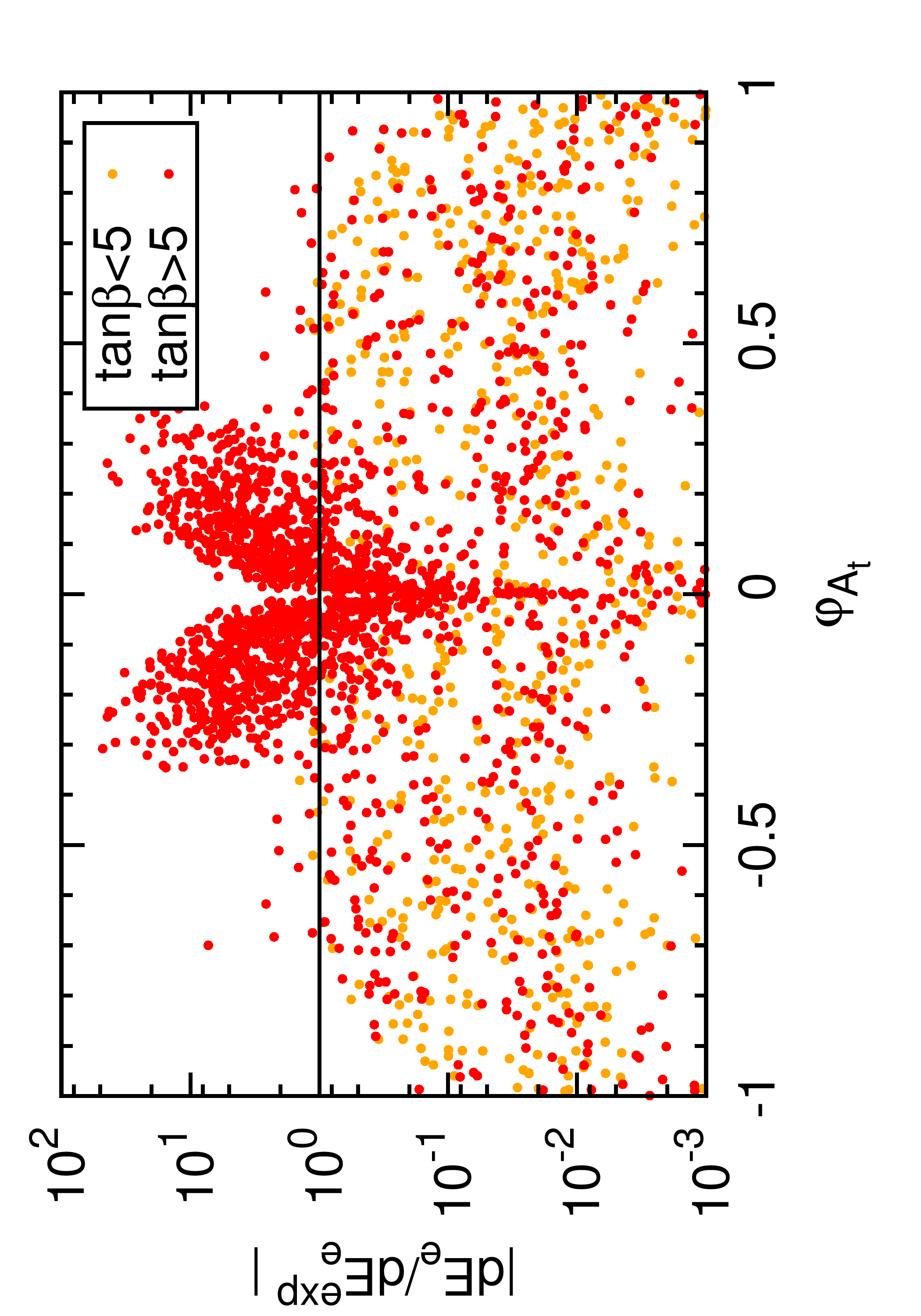}\includegraphics[height=0.45\textwidth,angle=-90]{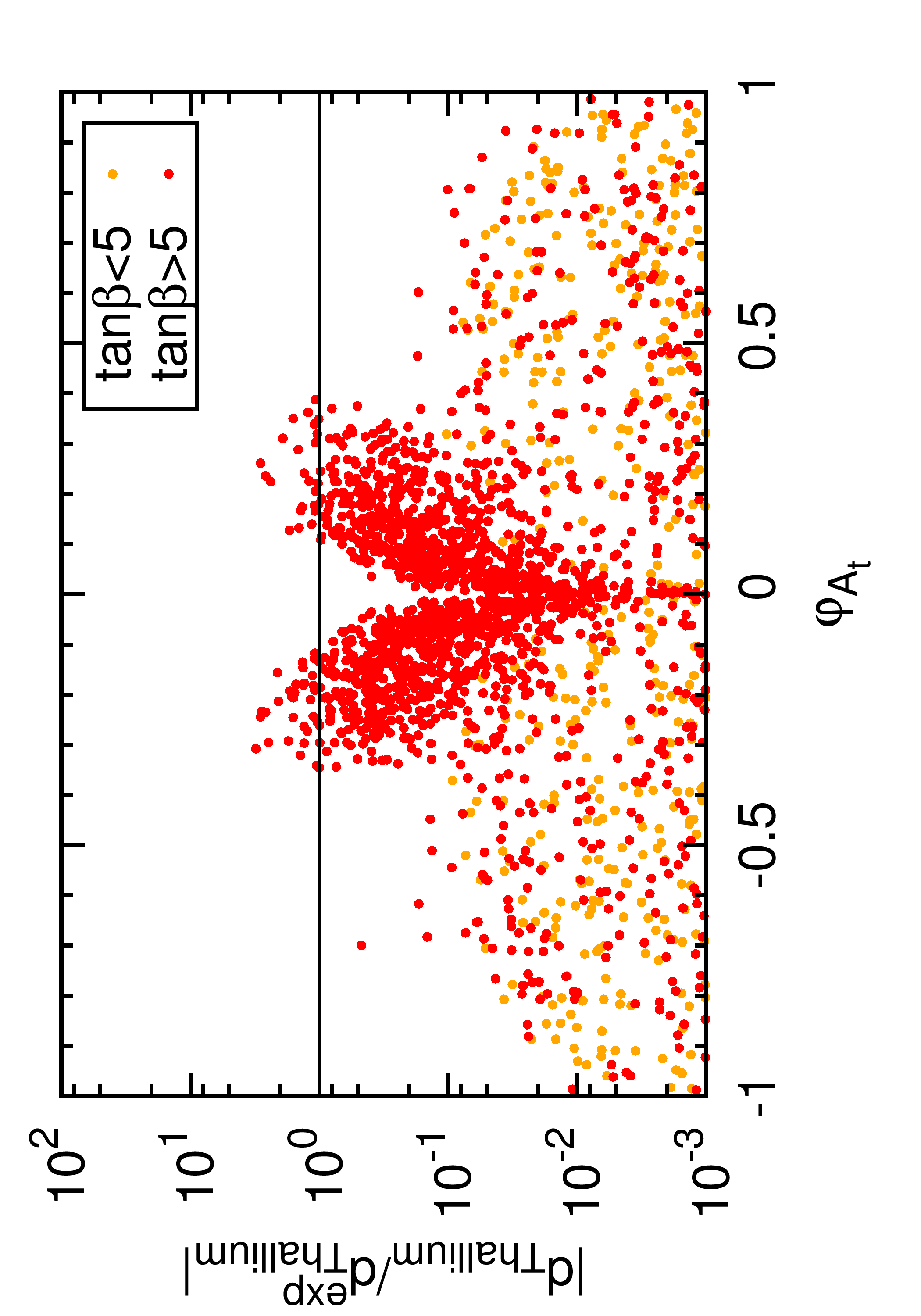} \\[-2mm]
\includegraphics[height=0.45\textwidth,angle=-90]{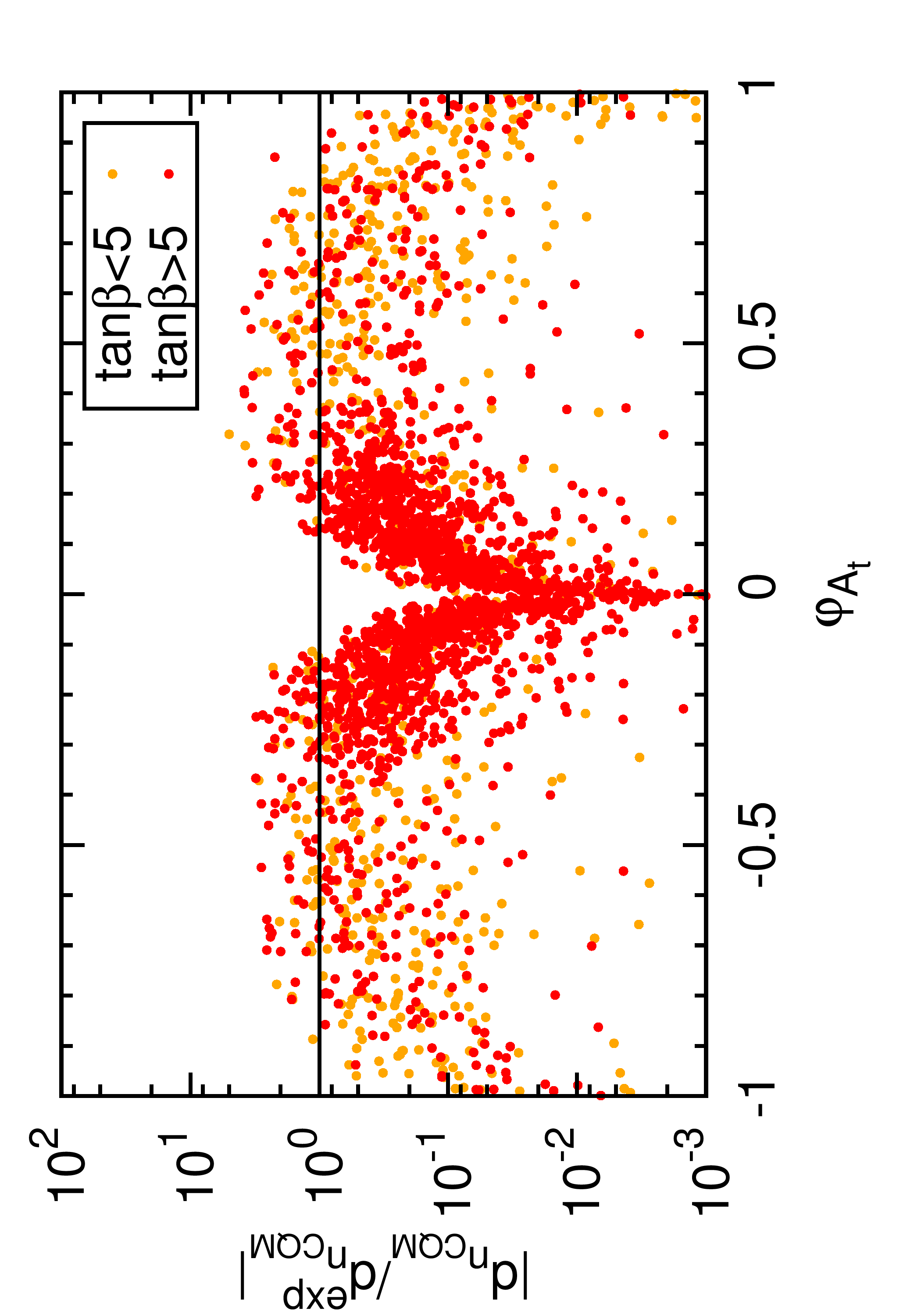}\includegraphics[height=0.45\textwidth,angle=-90]{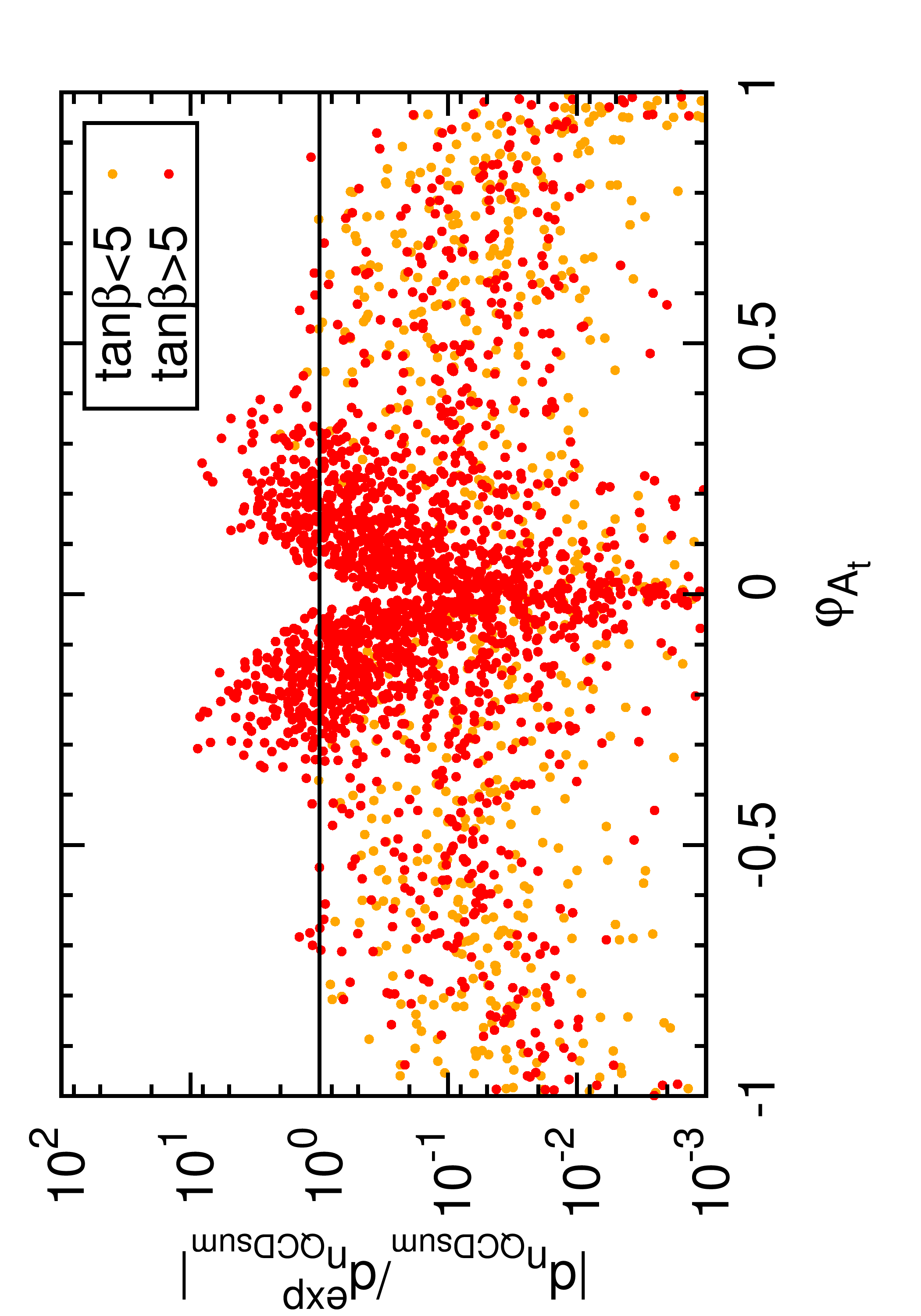} \\[-2mm]
\includegraphics[height=0.45\textwidth,angle=-90]{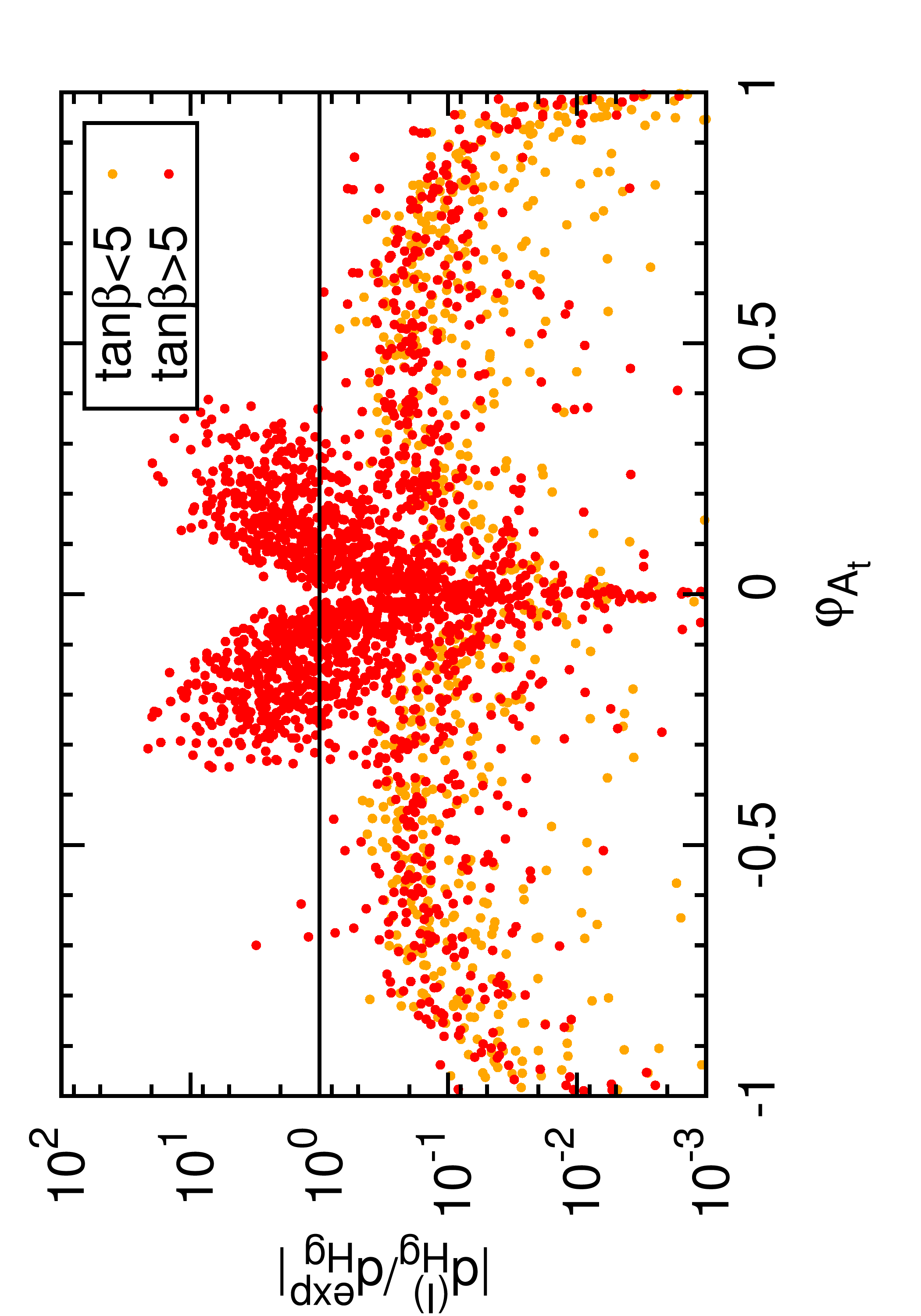}\includegraphics[height=0.45\textwidth,angle=-90]{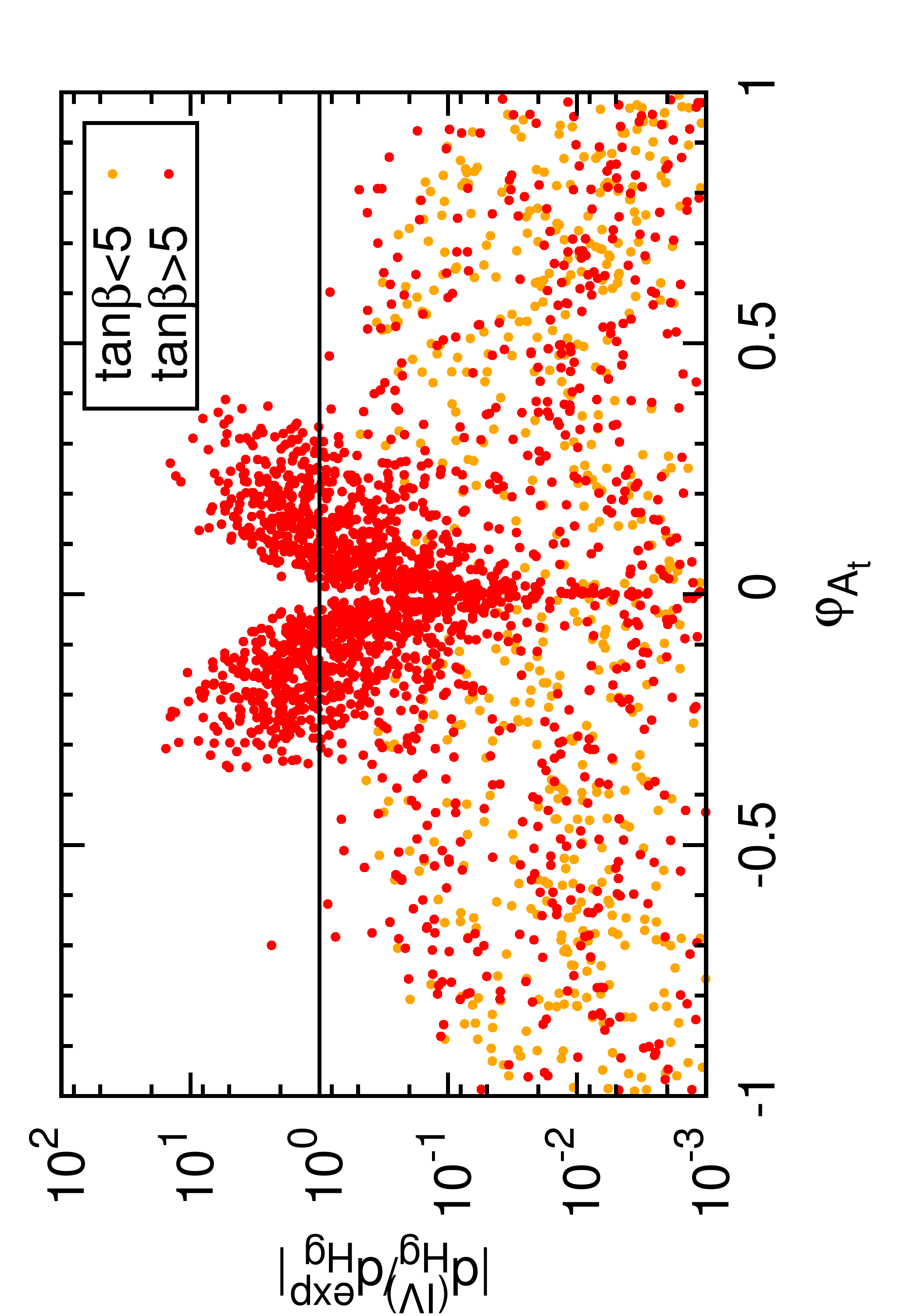} \\
\caption{Same as Fig.~\ref{fig:edmcompletelarge}, but for the variation of
  $\varphi_{A_t}$. All other CP-violating phases are set to zero. \label{fig:atedmlarge}}
\end{center}
\end{figure}

\underline{\it Variation of $\varphi_{A_t}$:} The effect of a
non-vanishing MSSM-like phase $\varphi_{A_t}$ is shown in
Fig.~\ref{fig:atedmlarge}. Apart from the neutron EDM based on the CQM approach
all EDMs display enhancements, sometimes by up to a factor of 50 for
the maximal values, due to the large parameter space. This is in
particular the case for 
small phase values, $|\varphi_{A_t}| \lsim
0.25~\pi$, and large values of $\tan\beta$, where now all EDMs
contribute to the exclusion bounds. \s 

The individual features of the EDMs as discussed already for the
Natural NMSSM, do not change. \s

\begin{figure}[h!]
\begin{center}
\vspace*{-0.2cm}
\includegraphics[height=0.45\textwidth,angle=-90]{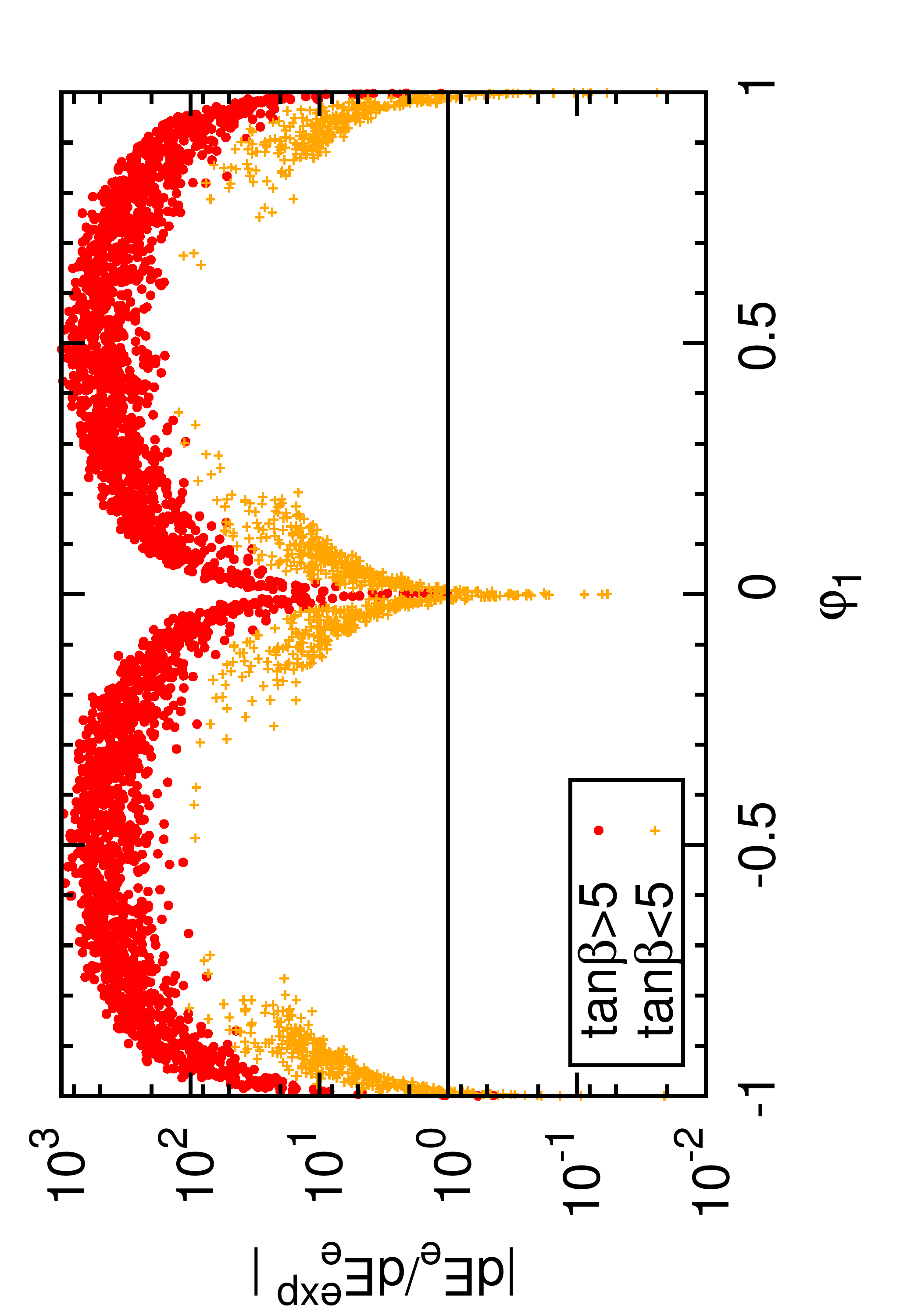}\includegraphics[height=0.45\textwidth,angle=-90]{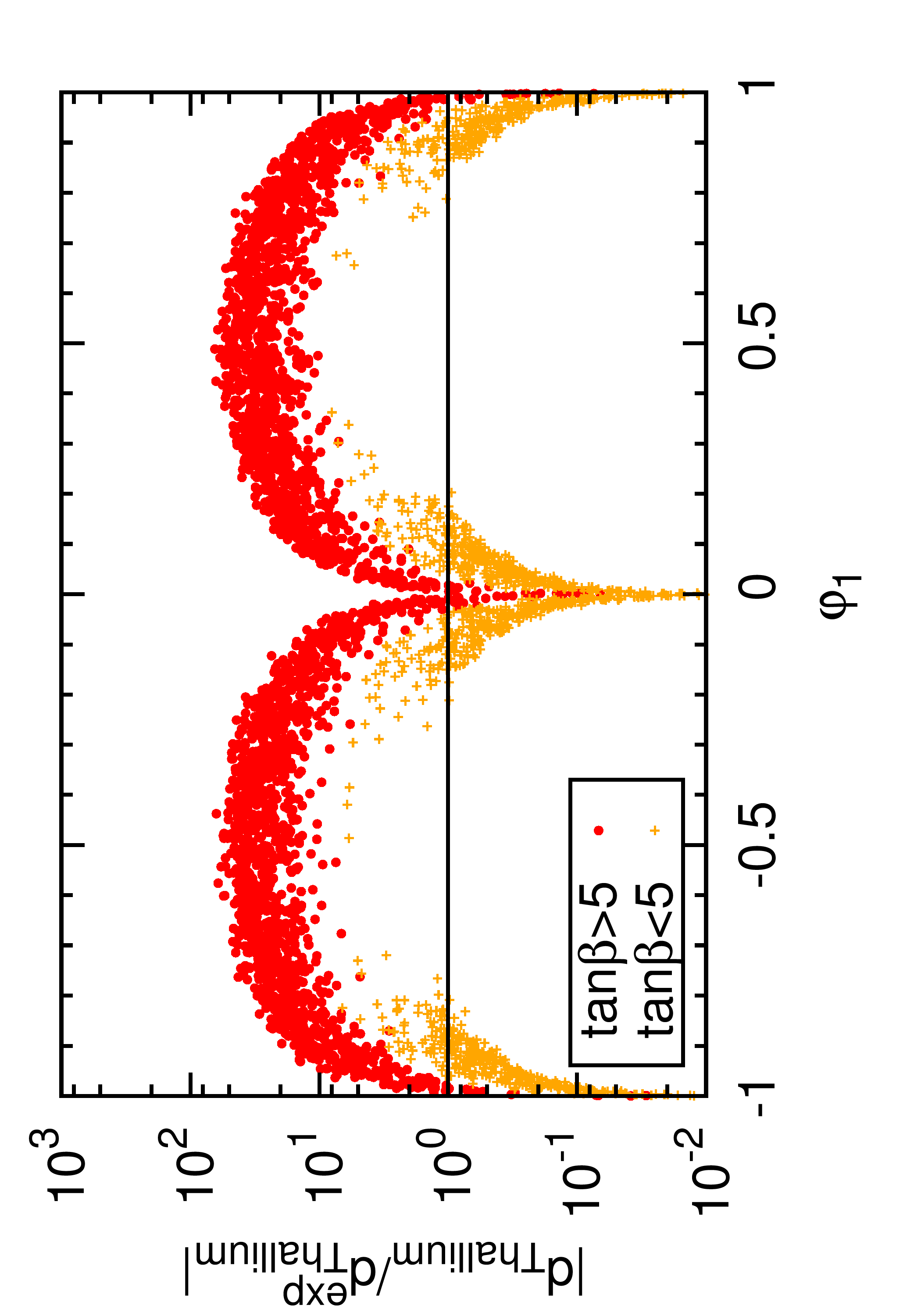} \\[-2mm]
\includegraphics[height=0.45\textwidth,angle=-90]{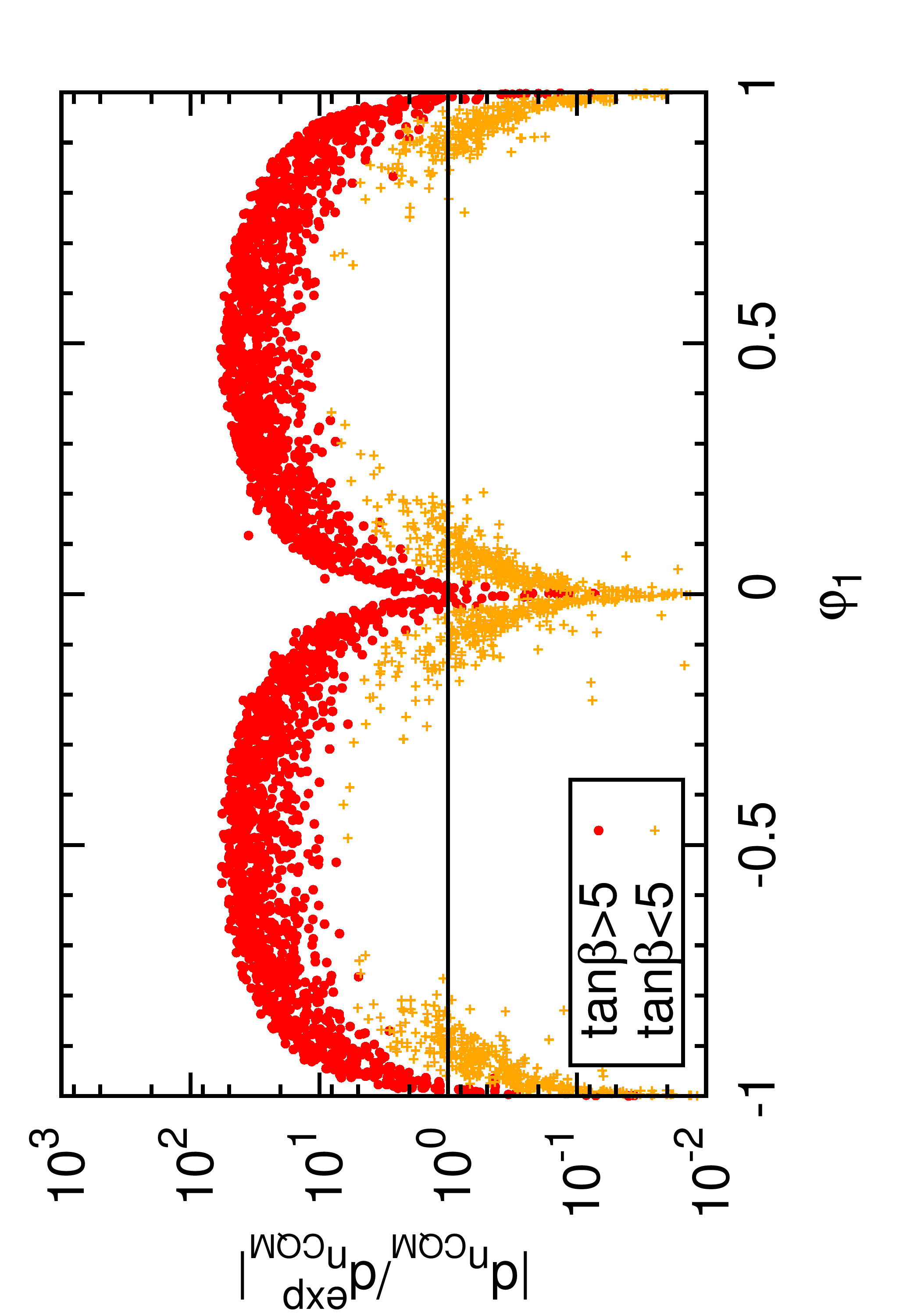}\includegraphics[height=0.45\textwidth,angle=-90]{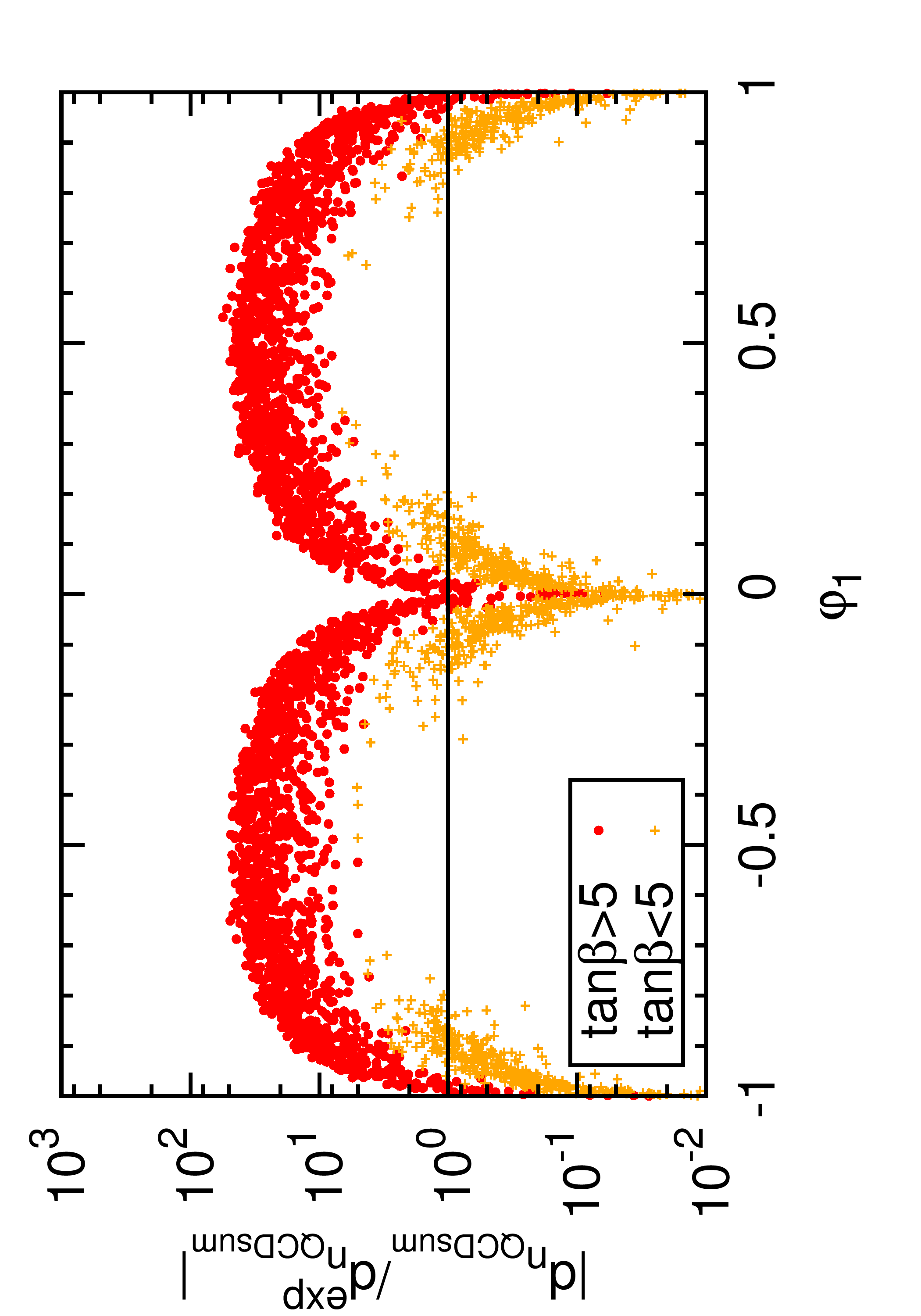} \\[-2mm]
\includegraphics[height=0.45\textwidth,angle=-90]{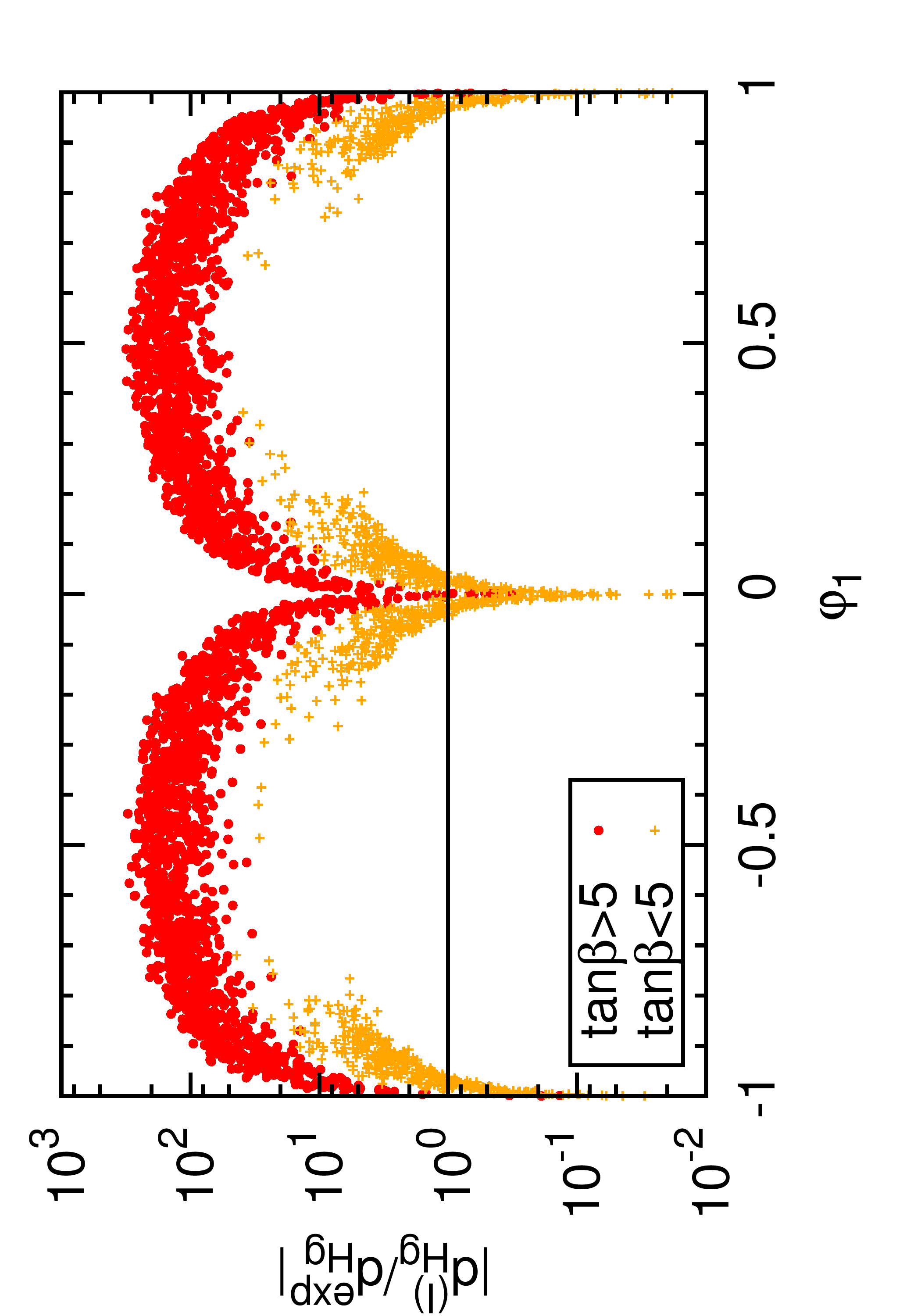}\includegraphics[height=0.45\textwidth,angle=-90]{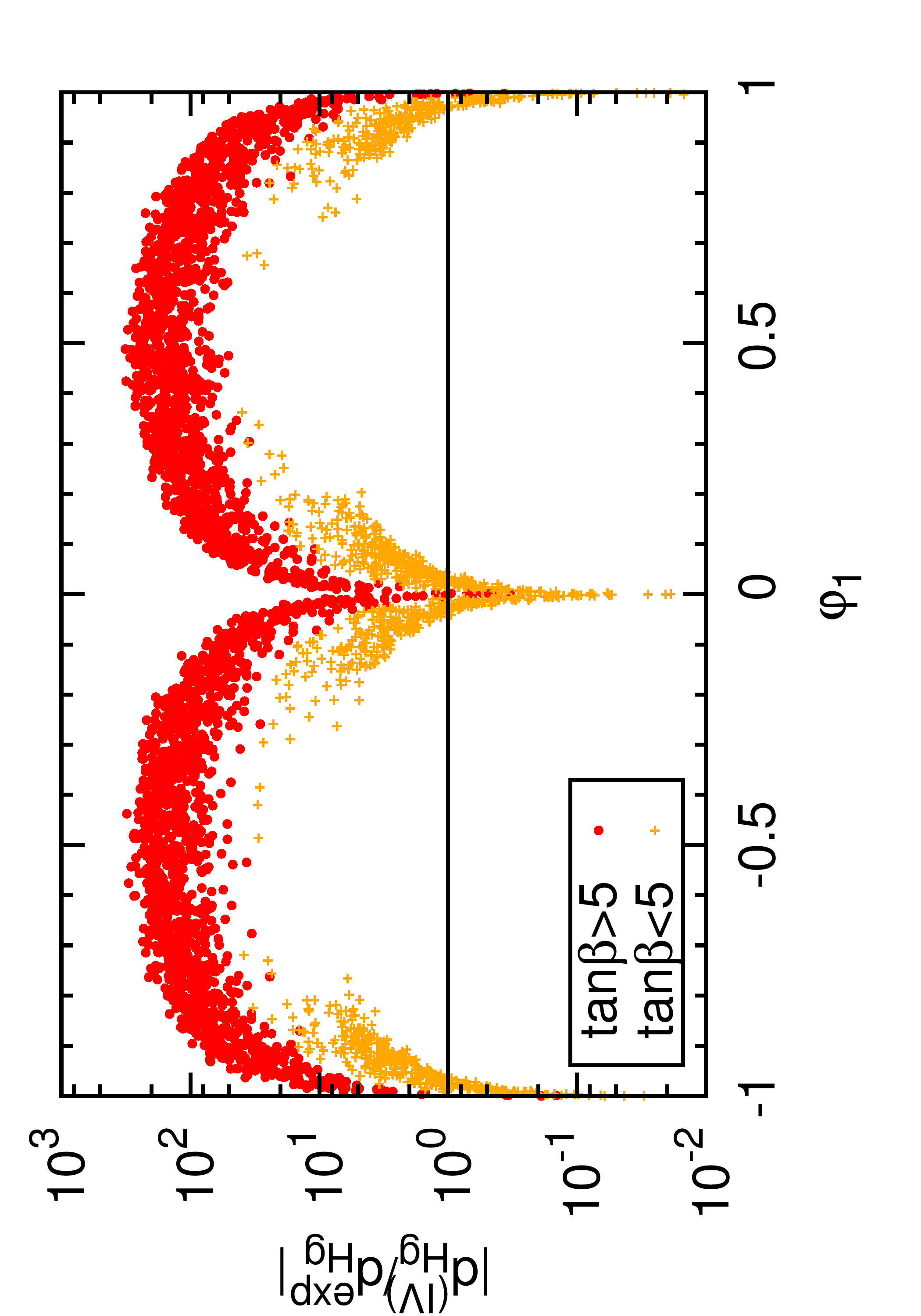} \\
\caption{Same as Fig.~\ref{fig:edmcompletelarge}, but for the variation of
  $\varphi_1$. All other CP-violating phases are set to
  zero. \label{fig:lambdaedmlarge}} 
\end{center}
\end{figure}

\underline{\it Variation of $\varphi_1$:} Finally, we show
the constraints arising from a non-vanishing phase $\varphi_1$. The
enlargement of the parameter space leads to an increase of the maximal
values of all EDMs by about a factor of 10, {\it
  cf.}~Fig.~\ref{fig:lambdaedmlarge}. The 
investigation of the EDMs shows, that for the variation of $\varphi_1$ they
are very sensitive to the value of $\tan\beta$. This can be inferred
from Fig.~\ref{fig:lambdaedmlarge} where the results for
$\tan\beta <5$ are indicated by the orange and those for $\tan\beta
> 5$ by the red colour. Overall, the non-vanishing
phase $\varphi_1$ leads to the strongest constraints on EDMs. This is
mainly due to the MSSM-specific CP violation from a complex $\mu$
parameter. \s

We also checked explicitly the behaviour of the EDMs with increasing
SUSY particle masses. As expected we observe a decoupling behaviour,
{\it i.e.}~the EDMs decrease with increasing SUSY particle masses in
the loops.

\section{Phenomenological Higgs Analysis\label{sec:pheno}}
The explicit verification of CP violation at the LHC is a non-trivial
task and requires high luminosities \cite{Heinemeyer:2013tqa}. For the
measurement of CP violation, observables can be constructed that are
sensitive to CP-violating effects in the Higgs couplings to gauge
bosons and fermions.\footnote{For a comprehensive list of the relevant
  literature, see {\it e.g.}~\cite{Heinemeyer:2013tqa,Godbole:2007cn}
  and references therein, complemented by recent investigations
  in
  \cite{Djouadi:2013qya,Harnik:2013aja,Sun:2013yra,Anderson:2013afp,Brod:2013cka,Buchalla:2013mpa,Dekens:2014jka,Chen:2014ona,Dolan:2014upa,Demartin:2014fia,Berge:2014sra,Godbole:2014cfa,Li:2014yda,Yue:2014tya,Arbey:2014msa,He:2014xla,Askew:2015mda,Boudjema:2015nda,Altmannshofer:2015qra,Zagoskin:2015sca,Buckley:2015vsa}.}
While the Higgs couplings to massive gauge bosons project on the
CP-even component of the Higgs state, the fermionic couplings have the
advantage to democratically couple to the CP-even and CP-odd components
of the Higgs bosons. The individual signal rates, on the other hand, 
do not allow for conclusions on 
CP violation, since specific parameter configurations in the
CP-conserving NMSSM can lead to the same rates as in the CP-violating
case within the experimental errors. Also the superposition of rates
stemming from CP-odd and CP-even Higgs bosons, that are close in mass,
can mimic CP-violating effects. The simultaneous measurement of Higgs
decay rates, however, that require a CP-even, respectively, a CP-odd
component, allow for conclusions on CP violation. \s

In the following two subsections we will discuss possible prospects for
accessing CP violation in two different approaches. These are on the
one hand the combined measurement of signal rates and on the other
hand the exploitation of decays into fermion pairs. 

\subsection{Hints towards CP violation in Higgs decays involving a $Z$ boson \label{sec:rates}}

One possibility to pin down CP violation in the Higgs sector is the
observation of a combination of decays that is not allowed in
CP-conserving scenarios. If a new scalar resonance is discovered there
are several decays that offer insights on its CP nature. For example
if the new resonance $H'$ decays into a pair of vector bosons, it has
to have a CP-even admixture. Furthermore one can make use of the
knowledge that the SM-like Higgs boson at 125~GeV is mostly
CP-even. Hence, the decay $H'\rightarrow hh$ is also an indication for
a large CP-even component of $H'$. However, if at the same time
$H'\rightarrow h Z$ can be observed, $H'$ must feature a CP-odd
component as well. In short, if we observe 
\beq
\begin{array}{llll}
&H' \to ZZ & \text{or}\quad H' \to hh&\qquad \widehat{=} \qquad \mbox{CP}_{H'}=+1 \\
\text{and}&H' \to Zh & &\qquad \widehat{=} \qquad \mbox{CP}_{H'}=-1 
\end{array}
\eeq
at the same time this proves CP violation in the Higgs
sector.\footnote{For a recent investigation in the complex 2-Higgs-Doublet
  Model,  see \cite{Fontes:2015xva}.} \s
 
To illustrate this idea we present an example scenario, taken from the scan we performed with {\tt NMSSMCALC} as described in Sec.~\ref{sec:edmconstr}. The parameters of the Higgs sector are given by\footnote{Note, that $\varphi_{A_\lambda}$ and $\varphi_{A_\kappa}$ are strictly speaking not input parameters, but derived quantities, determined via the tadpole conditions.}:
\begin{align}
 & |\lambda| = 0.635 \;, \quad |\kappa| = 0.288 \; , \quad |A_\kappa| = 210.70\,\text{GeV}\;,\quad 
|\mu_{\text{eff}}| = 178.65\,\text{GeV} \;, \qquad \quad\nonumber \\ 
&\varphi_{1}=0\;, \quad \varphi_{2}=0.0119\,\pi\;,\quad \varphi_{A_\lambda}=0.0037\,\pi\;,\quad \varphi_{A_\kappa}=0.983\,\pi\;,\nonumber \\ 
&\tan\beta = 1.88 \;,\quad M_{H^\pm} = 392.9 \,\text{GeV} \;. \label{eq:param3}
\end{align}
The other input parameters are 
\begin{align}
 &  m_{\tilde{u}_R,\tilde{c}_R} = 
m_{\tilde{d}_R,\tilde{s}_R} =
m_{\tilde{Q}_{1,2}}= m_{\tilde L_{1,2}} =m_{\tilde e_R,\tilde{\mu}_R} = 3\;\mbox{TeV}\, , \;  
m_{\tilde{t}_R}=844\,\text{GeV} \,,\; \nonumber \\ \nonumber
&  m_{\tilde{Q}_3}=844\,\text{GeV}\,,\; m_{\tilde{b}_R}=3\,\text{TeV}\,,\; 
m_{\tilde{L}_3}=1751\,\text{GeV}\,,\; m_{\tilde{\tau}_R}=1751\,\text{GeV}\,,
 \\ 
& |A_{u,c,t}| = 603\,\text{GeV}\, ,\; |A_{d,s,b}|=16\,\text{GeV}\,,\; |A_{e,\mu,\tau}| = 1086\,\text{GeV}\,,\; \\ \nonumber
& |M_1| = 764\,\text{GeV},\; |M_2|= 756\,\text{GeV}\,,\; |M_3|=2650\,\text{GeV} \,,\\ \nonumber
&  \varphi_{A_{u,c,t}}=\pi\,,\; 
\varphi_{A_{d,s,b}}=\varphi_{A_{e,\mu,\tau}}=\varphi_{M_1}=\varphi_{M_2}=\varphi_{M_3}=0
 \;. 
\end{align}
The resulting Higgs spectrum is relatively light, 
\begin{align}
 &M_{H_1}\approx 104\,\text{GeV} \;, \qquad M_{H_2}\approx 126.4\,\text{GeV} \;, \qquad M_{H_3}\approx258\,\text{GeV} \;, \qquad\\ 
& M_{H_4}\approx401\,\text{GeV} \;, \qquad M_{H_5}\approx405 \,\text{GeV} \;. \qquad
\end{align}
Possible candidates for $H'$ are either $H_3$, $H_4$ or
$H_5$. However, the two heavy mass eigenstates are very close in mass,
so that their individual signals cannot be
disentangled. In most scenarios of the CP-conserving
  case one of the heavy mass eigenstates is CP-even and the other
CP-odd. But since it is impossible to tell whether the decay of $H_4$
or $H_5$ is observed, no conclusion about CP violation can be
drawn. This leaves us with $H'=H_3$. For $H_3$ both the decay into a
pair of vector bosons as well as the decay into a $Z$ boson and the
SM-like Higgs boson $H_2$ are sizeable, 
\begin{equation}
 \text{BR}(H_3\to Z H_2)=5.7\%\;,\quad\text{BR}(H_3\to Z Z)=11.8\%\;,\quad\text{BR}(H_3\to WW)=27.5\%\;.
\end{equation}
These decays can only be observed if the production cross section for
$H_3$ is sufficiently large.  
At $\sqrt{s}=13$~TeV the production cross section through gluon fusion
at next-to-next-to-leading order\footnote{The cross section has been
  calculated with a private version of {\tt HIGLU}\cite{Spira:1996if},
  that has been adapted to the complex NMSSM.} is
$\sigma^{\text{13TeV}}_{ggH_3}=211.2\,\text{fb}$, 
which is not particularly large for a Higgs boson of this mass. The
reason is the suppression of the effective coupling of $H_3$ to a pair
of gluons due to the large singlet admixture to the mass
eigenstate. 
\begin{table}{t}
\begin{center}
\small
 \begin{tabular}{|l|l|}
\hline
  \rule{0cm}{0.45cm} $\sigma{(ggH_3)}\;\text{BR}\big(H_3\to Z Z\big)$&24.8\text{~fb}  \\[0.5mm]\hline
  \rule{0cm}{0.45cm} $\sigma{(ggH_3)}\;\text{BR}\big(H_3\to W W\big)$&58.1\text{~fb}\\[0.5mm]\hline
 \rule{0cm}{0.45cm}  $\sigma{(ggH_3)}\;\text{BR}\big(H_3\to H_2 Z\big)$&12.1\text{~fb}\\[0.5mm]\hline
  \rule{0cm}{0.45cm} $\sigma{(ggH_3)}\;\text{BR}\big(H_3\to H_2 Z\to (b b) Z\big)$&5.5\text{~fb}\\[0.5mm]\hline
  \rule{0cm}{0.45cm} $\sigma{(ggH_3)}\;\text{BR}\big(H_3\to H_2 Z \to (\gamma \gamma) Z\big)$&0.04\text{~fb}\\[0.5mm]\hline
  \rule{0cm}{0.45cm} $\sigma{(ggH_3)}\;\text{BR}\big(H_3\to H_2 Z\to (Z Z) Z\big)$&0.45\text{~fb}\\[0.5mm]\hline
  \rule{0cm}{0.45cm} $\sigma{(ggH_3)}\;\text{BR}\big(H_3\to H_2 Z \to (W W) Z\big)$&3.5\text{~fb}\\[0.5mm]\hline
  \rule{0cm}{0.45cm} $\sigma{(ggH_3)}\;\text{BR}\big(H_3\to H_1 Z)$&76.0\text{~fb}\\[0.5mm]\hline
  \rule{0cm}{0.45cm} $\sigma{(ggH_3)}\;\text{BR}\big(H_3\to H_1 Z \to (b b) Z\big)$&66.0\text{~fb}\\[0.5mm]\hline
  \rule{0cm}{0.45cm} $\sigma{(ggH_3)}\;\text{BR}\big(H_3\to H_1 Z \to (\gamma\gamma) Z\big)$&0.04\text{~fb}\\[0.5mm]\hline
 \rule{0cm}{0.45cm} $\sigma{(ggH_3)}\;\text{BR}\big(H_3\to H_1 Z \to (Z Z) Z\big)$&0.06\text{~fb}\\[0.5mm]\hline
 \rule{0cm}{0.45cm} $\sigma{(ggH_3)}\;\text{BR}\big(H_3\to H_1 Z \to (W W) Z\big)$&0.65\text{~fb}\\[0.5mm]\hline
  \rule{0cm}{0.45cm} $\sigma{(ggH_1)}\;\text{BR}\big(H_1\to Z Z\big)$&1.53\text{~fb}\\[0.5mm]\hline
 \end{tabular}
\caption{Signal rates for $H_3$ production in gluon fusion at
  $\sqrt{s}=13$~TeV with subsequent decay into various final
  states.}\label{tab:CPvsignals} 
\end{center}
\end{table}
In Table~\ref{tab:CPvsignals} we list the obtained signal rates for
several final states. The rates in the vector boson final states $ZZ$
and $WW$ are with $24.8$~fb and $58.1$~fb of moderate size. For the
final state $Z H_2$ the signal only amounts to $12.1$~fb. If the decay
of the SM-like $H_2$ is taken into account as well the overall signal
becomes rather small. The decay of $H_3$ into a $Z$ boson and $H_1$,
on the other hand yields a signal rate of $76$~fb. Together with the
decay of the lightest Higgs boson into a pair of $Z$ bosons, which
establishes a CP-even admixture to $H_1$, this could also be used to
search for CP violation in the Higgs sector. The signals are
rather small though so that the experimental observation will be
challenging. \s

Note that in this scenario the EDM bounds are respected. As we
observed in Sec.~\ref{sec:edmconstr}, this is not uncommon for scenarios
that feature $\varphi_2\neq 0$ and $\varphi_1=0$. 

\subsection{CP violation in the Higgs to $\tau^+\tau^-$ decays \label{sec:fermions}}

\begin{figure}
\begin{center}
 \includegraphics[height=0.85\textwidth,angle=-90]{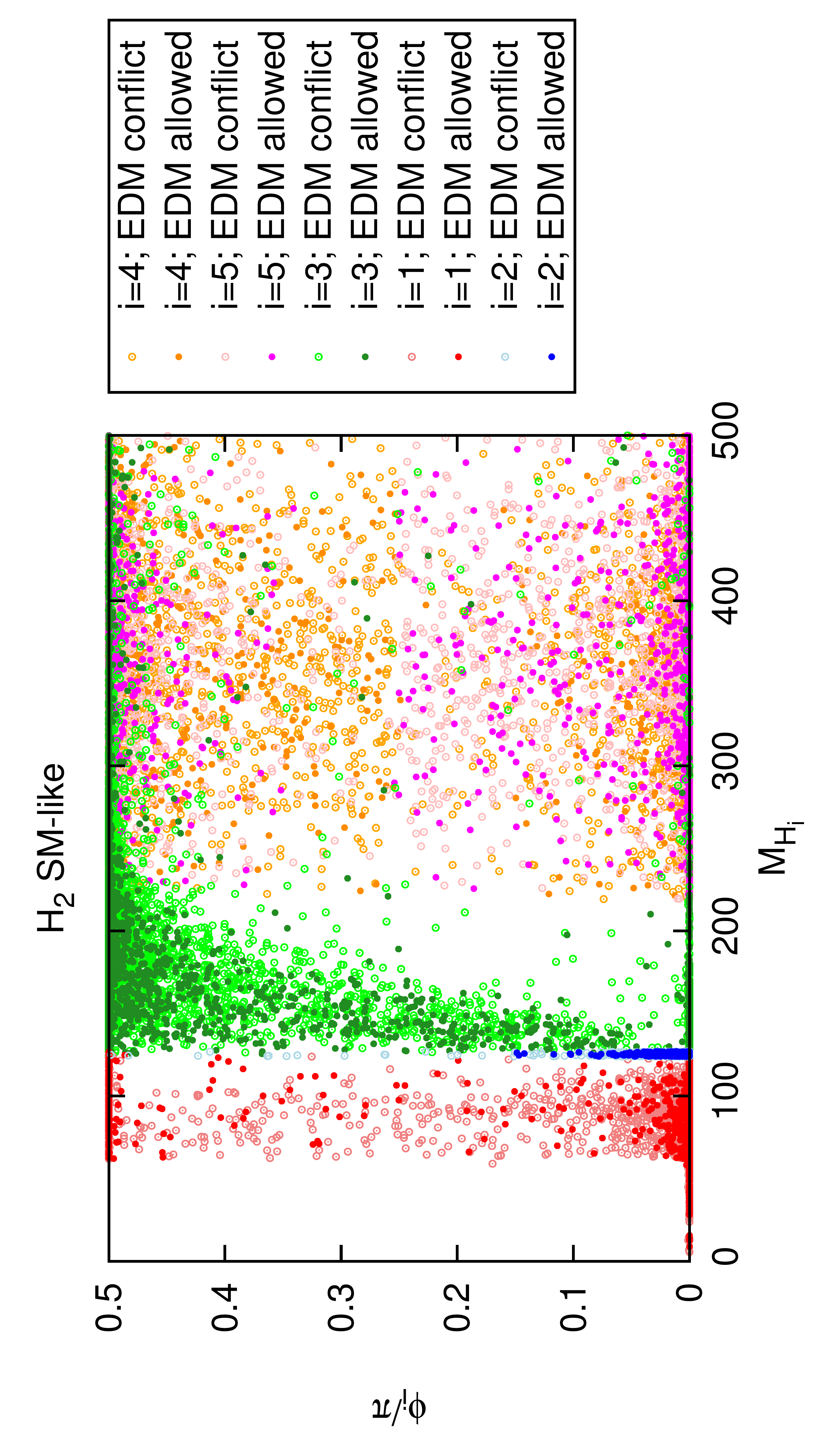}
\end{center}
\vspace*{-0.8cm}
\caption{The phase $\phi_i$, which measures the CP violation in the
  $H_i\tau^+\tau^-$ coupling, as a function of the mass of the Higgs
  boson $H_i$. Only scenarios that feature $H_2$ as the SM-like Higgs
  boson with a mass of 125~GeV are included. Light colored open points are
  in conflict with the EDM bounds, whereas dark colored full points
  respect the bounds.}\label{fig:cpphase} 
\end{figure}
\begin{figure}[h!]
\includegraphics[height=0.5\textwidth,angle=-90]{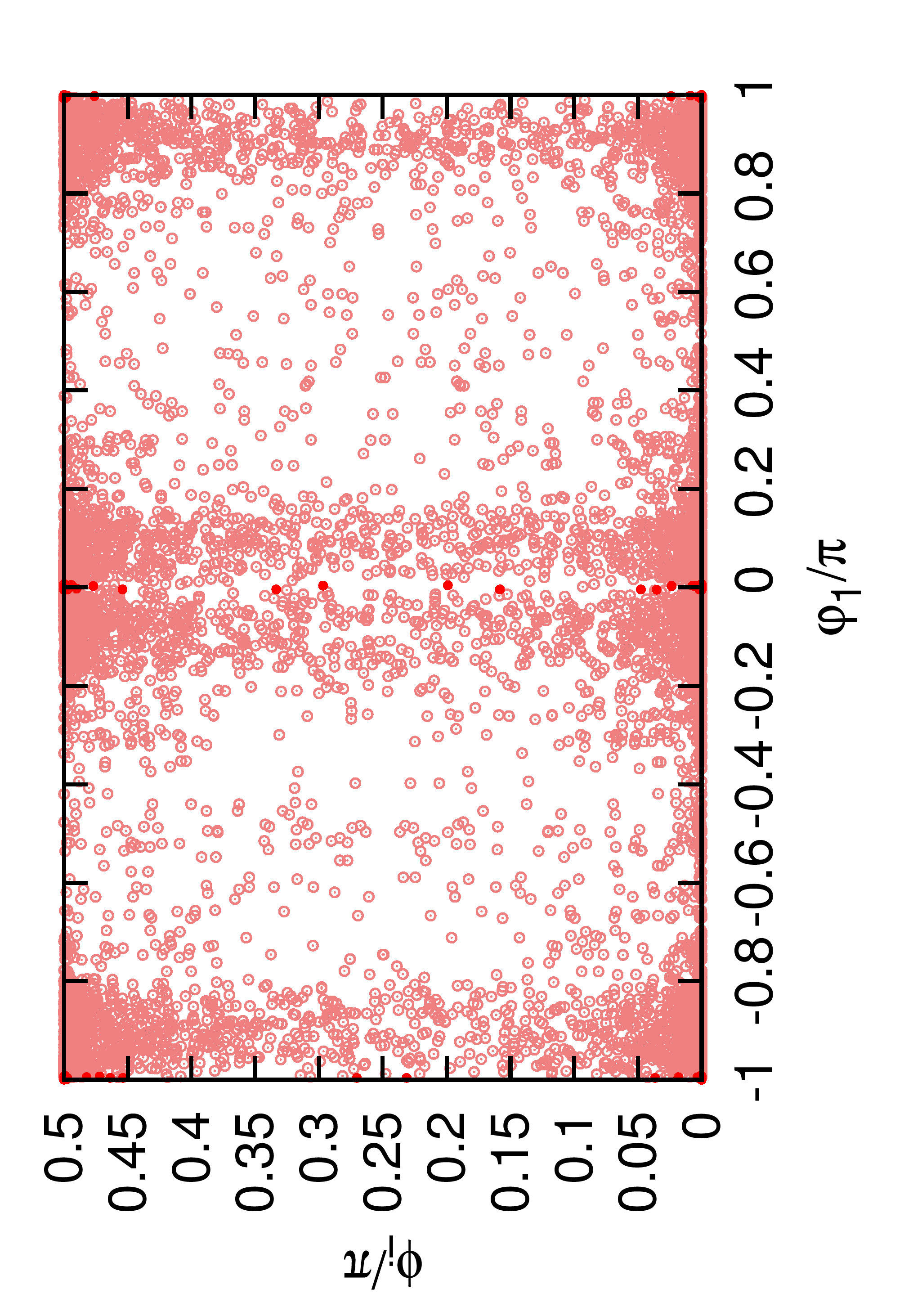}\includegraphics[height=0.5\textwidth,angle=-90]{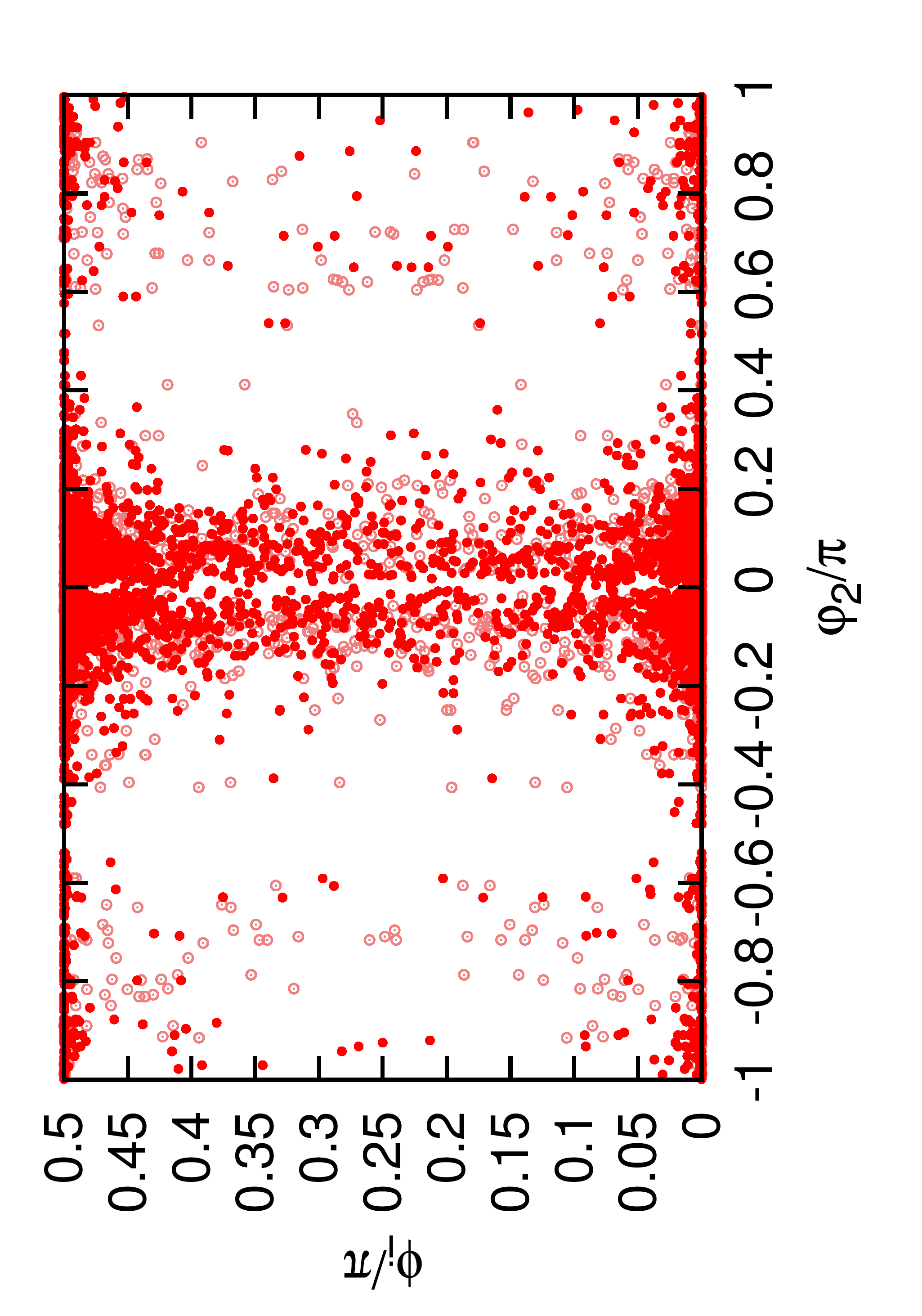}\\[-2mm]
\includegraphics[height=0.5\textwidth,angle=-90]{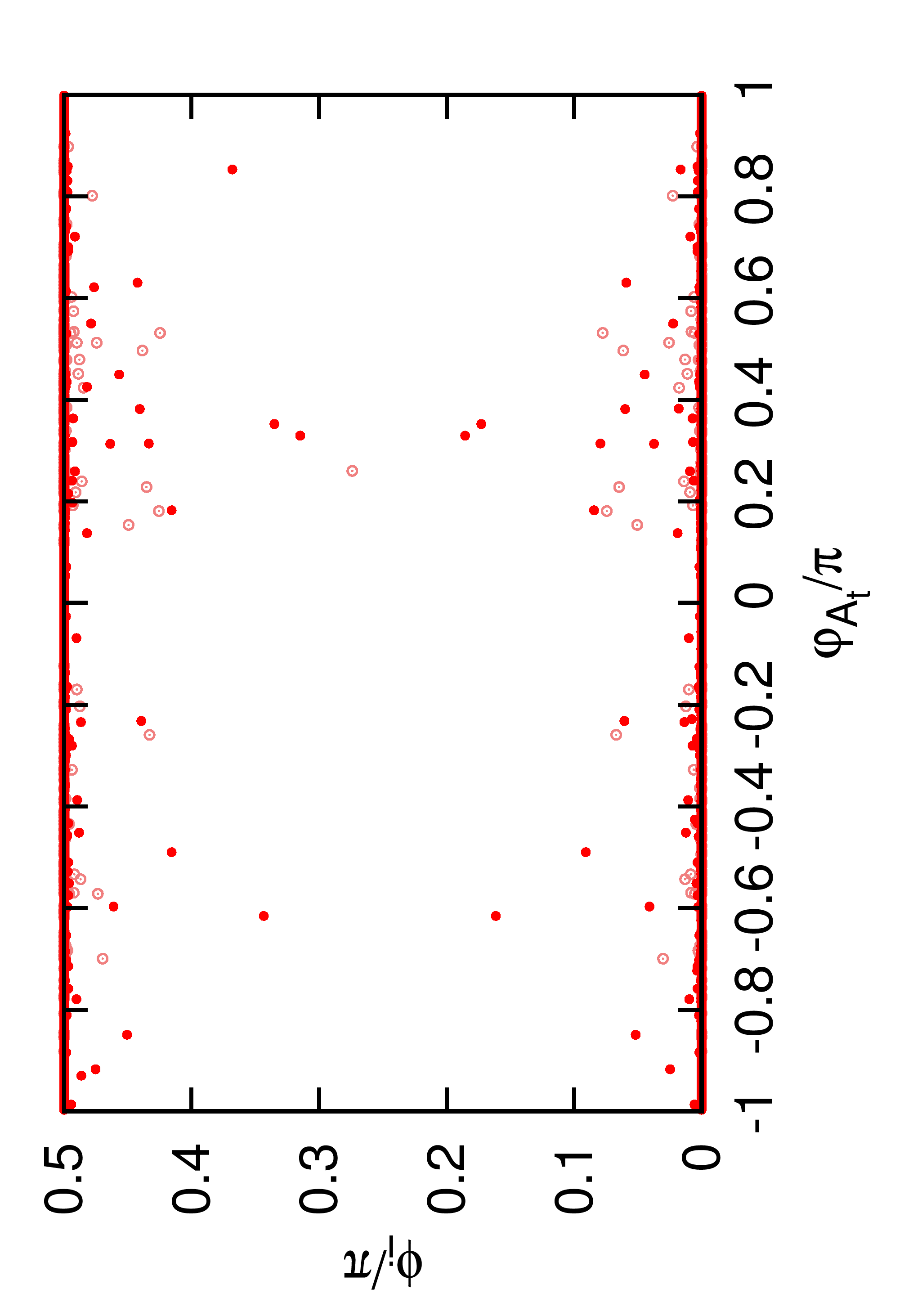}\includegraphics[height=0.5\textwidth,angle=-90]{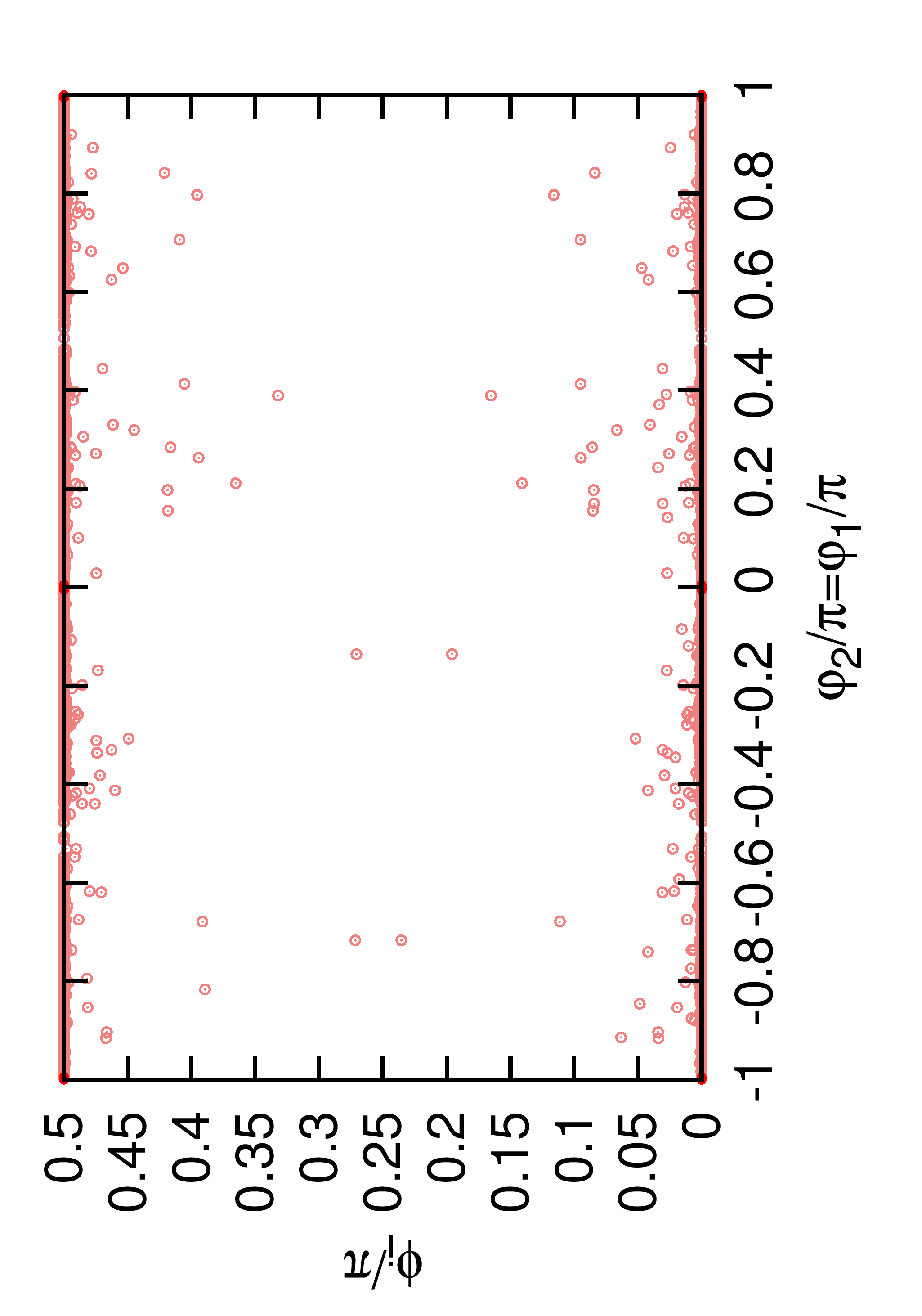}
\caption{The angle $\phi_i$ as a function of various phases of the
 input parameters. All other phases are set to zero. Light colored
 open points are in conflict with the EDM constraints, whereas dark
 colored full points respect the bounds.}\label{fig:cpphase2}
\end{figure}

In \cite{Berge:2014sra,Berge:2008wi,Berge:2008dr,Berge:2011ij} and
\cite{Harnik:2013aja} the possibility to make use of 
the $\tau^+ \tau^-$ decay mode for the determination of the CP properties of a
Higgs boson has been suggested. We therefore investigate here the
coupling of Higgs bosons to a $\tau$ pair. If a mass eigenstate $H_i$
is an admixture of CP-even and CP-odd components, it can feature both
a scalar coupling (denoted by $c_\tau^S$) and a pseudoscalar coupling
(denoted by $c_\tau^P$). The corresponding term in the Lagrangian
reads 
\begin{align}
 \mathcal{L}_{H_i \tau\tau}&=-g_{\tau}\;\tau^+\,\big(c^S_{\tau}+ic^P_{\tau}\gamma_5\big)\,\tau^-\,H_i
\quad\qquad\qquad\qquad\qquad\quad\quad\text{with }
g_{\tau}=\frac{m_{\tau}}{v}\\ 
&=-g_{\tau}\sqrt{(c^S_{\tau})^2+(c^P_{\tau})^2}\,\;\tau^+\,\big(\cos\phi_i+i\sin\phi_i\,\gamma_5\big)\,\tau^-\,H_i
\quad \text{with } \tan\phi_i=\frac{c^P_{\tau}}{c^S_{\tau}}\,. 
\end{align}
In the second line we followed \cite{Berge:2014sra,Berge:2013jra} and
introduced the CP-violating angle $\phi_i$ which parame\-tri\-zes the
CP-mixing of the Higgs boson $H_i$ which couples to the $\tau$. For a
CP-even Higgs boson 
$\phi_i=0$, and for a CP-odd one
$\phi_i=\pi/2$. Figure~\ref{fig:cpphase} shows $\phi_i$
plotted against the mass of the respective Higgs boson. We included
all points of the previously described scans that feature the
next-to-lightest Higgs boson as the 125~GeV Higgs state. The light
colored open points indicate conflict with the EDM constraints, while
dark colored full points respect the EDM
constraints. Figure~\ref{fig:cpphase} shows that $H_1$ and $H_2$
couple mostly scalar-like to the $\tau$ pair, whereas $H_3$ features a
pseudoscalar-like coupling in most cases. For the heavier Higgs states
no such tendency can be observed. Especially for the three heavier
Higgs states there are points that display CP violation in the
$H_i \tau^+ \tau^-$ coupling and that are not in conflict with the EDM
bounds. \s

In the following we investigate by which phases of the input
parameters these CP-violating effects in the  $H_i \tau^+ \tau^-$
coupling can be generated. Figure~\ref{fig:cpphase2} shows the phase
$\phi_i$ plotted versus several phases of the input parameters, namely
$\varphi_1$ (upper left), $\varphi_2$ (upper right), $\varphi_{A_t}$
(lower left) and $\varphi_{1}=\varphi_{2}$ (lower right). Even
relatively small phases of $\varphi_1$ and $\varphi_2$ can generate
considerable CP violation in the $H_i\tau^+ \tau^-$
couplings that should be accessible in the experiment. This is due to
CP violation already at 
tree level in the Higgs sector, that can occur if solely $\varphi_1$ or
$\varphi_2$ are set to non-trivial values. As we already saw earlier
the EDM bounds are exceeded, however, if 
$\varphi_1$ takes on non-trivial values, whereas $\varphi_2$ leads to
scenarios, which respect the bounds. If the CP violation is only
induced by loop effects, as it is the case for $\varphi_{A_t}$ or if
$\varphi_1$ and $\varphi_2$ are chosen equal, hardly any CP-violating
effect is visible in the $H_ i \tau^+ \tau^-$ coupling. The prospects
of measuring such a small CP violation are less good in this case. 

\section{Conclusions
\label{sec:concl}}
Supersymmetric theories feature many new sources for CP violation. In
particular in the NMSSM, CP violation can already be induced at
tree level in the Higgs sector. The upper bounds on the EDMs, on the
other hand, pose stringent constraints on possible CP-violating phases. We
have investigated the allowed ranges for CP-violating phases in the
NMSSM by taking into account the current limits on EDMs and the latest
Higgs data from the LHC. On the one hand the phase $\varphi_1-\varphi_2$,
  which induces the tree-level CP violation in the Higgs sector, is of interest. On the other hand radiative corrections to the Higgs masses are indispensable to
achieve the measured mass value of 125~GeV for the SM-like Higgs
boson. In particular the top/stop sector, which delivers the
dominant Higgs mass corrections, introduces the additional phase
$\varphi_{A_t}$. Furthermore the phase $\varphi_1$ appears on its own 
(not in combination with $\varphi_2$) in the stop sector and also enters
in the chargino sector. \s

Our analysis has shown that the EDMs induced by the NMSSM specific
phase $\varphi_2$, that appears already at tree level in the NMSSM Higgs sector
in contrast to the MSSM, leads to small contributions to the
EDMs. In this case, non-trivial CP-violating phases can still be
compatible with the EDMs. The most stringent constraints on the phases
are then due to the LHC Higgs data, as the possible large CP admixture
in the 125~GeV Higgs boson leads to signal rates that are not
compatible with the experiment any more. On the other hand chargino
contributions to the EDMs through a complex phase of the effective
$\mu$ parameter generate EDMs that are above the experimental constraints. 
The EDM contributions stemming from a non-vanishing phase
$\varphi_{A_t}$ can be important for small values of the phases and
large values of $\tan\beta$. Otherwise, this phase is hardly constrained by
the EDMs. Overall, we observe that the induced EDMs are more important
for larger values of $\tan\beta$. Finally the phases of the gaugino mass
parameters can induce sizeable EDM contributions. We have also
verified, that the CP-violating effects in the  
EDMs decouple with rising masses of the SUSY particles in the
loops as expected. \s

The experimental verification of CP violation in the Higgs sector is a
non-trivial task. Higgs signal rates can be used for the verification
of CP violation only in the simultaneous measurement of decay rates
requiring a dominantly CP-even, respectively, CP-odd admixture in the Higgs mass
eigenstate. We have shown one example where such measurements should be
feasible, although the signal rates are challenging. Besides other
observables, Higgs decays into fermions provide observables sensitive to
CP violation. Taking into account the EDM and LHC Higgs constraints we
have shown, that in particular for the three heavier Higgs bosons
there are scenarios that display CP violation in the Higgs couplings to
tau leptons and are not in conflict with the bounds on the EDMs. Our
investigation of the origin of the CP effects has revealed, that CP
violation induced by loop effects hardly leads to CP-violating Higgs
couplings to fermions. Large CP-violating effects are generated by the
CP-violating phases present already at tree level in the 
NMSSM Higgs sector. While the parameter points for non-vanishing
values of $\varphi_1$ are to a large extent excluded by the EDM
bounds, those for non-trivial $\varphi_2$ values respect the EDM
constraints. In this case, CP violation in the Higgs couplings to $\tau$
leptons may be accessible experimentally.

\subsubsection*{Acknowledgments}
SFK acknowledge partial support from the STFC Consolidated ST/J000396/1 and 
the European Union FP7 ITN-INVISIBLES (Marie Curie Actions, PITN-
GA-2011-289442). The work by RN was supported by the Australian Research 
Council through the ARC Center of Excellence in Particle 
Physics at the Terascale. KW has been supported in part by the
Graduiertenkolleg ``GRK 1694: Elementarteilchenphysik bei h\"ochster
Energie und h\"ochster Pr\"azision''. The authors thank Rui Santos for
helpful discussions. 


\vspace*{0.5cm}

\end{document}